\documentclass[lettersize,journal]{IEEEtran}

\usepackage{cite}
\usepackage{amsmath,amssymb,amsfonts}
\usepackage{algorithmic}
\usepackage{graphicx}
\usepackage{textcomp}
\usepackage{xcolor}
\usepackage{babel}
\usepackage{gensymb}
\usepackage{subcaption} 
\usepackage{tikz}
\usepackage{siunitx}
\usepackage{booktabs}
\usepackage{multirow}
\usepackage{microtype}
\usepackage{wasysym}

\usepackage{eso-pic}

\usepackage{eso-pic}

\newcommand{\FirstPageHeader}{%
  \AddToShipoutPictureFG*{%
    \AtPageUpperLeft{%
      \raisebox{-12mm}{%
        \hspace*{12mm}
        \parbox{\dimexpr\paperwidth-24mm\relax}{
          \centering\footnotesize
          \textcopyright\ 2024 IEEE. Personal use of this material is permitted.
          Permission from IEEE must be obtained for all other uses, in any current or future media,
          including reprinting/republishing this material for advertising or promotional purposes,
          creating new collective works, for resale or redistribution to servers or lists,
          or reuse of any copyrighted component of this work in other works.
          DOI: 10.1109/COMST.2023.3347145.
        }%
      }%
    }%
  }%
}
\newcommand\doublecheck{\textcolor{black}{\checked\kern-1em\checked}}
\def\checkmark{\tikz\fill[scale=0.4](0,.35) -- (.25,0) -- (1,.7) -- (.25,.15) -- cycle;} 

\newcommand{\revision}[1]{{\color{black} {#1}}}
\newcommand{\newrevision}[1]{{\color{black} {#1}}}

\begin{document}

\title{Revolutionizing Future Connectivity: A Contemporary Survey on AI-empowered Satellite-based Non-Terrestrial Networks in 6G}

\author{Shadab Mahboob and Lingjia Liu 
\thanks{S. Mahboob and L. Liu are with Wireless@Virginia Tech, Bradley Department of Electrical and Computer Engineering, Virginia Tech, Blacksburg,
VA, 24060 USA.
The work is supported by US National Science Foundation under grant CNS-2148212. 
The corresponding author is ljliu@vt.edu.}} 

\FirstPageHeader
\maketitle

\vspace{6mm}

\begin{abstract}

Non-Terrestrial Networks (NTN) are expected to be a critical component of 6th Generation (6G) networks, providing ubiquitous, continuous, and scalable services. Satellites emerge as the primary enabler for NTN, leveraging their extensive coverage, stable orbits, scalability, and adherence to international regulations. However, satellite-based NTN presents unique challenges, including long propagation delay, high Doppler shift, frequent handovers, spectrum sharing complexities, and intricate beam and resource allocation, among others. The integration of NTNs into existing terrestrial networks in 6G introduces a range of novel challenges, including task offloading, network routing, network slicing, and many more. To tackle all these obstacles, this paper proposes Artificial Intelligence (AI) as a promising solution, harnessing its ability to capture intricate correlations among diverse network parameters. We begin by providing a comprehensive background on NTN and AI, highlighting the potential of AI techniques in addressing various NTN challenges. Next, we present an overview of existing works, emphasizing AI as an enabling tool for satellite-based NTN, and explore potential research directions. Furthermore, we discuss ongoing research efforts that aim to enable AI in satellite-based NTN through software-defined implementations, while also discussing the associated challenges. Finally, we conclude by providing insights and recommendations for enabling AI-driven satellite-based NTN in future 6G networks.

\end{abstract}


\section{Introduction}
\label{sec:intro}

The Third Generation Partnership Project (3GPP) has already started the standardization towards the 5th Generation (5G)-Advanced in Release 17 and 18 to facilitate its worldwide deployment\cite{5G-adv-1,5G-adv-2}. 
5G-Advanced provides much higher data rates, lower latency, increased capacity, and more efficient spectrum utilization than any of its predecessors. 
It supports a wide range of applications encompassing all 5G use cases such as Ultra Reliable Low Latency Communications (URLLC), massive Machine Type Communication (mMTC), and enhanced Mobile Broadband (eMBB) communication with different Key Performance Indicator (KPI) requirements \cite{5G-adv}.
Nevertheless, future applications such as Augmented Reality (AR), Virtual Reality (VR), Tactile Internet, Holographic Type Communication (HTC), remote health and surgery, etc. require extremely high throughput, low latency, high reliability, and ubiquity at the same time which cannot be met with current technological standards \cite{6g-1}. 
Consequently, the next-generation global wireless standard, namely, 6th Generation (6G) has become the current research focus for the industry and research community \cite{6g-ppf}. 

6G is expected to provide an extremely high data rate (peak data rate up to 1 Tbps and user-experienced data rate up to 10 Gbps, around 100 times higher than 5G), very low latency (in the order of $\mu$s), high reliability (around 100 times better than 5G) and extreme coverage to support the diverse set of future applications \cite{6g-huawei,6g-eric, 6g-samsung}.
Due to the limited coverage area and geographical constraints, it is not possible to guarantee ubiquitous connectivity with existing terrestrial-only network infrastructures. 
Non-Terrestrial Networks (NTNs), networks involving space and aerial platforms, can provide us with multicast opportunities over very large areas as well as can serve users even in remote areas or during times of natural calamities \cite{6g-1}.
Furthermore, the launching and maintenance costs for satellites have significantly decreased as they are deployed at lower heights (typically around 600 km). These satellites can provide much higher throughput and lower latency compared to legacy satellites and potentially can support different use cases of 6G.
So NTN is considered to become one of the major technological enablers of future 6G networks visioning connectivity anywhere and anytime \cite{6g-1,6g-2,6g-3, 6g-4}. 
Tech giants such as SpaceX Starlink, Amazon Kuiper, and OneWeb have already begun to invest billions of dollars in this field, reflecting its massive potential for future growth \cite{6g-sat}.

Although NTN presents numerous potential benefits for the development of future 6G networks, it also entails several challenges that need to be addressed, primarily due to the unique characteristics of its mobility and propagation environments \cite{NTN_1, NTN_2}. 
Due to the long distances between the space-borne Base Stations (BS) and the ground User Equipment (UE), the propagation delay is usually higher in NTN environments. 
Additionally, high-speed air or space-borne platforms necessitate modifications to existing handover and paging protocols, as well as introduce a significant Doppler shift in carrier frequencies. 
The large path loss also increases the minimum power requirement for reliable transmission, initiating the need for novel beam and resource allocation strategies. 
Spectrum sharing in the same frequency band with existing terrestrial or other services also requires further study in order to avoid interference between terrestrial and non-terrestrial users.
Even though currently there are some stand-alone satellite network deployments, the ultimate goal is the convergence of terrestrial and non-terrestrial environments for extreme network performance in 6G. \cite{6g-1}. 
This potential integrated environment requires efficient computing, routing, and slicing algorithms for meeting the expected KPI requirements of 6G. 

Artificial intelligence (AI) is currently having a profound and revolutionary impact on a multitude of industries, including but not limited to healthcare, military, transportation, and e-Commerce \cite{ai}. AI encompasses a wide array of smart machines, while Machine Learning (ML) is a popular subset of AI that allows machines to learn from large amounts of data and make decisions without the need for explicit programming \cite{ai-ml}. Deep Learning (DL) is a special subset of ML that studies Artificial Neural Networks (ANNs) which contain more than one hidden layer, often implemented to simulate the human brain \cite{ai-dl}. DL is currently being leveraged in various applications, such as computer vision, speech recognition, and bioinformatics, outperforming human-level performance in these particular domains.
The cellular domain is still in its infancy in terms of AI integration \cite{ai-wireless} compared to other fields due to the complex and dynamic nature of wireless networks. 
As an integral part of 6G networks, challenges associated with NTN deployment provides an enticing field for AI applications. 
However, while deploying algorithms in a real environment, practical implementation difficulties may arise to provide reliable vertical connectivity between the ground and space networks. 
To reach optimal performance, theoretical advancements in communication system design must be complemented by appropriate AI solutions for NTN integration into 6G.

\begin{figure*}[ht]
  \centering
  \includegraphics[width=1.0\textwidth]{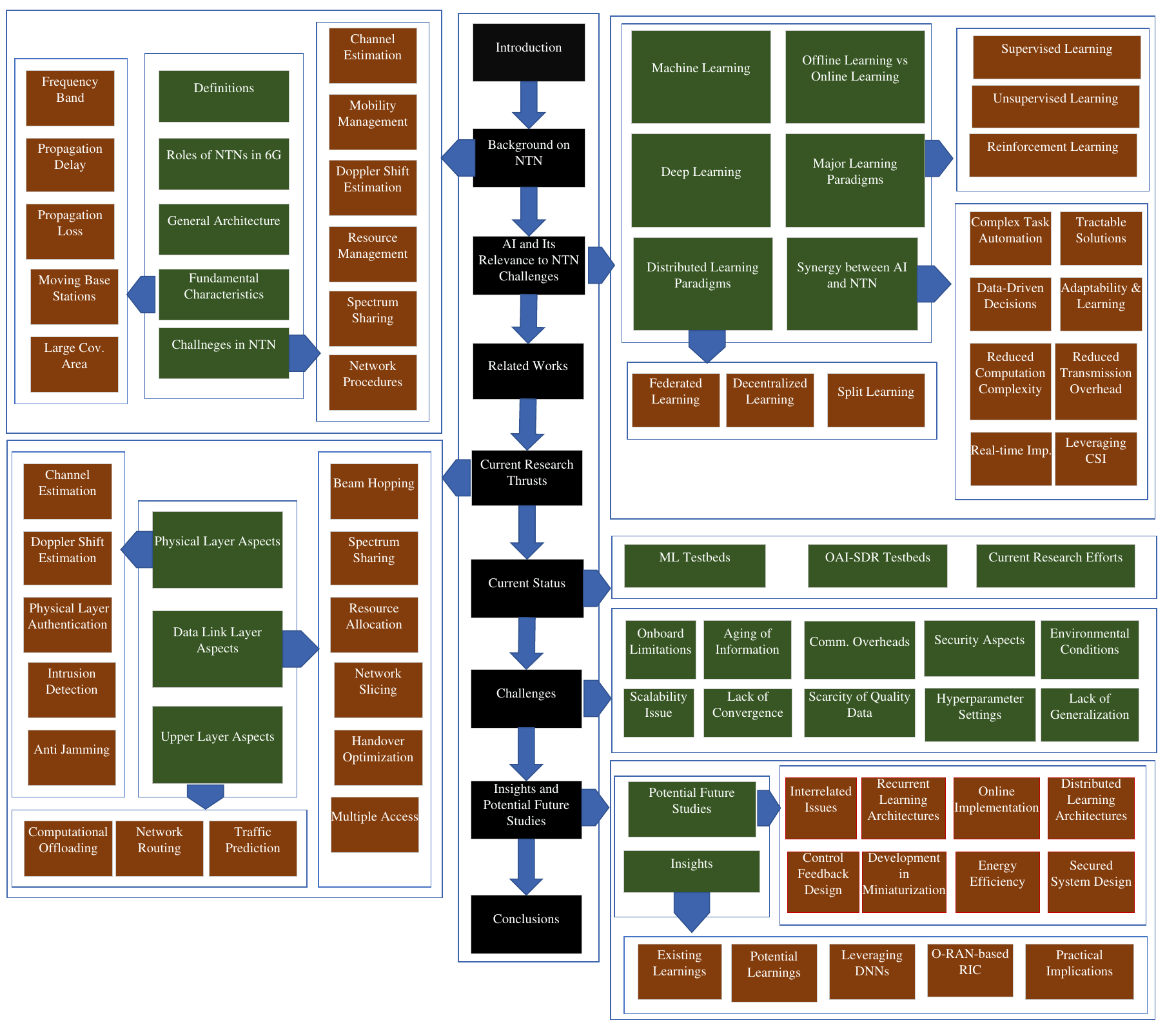}
  \caption{Structure of the paper.}
  \label{fig:structure}
\end{figure*}

\begin{table*}[!htb]
\centering
\caption{List of acronyms.}
\label{tab:acronyms}
\begin{subtable}{.5\linewidth}
\begin{tabular}{p{.2\linewidth}|p{.71\linewidth}|}
    \hline
    \textbf{Acronyms} &\textbf{Definitions} \\ \hline
    3GPP & Third Generation Partnership Project  \\ 
    5G & 5th Generation  \\
    6G & 6th Generation  \\ 
    AC & Actor-Critic  \\
    ACO & Ant-Colony based Optimization  \\
    AI & Artificial Intelligence  \\ 
    ANN & Artificial Neural Networks  \\
    AoA & Angle of Arrival  \\
    AoD & Angle of Departure  \\
    AR & Augmented Reality  \\ 
    ARIMA & Auto-Regressive Moving Average \\
    ARMA & Auto-Regressive Integrated Moving Average \\
    ATM & Asynchronous Transfer Mode  \\
    BS & Base Station  \\
    CCI & Co-Channel Interference  \\
    CDMA & Code Division Multiple Access \\
    CG & Coordinate Graph  \\
    CNN & Convolutional Neural Network  \\
    CSD & Cyclo-Stationary Detection  \\
    CSI & Channel State Information  \\
    CU & Central Unit \\
    DBN & Deep Belief Network \\
    DcL & Decentralized Learning \\
    DDPG & Deep Deterministic Policy Gradient  \\
    DDQN & Double Deep Q-Learning Network  \\
    DL & Deep Learning \\
    DN  & Deconvolutional Network \\
    DoS & Denial-of-Service \\
    DP & Dynamic Programming  \\
    DPG & Deterministic Policy Gradient  \\
    DQN & Deep Q-Learning Network  \\
    DRL & Deep Reinforcement Learning \\
    DRN & Deep Residual Network  \\
    DRQN & Deep Recurrent Q-Learning Network  \\
    DU & Distributed Unit \\
    ED & Energy Detection  \\
    eMBB & enhanced Mobile Broadband  \\ 
    ELM & Extreme Learning Machine  \\
    ESA & European Space Agency \\
    ESN & Echo-State Network \\
    EVD & Eigen Value-based Detection  \\
    FCNN & Fully Connected Neural Network  \\
    FDMA & Frequency Division Multiple Access \\
    FL & Federated Learning \\
    FlexRIC & Flexible RIC \\
    GA & Genetic Algorithm  \\
    GDM & Generative Diffusion Model \\
    GMM & Gaussian Mixture Model \\
    GAN & Generative Adversarial Network  \\
    GNN & Graph Neural Network  \\
    GEO/GSO & Geostationary Earth Orbit  \\
    GNSS & Global Navigation Satellite System  \\
    GPS & Global Positioning System  \\
    GRU & Gated Recurrent Unit  \\
    HAPS & High Altitude Platform System \\
    HARQ & Hybrid Automatic Repeat Request  \\
    HIBS & High-altitude International Mobile Base Station  \\
    HTC & Holographic Type Communication \\
    IC & Integrated Circuit \\
    ICI & Inter-Carrier Interference  \\
    IoT & Internet of Things  \\
    IP & Internet Protocol  \\
    ISL & Inter-Satellite Link  \\
    ITU & International Telecommunication Union  \\
    KKT & KarushKuhnTucker \\
    KNN & k-Nearest Neighbor \\
    KPI & Key Performance Indicator \\
    LEO & Low Earth Orbit  \\
    LoS & Line of Sight \\
    LSM & Liquid State Machine  \\
    LSTM & Long-Term Short Memory  \\ 
        \hline
\end{tabular}
\end{subtable}%
\begin{subtable}{.5\linewidth}
\begin{tabular}{p{.2\linewidth}|p{.7\linewidth}}
    \hline
    \textbf{Acronyms} &\textbf{Definitions} \\ \hline
    MADRL & Multi-Agent Deep Reinforcement Learning  \\
    MAP & Maximum A Posteriori  \\
    MARL & Multi-Agent Reinforcement Learning  \\
    MC & Monte Carlo method \\
    MDP & Markov Decision Process  \\
    MEMS & Micro-Electro-Mechanical Systems \\
    MEO & Medium Earth Orbit  \\
    MiM & Man-in-the-Middle \\
    ML & Machine Learning  \\ 
    MLP & Multi-Layer Perceptron  \\
    MMSE & Minimum Mean Square Error  \\
    mMTC & massive Machine Type Communication  \\ 
    MNL & Minimum Network Load  \\
    MSQ & Maximum Signal Quality  \\
    MST & Maximum Service Time  \\
    mULC & massive Ultra-reliable low-Latency Communication \\
    NGEO/NGSO & Non-Geostationary Earth Orbit  \\
    NN & Neural Network  \\
    NOMA & Non-Orthogonal Multiple Access \\
    NTN  &  Non-Terrestrial Networks \\ 
    OAI & OpenAirInterface \\
    OFDM & Orthogonal Frequency Division Multiplexing  \\
    O-RAN & Open Radio Access Network \\
    OSI & Open Systems Interconnection \\
    OSPF & Open Shortest Path First  \\
    PCA & Principal Component Analysis  \\
    PG & Policy Gradient \\
    PGM & Probabilistic Graph Model \\
    PNN & Probabilistic Neural Network  \\
    PSD & Power Spectral Density  \\
    PSO & Particle Swarm Optimization  \\
    QoE & Quality of Experience  \\
    QoS & Quality of Service  \\
    RC & Reservoir Computing  \\
    RAN & Radio Access Network \\
    RBM & Restricted Boltzmann Machine \\
    RF & Random Forest \\
    RIC & RAN Intelligent Controller \\
    RIS & Reflective Intelligent Surface \\
    RL & Reinforcement Learning  \\
    RMS & Root Mean Square  \\
    RNN & Recurrent Neural Network  \\
    RSMA & Rate-Splitting Multiple Access \\
    RSRP & Reference Signal Received Power  \\
    RSRQ & Reference Signal Received Quality  \\
    RU & Radio Unit \\
    SA & Simulated Annealing  \\
    SAGIN & Space-Air-Ground Integrated Networks \\
    SARSA & State-Action-Reward-State-Action \\
    SDMA & Space Division Multiple Access \\
    SDN & Software Defined Network \\
    SDR & Software Defined Radio \\
    SINR & Signal to Interference and Noise Ratio  \\
    SL & Supervised Learning \\
    SNR & Signal to Noise Ratio  \\
    SoC & System-on-Chip \\
    SOM & Self-Organizing Map \\
    SpL & Split Learning \\
    SVM & Support Vector Machine  \\
    TDD & Time Division Duplexing  \\
    TDMA & Time Division Multiple Access \\
    TNTN & integrated Terrestrial and Non-Terrestrial Network  \\
    UE & User Equipment  \\
    UL & Unsupervised Learning  \\
    ULBC & Ultra-reliable low Latency Broadband Communication \\
    uMBB & ubiquitous Mobile BroadBand \\
    URLLC & Ultra Reliable Low Latency Communication  \\
    V2X & Vehicle-to-Everything \\
    VR & Virtual Reality  \\
    VSAT & Very Small Aperture Terminal  \\
    WMMSE & Weighted Minimum Mean Square Error  \\
    \hline
\end{tabular}
\end{subtable}
\end{table*}

\subsection{Contribution}

\label{sec:intro-cont}

Most of the existing articles either focus on a discussion of architecture and challenges associated with NTN or AI approaches for wireless communications from a broader point of view. 
Although some research articles also discuss the potential research scopes for AI-powered NTNs to some extent, those discussions are either generally not very comprehensive or do not capture the role of AI in NTN-integrated 6G networks in a complete manner. 
Also, the current research efforts and practical complications related to AI-empowered NTN-integrated 6G networks are not covered.
\revision{This survey article aims to provide a comprehensive survey into different AI methods used to overcome the specific challenges of NTN. To help our readers understand better, we also provide a necessary relevant background discussion on NTN and its challenges in the context of 6G. We also discuss different AI approaches and how they can help solve NTN challenges. Additionally, we explore ongoing research efforts and the difficulties of using AI methods in real integrated TNTN setups in 6G.}
The main contributions of this article can be summarized as follows: 

\begin{enumerate}
    \item We \revision{present} a systematic survey of existing and relevant research works in each research thrust to organize the current research progress in these fields. This helps us to get an insight into the current status and potential future research scopes of different relevant research fields in this domain.
    \item We summarize the current AI testbeds for satellite networks and potential integration efforts to current 5G software-defined testbeds for implementing integrated satellite-terrestrial networks. 
    \item We \revision{explore} various practical complications associated with applying AI approaches to NTN as future open issues. This helps us access the maximum potential of AI techniques while being mindful of the practical constraints of NTN integration into next-generation wireless networks.
    \item We \revision{provide} insights and recommendations on various aspects of applying AI techniques to satellite-based NTNs for future 6G networks. 
\end{enumerate}

\subsection{Paper Organization}

\label{sec:intro-org}

The rest of the paper is organized as follows. In Section \ref{sec:ntn}, we provide a compact overview discussion of NTN and its platforms, use cases, architecture, and characteristics; and discuss potential challenges associated with its deployment in 6G. In Section \ref{sec:ai}, we introduce different types of AI approaches to provide a brief overview of relevant AI techniques to solve various challenges associated with NTNs. 
\revision{We then discuss the related surveys on AI-enabled satellite-based NTNs empowering future 6G networks in Section \ref{sec:related}.}
We then summarize the existing AI approaches to address various NTN challenges categorizing them into different NTN research thrusts in Section \ref{sec:ai-ntn}. Furthermore, We summarize the current research efforts from the industrial and research community to \revision{apply AI into} satellite-based NTN in future 6G networks in Section \ref{sec:progress}. 
We also discuss the technical challenges associated with the integration of AI to NTNs in Section \ref{sec:chal}.
Finally, we provide a discussion on insights and \revision{ potential future studies for ensuring the proper } AI-enabled satellite-based NTN in future 6G networks in Section \ref{sec:insights}.
We illustrate the structure of the paper showing the major components in Figure \ref{fig:structure} for better understanding. 
We also provide the list of acronyms in Table \ref{tab:acronyms} for the convenience of the readers. 

\section{Background on NTN}
\label{sec:ntn}

\begin{figure*}[ht]
  \centering
  \includegraphics[width=0.8\textwidth]{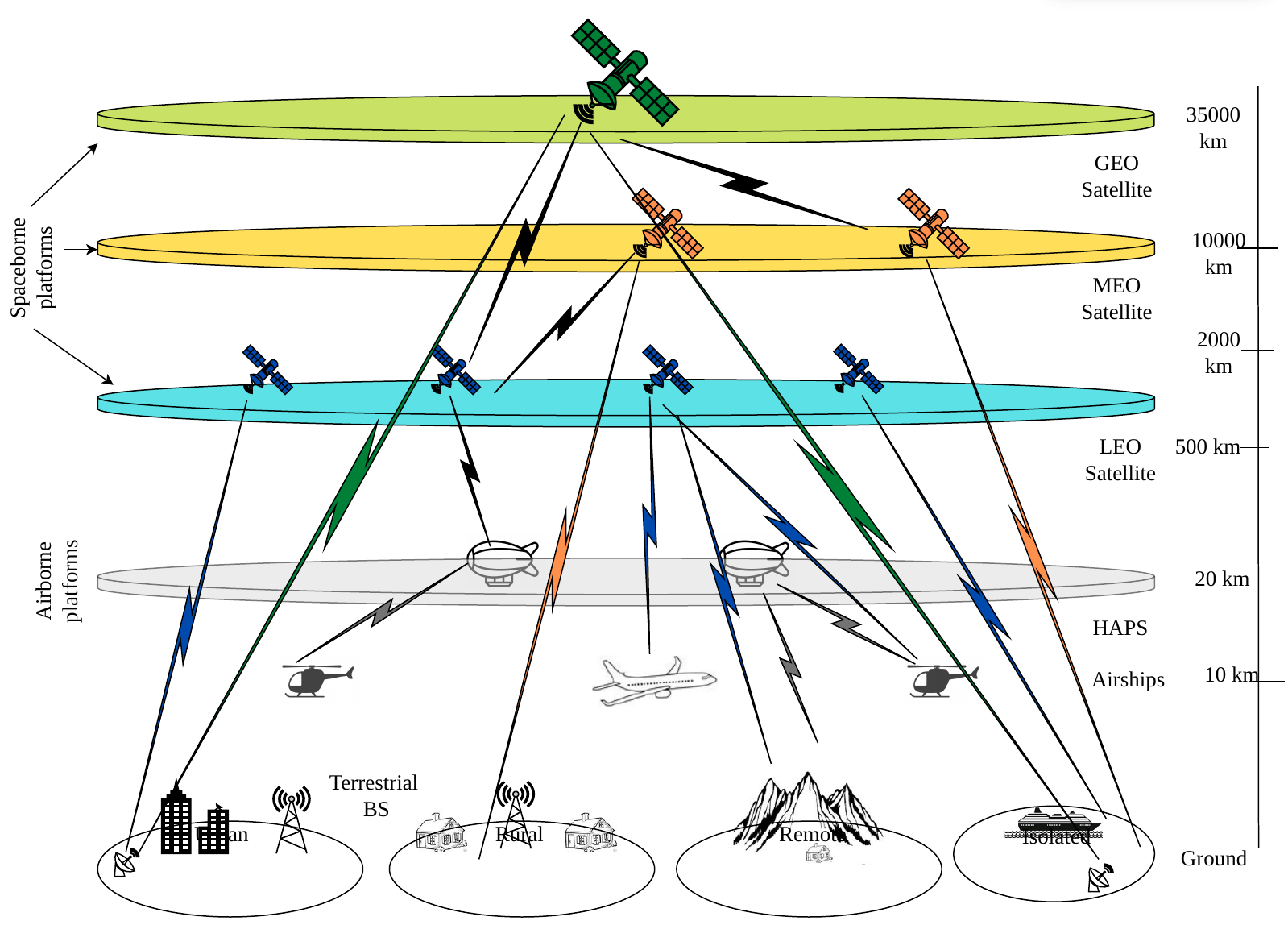}
  \caption{An illustration of different NTN components in 6G.}
  \label{fig:NTN-types}
\end{figure*}

To understand the role of AI in enabling NTNs in 6G, we provide a concise background discussion on NTNs and the challenges associated with NTNs to realize them in 6G in this section.
First, we familiarize the readers with various space and air-borne NTN components along with the general architectures and use cases in 6G.
We clarify that we focus on satellite-based NTN while discussing NTNs for the rest of the paper due to their critical role in enabling 6G with ubiquitous coverage, predictable trajectory, and scalability. 
Then we emphasize on unique characteristics of NTNs which pose new challenges for integrating them into existing terrestrial networks for 6G. 
Depending on the nature of these challenges, we present the current research trends in this domain in \revision{Section} \ref{sec:ai-ntn} by combining them with the AI techniques discussed in \revision{Section} \ref{sec:ai}. 

\subsection{Definition}

\label{sec:ntn-def}

Non-Terrestrial Network (NTN) refers to any network operating through the air or space-borne vehicle(s) for communication \cite{5G-space}. 
This definition implies that two distinct types of NTN platforms (space-borne and air-borne) can be utilized for NTN at different heights which is illustrated in Figure \ref{fig:NTN-types}.

\subsubsection{Space-borne platforms}

Space-borne platforms, such as satellites, are deployed in space for communication \cite{5G-space}.
They move around the Earth in specific orbits with varying angular velocities, relying on gravity to provide the necessary centripetal force to maintain their orbits.
The orbital period of a satellite refers to the time required for the satellite to complete one full revolution around the Earth. 
Due to differences in orbital periods, some satellites may not be visible to ground observers all the time. To characterize this, another term is used to denote the duration of direct visibility for a satellite. 
This is known as the horizon time, which refers to the maximum duration during which the satellite is within the line of sight of a given ground station or receiver.
Depending on their mobility with respect to the Earth, satellites can be classified into two broad categories: Geostationary (GEO/GSO) and Non-Geostationary (NGEO/NGSO) Earth Orbit satellites. 
We discuss these two types of satellites below and summarize their key features in Table \ref{tab:ntn-types}.

\begin{itemize}
    \item \textbf{Geostationary Earth Orbit (GEO or GSO) Satellites:} These satellites have an orbital period of 24 hours which is the same as the time required for the Earth to complete a full rotation on its axis. 
    As a result, these satellites appear stationary from the ground and are named Geostationary Earth Orbit (GEO or GSO) Satellites. 
    These satellites orbit on the Earth’s equatorial plane at an altitude of about 35,786 km to maintain this orbital period.
    Due to this high altitude, it has an extremely large beam footprint (typically the diameter ranges from 200 to 1000 km) covering a pretty wide area. 
    However, it also incurs an extremely long propagation delay (typically around 270 ms)  \cite{3gpp1} which makes it infeasible for low-latency communications.
    These satellites have been used in broadcasting services for a very long time, but are not very suitable for low-latency emerging applications.
    
    \item \textbf{Non-Geostationary Earth Orbit (NGEO or NGSO) Satellites:} As the name suggests, these satellites orbit around the Earth at a period lower than 24 hours, so they are not stationary with respect to a ground observer.  
    As the orbital period is smaller, their angular velocity is also higher but the altitude is lower compared to GEO satellites.
    Depending on the heights, they can be divided into two categories: Low Earth Orbit (LEO) and Medium Earth Orbit (MEO) satellites.
    Typically they are deployed at a height ranging from 200 to 2000 km for LEO and 2000 to 25000 km for MEO satellites.
    The horizon time is much smaller for NGEO satellites due to smaller orbital periods, for example, the LEO satellites deployed at a height of around 500-600 km with an orbital period of 1.5-2 hours can have a horizon time of 5-10 minutes depending on channel conditions.  
    Due to smaller heights, these satellites have a smaller beam footprint (diameter ranges from 5 to 500 km) with a much smaller propagation delay (typically around 20 ms for LEO satellites and 94 ms for MEO satellites)\cite{3gpp1} compared to GEO satellites.
    With their proximity to Earth and lower cost of launch and maintenance, these satellites, especially the LEO satellites, have gained significant attention in recent years. 
    Their reduced propagation delay and path loss make them an attractive choice for facilitating high-speed data transfer and real-time communication, so as to transform the future 6G connectivity.
\end{itemize}

\begin{table}[htb]
\centering
\caption{Key features of different types of satellites.}
\label{tab:ntn-types}
\begin{tabular}{|p{27mm}|p{15mm}|p{15mm}|p{15mm}|}
    \hline
    \textbf{Attribute} &\textbf{GEO} &\textbf{MEO} &\textbf{LEO} \\ \hline
    Orbital height (km)   &  35786 & 2000-25000 & 200-2000\\ \hline
    Typical diameter of beam footprint (km) & 1000 & 500 & 100\\ \hline
    Propagation delay (ms) & 270 & 94 & 12-25 \\ \hline
    Orbital period (hours) & 24 & 12 & 1.5-2 \\ \hline 
    Horizon time & 24 hours & 1-2 hours & 5-10 minutes\\
    \hline 
\end{tabular}
\end{table}

\subsubsection{Air-borne platforms}

High Altitude Platform Systems (HAPS) refer to air-borne platforms that can be used for wireless communication. Airships, balloons, and airplanes are the most prominent types of air-borne platforms in NTN. They are viewed as air-borne counterparts of terrestrial base stations serving as High-altitude International Mobile Base Stations (HIBS) \cite{haps}. They usually operate at the stratosphere region with an altitude of around 20 km and a beam footprint size with a diameter of several km. Despite it having a lot smaller propagation delay compared to space-borne platforms, it has some additional challenges related to stabilization on air and refueling.

While both satellites and airborne platforms can be utilized in the development of NTNs, satellites are often considered more critical for discussions related to NTNs. This is due to their global coverage, stable and predictable orbits, high scalability, and the existence of international regulations that govern satellites. As such, satellite networks comprise a significant portion of future NTN-enabled communication networks. Therefore, for the purposes of this article, we will primarily focus on satellite-enabled NTNs in the context of 6G communication technology. 

\begin{figure*}[ht]
  \centering
  \includegraphics[width=1.0\textwidth]{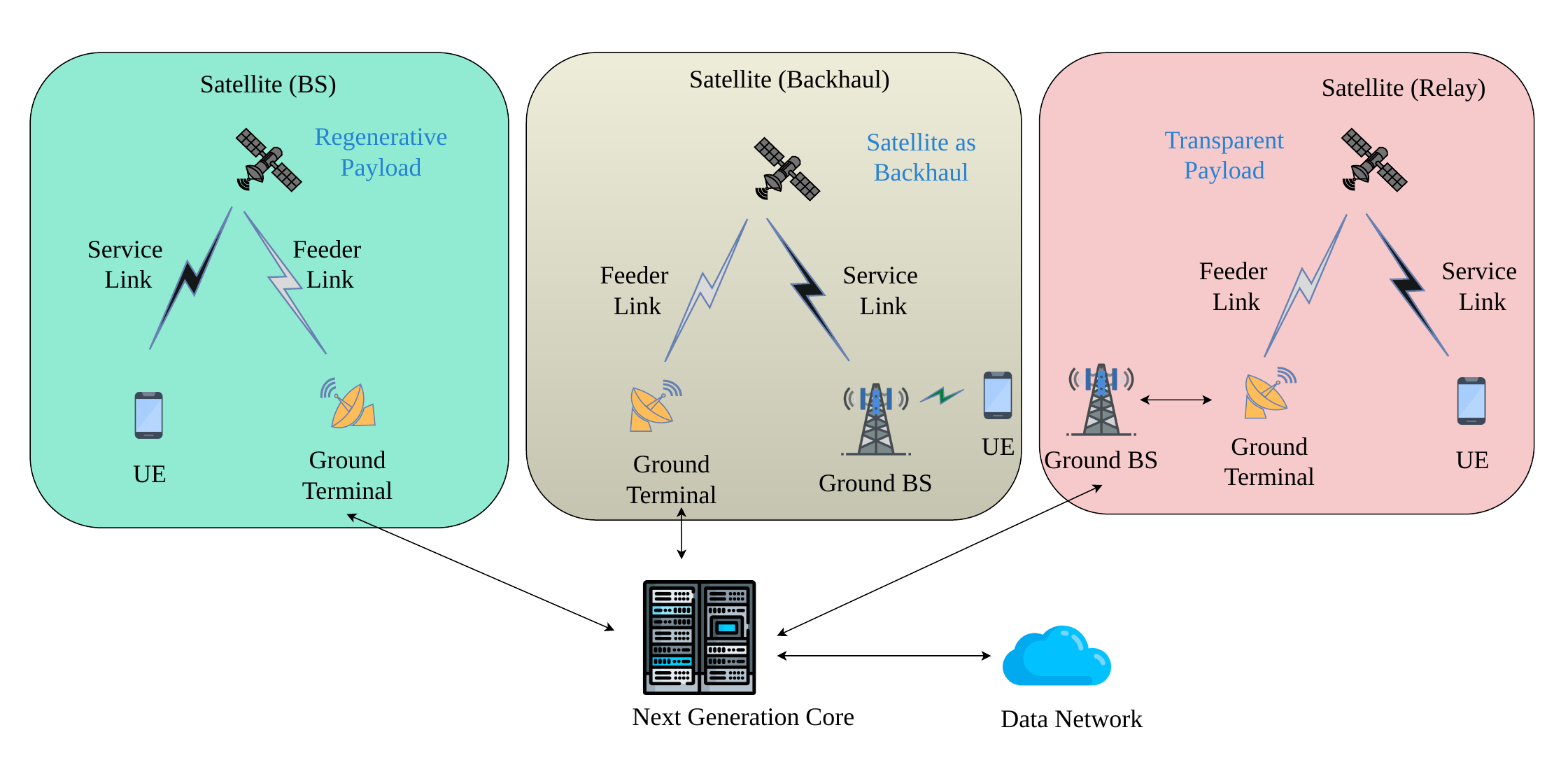}
  \caption{General communication architecture for satellite-based NTN.}
  \label{fig:NTN-arch}
\end{figure*}

\subsection{\revision{Role of NTN in 6G}}

\label{sec:ntn-use}

NTNs are anticipated to be a major component of 6G communication systems, providing a wide array of vertical services, such as transport, health, energy, automotive, public safety, and many more. The International Telecommunication Union (ITU) has identified three major categories of applications for 5G that are based on network performance and user Quality of Experience (QoE): (1) eMBB: extremely high bandwidth with moderate latency requirements, for example, multimedia applications; (2) mMTC: low power and bandwidth and no strict delay requirements, for example, IoT; and (3)  URLLC: low latency and high-reliability requirements, for example, remote medical surgery. However, future applications such as AR, VR, Tactile Internet, HTC, intelligent transport and automation, multi-sense communication, global ubiquitous connectivity, etc. require extremely high throughput, low latency, high reliability, and ubiquity at the same time which cannot be met with current 5G standards \cite{6g-1}. Based on the characteristics, these new applications are classified into three more new groups, 

\subsubsection{Ubiquitous MBB (uMBB)}
High throughput and extreme coverage requirements, combining both eMBB and mMTC. Examples: Digital twins, pervasive intelligence, global ubiquitous connectivity, etc. 

\subsubsection{Ultra-reliable low Latency Broadband Communication (ULBC)}
High throughput and low latency requirements, combining both eMBB and URLLC. Examples: HTC, AR, VR, Tactile Internet, multi-sense experiences, etc. 

\subsubsection{massive Ultra-reliable low-Latency Communication (mULC)}
Extreme coverage and low latency requirements, combining both mMTC and URLLC. Example: Vehicle-to-Everything (V2X), intelligent transport and automation, etc. 

The principal strength of NTNs lies in their extreme coverage. 
\revision{As discussed in Hexa-X project \cite{hexax1}, a flagship for B5G/6G vision and intelligent fabric of technology enablers connecting human, physical, and digital worlds, the vision of enabling 6G networks towards provisioning service everywhere and always through NTN is presented.}
Due to its extreme coverage, satellites can reach underserved or unserved areas such as islands, remote locations, ships, airplanes, etc. where terrestrial communication is either difficult or impossible to some extent. 
In times of natural disaster, terrestrial links can be unavailable, in which case users can benefit from the reliable backup of non-terrestrial links. 
This ensures resilient and robust communication with global connectivity which is considered to be one of the main features of future 6G networks. 
With the advancements in antenna techniques and miniaturization, high throughput satellites are also deployed in low earth orbits. 
\revision{Consequently, current 5G use cases such as mMTC and eMBB as well as future 6G use cases such as uMBB can be the most important use cases for NTNs.}
Furthermore, the considerably low latency for LEO satellite systems makes the satellite useful even for low-latency applications. 
\revision{However, 5G URLLC or 6G new use cases with extremely low latency may not be directly applicable for NTN use cases. 
Nevertheless, NTNs can still be beneficial for these use cases in conjunction with terrestrial networks to improve network efficiencies and reliability.}
Combining all these, satellites are expected to be one of the major driving forces toward revolutionizing the future 6G applications extensively. 

\subsection{General Architecture}

\label{sec:ntn-arch}

\revision{Satellites can employ a transparent payload configuration, acting as a relay that performs RF filtering, frequency conversion, and amplification to facilitate communication between UEs and ground stations. 
Alternatively, they can utilize a regenerative payload configuration, which involves payload processing after modulation and coding, and act as base stations with additional onboard processing capabilities. 
Besides, the satellites can provide backhaul support for the core networks of terrestrial networks. 
The general architecture for a satellite-based NTN for the above-mentioned different configurations as per release 16 and 17 is discussed below \cite{3gpp1, 3gpp3}:}

\begin{enumerate}
    \item \textbf{Satellite:} Satellite is the key component of this architecture. 
    It carries the payload between the UE and the ground station as shown in Figure \ref{fig:NTN-arch}. 
    In the case of a transparent payload, it works as a simple relay that transmits the payload after RF filtering, frequency conversion, and amplification to the ground station (or UE). 
    Conversely, in the case of a regenerative payload, it processes the payload after modulation and coding on top of these actions, so it works like a BS that needs onboard processing capabilities.
    \revision{Also as per \cite{3gpp3}, satellites can provide backhaul by providing a connection between ground BS and the core network as illustrated in Figure \ref{fig:NTN-arch}.}
    \item \textbf{Gateway:} Gateway refers to the ground station that connects NTN to the public data network. 
    In the case of a transparent payload, the ground terminal needs to be equipped with a terrestrial base station. 
    In the case of a regenerative payload \revision{and satellite backhaul support}, the ground terminal only relays the received information to the core networks.
    \item \textbf{User Equipment (UE):} User equipment is either handheld or Very Small Aperture Terminal (VSAT) within the coverage area of the satellite.
    \item \textbf{Feeder Link:} Feeder link connects a satellite to the gateway.
    \item \textbf{Service Link:} Service link connects UEs to the serving satellite. 
    \item \textbf{Inter-Satellite Links (ISLs):} ISLs provide connectivity between multiple satellites deployed in NTN so that a payload can be delivered to other cells.
\end{enumerate}

\subsection{Fundamental Characteristics of NTN}

\label{sec:ntn-feat}

NTN presents us with several potentially promising use cases for the next generation of wireless networks as discussed in \revision{Section} \ref{sec:ntn-use}. However, as can be seen from \revision{Section} \ref{sec:ntn-def}, it also has a number of unique characteristics due to the large distances between satellites and ground transceivers, the high mobility of NGEO satellites, and the proposed frequency range for operation. In this subsection, we will delve into these features of NTN and discuss their impact on network performances and procedures.

\subsubsection{Target Frequency Band}

The allowable frequency range of operation is 0.5-100 GHz \cite{spec-shar}. Traditionally, six major frequency bands within this range are used in satellite communications, which are listed below:

\begin{enumerate}
    \item \textbf{L-Band (1-2 GHz):} Global Positioning System (GPS) carriers and satellite mobile phones, e.g. Iridium, use this band.
    \item \textbf{S-Band (2-4 GHz):}  This frequency band is used for weather radar, marine radar, etc. satellite communications.
    \item \textbf{C-Band (4-8 GHz):} Primarily used for satellite television broadcasting.
    \item \textbf{X-Band (4-8 GHz):} Primarily used in military communications.
    \item \textbf{Ku-Band (12-18 GHz):} Primarily used for satellite broadcasting services.
    \item \textbf{Ka-Band (26-40 GHz):} This frequency band is used for high-speed data transmission, including broadband internet access via communication satellites.
\end{enumerate}

However, as per 3GPP, currently, two frequency bands (S and Ka-band) are targeted in particular for integrated TNTN environments considering performance and regulatory concerns \cite{3gpp1}. The two target frequency bands are: 

\begin{itemize}
    \item \textbf{S-Band:} The downlink frequency band is 2170-2200 MHz and the uplink frequency band is 1980-2010 MHz.
    \item \textbf{Ka-band:} The downlink frequency band is 19.7-21.2 GHz and the uplink frequency band is 29.5-30 GHz.
\end{itemize}

\subsubsection{Propagation Delay}

Propagation delay is the time duration taken for a signal to reach its destination. For communication signals, we can calculate the propagation delay for a signal by using the equation: $t = \frac{d}{c}$ where $d$ is the distance between the source and destination and $c = 3 \times 10^8$ m/s is the speed of light. Considering the speed of light as a constant, we observe the propagation delay for a signal is proportional to the propagation distances. Satellites are located very far from the surface of the earth as discussed in \revision{Section} \ref{sec:ntn-def}. Consequently, the propagation delay is going to be extremely large for NTNs. The GEO, MEO, and LEO satellites can have a one-way propagation delay of about 270, 94, and 20 ms respectively as shown in Table \ref{tab:ntn-types}. These values are much larger, especially for GEO and MEO cases compared to conventional terrestrial networks, which generally have a very negligible propagation delay of around a few $\mu s$ \cite{3gpp1}. This extended propagation delay has an effect on different network procedures and performances for communication systems. 

\subsubsection{Propagation Loss}

The propagation loss, or path loss, refers to the reduction in power density that an electromagnetic signal experiences as it travels through space. The most significant component of this path loss is the free space path loss, which is proportional to the distance between the source and destination and the frequency of the signal \cite{ntn-fspl}. For NTNs, this free space path loss is much higher (around 60-120 dB) than it is for terrestrial networks, due to the greater distances between satellites and the use of higher carrier frequencies. In fact, the Ka-band is not suitable for GEO satellites, as it does not meet the minimum link budget for them. The basic path loss component also includes shadow fading \cite{ntn-sf}, as with traditional terrestrial networks.

\begin{figure}[ht]
  \centering
  \includegraphics[width=0.5\textwidth]{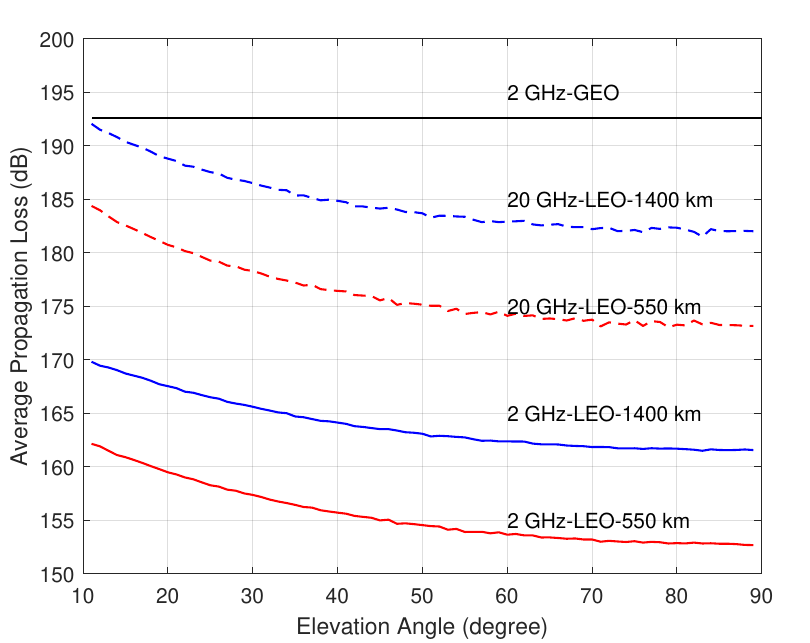}
  \caption{Propagation loss for satellites at different heights and with different carrier frequencies. }
  \label{fig:pl-satellite}
\end{figure}

In addition to that, there is attenuation due to atmospheric gases that depends on frequency, elevation angle, altitude above sea level, and water vapor density\cite{ntn-atm}. Another important component is attenuation due to rain and fog, which is typically significant for frequencies above $6$ GHz \cite{ntn-rf}. Additionally, scintillation corresponds to rapid fluctuation in amplitude and phase of the propagating signal in the ionosphere (for Sub-$6$ GHz) and troposphere (for above $6$ GHz) \cite{ntn-sci}. Depending on different scenarios, either flat fading based on ITU two-state model \cite{ntn-ts} or fast fading \cite{3gpp1} can be considered. The average propagation loss for different types of satellites in different frequency bands is illustrated in Figure \ref{fig:pl-satellite}. This high path loss necessitates the need for efficient power allocation strategies in NTN.

\subsubsection{Moving Base Stations}

As discussed in \revision{Section} \ref{sec:ntn-arch}, for regenerative payloads, satellites can be used as base stations for improved network performances. The terrestrial base stations are located at fixed locations. As the GEO satellites do not change their relative positions with respect to the ground terminal, they appear static in nature with respect to the earth's surface, so the scenario is similar to terrestrial ones. However, the scenario is very different for NGEO satellites where they need to maintain a lower height and higher angular velocity compared to GEO satellites as discussed in \revision{Section} \ref{sec:ntn-def}, so they do not appear static from the earth's surface. Due to the dynamic nature of NGEO satellites, they turn into moving base stations in case of regenerative payloads. Due to this high-speed movement of NGEO satellites, different mobility issues arise for NTN platforms. 

\subsubsection{Coverage Area}

One of the most important features of the satellites is the large beam footprint associated with them due to their long distances from the earth's surface. This enables the network coverage of very large areas compared to the coverage area of terrestrial counterparts. It provides us with ubiquitous network coverage including remote, even isolated areas. However, this also creates the necessity for modifications in existing timing and synchronization procedures for conventional terrestrial networks. The cell area is much larger, so the UEs situated at the farthest side of the cells experience a larger delay compared to the UEs situated closer to the satellites \cite{3gpp1}. So the timestamps for different network procedures need to be modified according to the distances of the users as we will see in the next subsection. 

\subsection{Challenges Associated with NTN}

\label{sec:ntn-chal}

NTN offers a range of unique features due to the large distances between the transceivers and the high mobility of NGEO satellites, as outlined in \revision{Section} \ref{sec:ntn-feat}. These features open up possibilities for new use cases, taking advantage of the extensive coverage offered by the satellites. The high mobility of the satellites also allows for the deployment of satellites across the globe to provide global network coverage. However, NTN also presents a number of new challenges that must be tackled due to these characteristics, which are discussed in detail below:

\subsubsection{Channel Estimation}

In wireless communications, Channel State Information (CSI) refers to the information which represents the state of a communication channel between the transmitter(s) and the receiver(s); the process of obtaining this information is known as channel estimation. By having access to CSI, it is possible to adjust transmissions to the current channel conditions, which is essential for achieving reliable communication with high data rates in multi-antenna systems with effective channel resources and interference management. There are numerous advanced approaches, such as Maximum Likelihood estimation and Minimum Mean Square Error (MMSE) estimation, for effective channel estimation in traditional terrestrial cellular networks. Nevertheless, these methods are not well-suited for NTN, particularly for LEO satellites due to the inherent time-variant nature of the satellite communication channels. LEO satellites usually move from horizon to horizon in approximately 5-10 minutes, so a UE remains within the coverage of a specific LEO satellite for a very short time period. Furthermore, the propagation delay for satellite networks, especially in the case of GEO satellites, is considerably larger (250 ms RTT) in comparison to general terrestrial networks. Therefore, the CSI estimated by the LEO satellites can be outdated \cite{3gpp2}. Because of these reasons, the CSI estimation in NTN necessitates new efficient techniques in addition to the traditional terrestrial estimation methods.

\subsubsection{Mobility Management}

Since an NGEO satellite operates at a lower altitude, the coverage area of each NGEO satellite is smaller than that of a GEO satellite. Typically around 5-20, NGEO satellites form complex mega-constellations to sustain global coverage across the earth. The NGEO satellite needs to move at a much higher speed than the earth's rotational speed (can be up to around 7.8 km/s) to get the necessary centripetal force to move around the earth at that low altitude. As a result, these satellites typically orbit around the earth pretty fast (usually within around 2-10 hours) as discussed in \revision{Section} \ref{sec:ntn-def}. This quick orbital motion poses a great challenge for integrating NGEO satellites into traditional wireless communication systems.  Due to the smaller orbital period, any specific terrestrial UE can be only visible to an NGEO satellite for a very short span of time, typically several minutes. So the UE needs to undergo multiple handovers within a short span of time interval regardless of its mobility \cite{NTN_2}. If the satellite covers an area using multiple spot beams, the scenario is worse because the spot beam is much smaller compared to a total coverage area of an NGEO satellite. So the UEs need to through multiple (beam) handovers within a few minutes, even when they are stationary, for seamless continuation of data sessions. This frequent handover phenomenon in NGEO satellite networks creates a lot of overhead in communication channels, leading to an overall degradation in network performance.
 
\subsubsection{High Doppler Shift}

Doppler shift is the shift in the signal frequency due to the motion of the transceivers. In the case of NGEO satellites, satellites are moving at a very high speed under a specific constellation. Due to this relative motion between UEs and satellites, Doppler shift happens in the original signal frequency. Due to frequency offsets, UEs tune to different carrier frequencies than the original carrier frequencies. So the frequency synchronization is lost, and the UEs may interfere with the other users. This is known as Inter-Carrier Interference (ICI) between multiple UEs. Generally, even for the high mobility scenarios in terrestrial networks, the frequency shift is pretty negligible, and so is the Doppler shift. However, the frequency offset is pretty significant in NTN due to the much higher speed of the NGEO satellites. The Doppler shift value mainly depends on the carrier frequency and height of the satellites. For NGEO satellites operating at Ka-Band, the Doppler shift can go from 225 kHz to even 720 kHz \cite{3gpp1} depending on the heights. This can cause significant ICI among NTN users which requires efficient strategies for compensation of the Doppler effect. 

\subsubsection{Resource Management}

Spectrum and power are the two fundamental resources for any communication system. In NTN, the allocation of these two resources becomes an even more complex problem due to the high path loss and limited spectrum availability. As discussed in \revision{Section} \ref{sec:ntn-feat}, the path loss associated with Non-Terrestrial Networks is much higher compared to terrestrial networks. To correctly decode the transmitted symbols from the received signals, the received signal needs to meet the minimum RSRP requirement. That means the transmitted signal power needs to be much higher (typically at least 10 times the terrestrial transmitted signals) than terrestrial signals. This poses a great obstacle for traditional UEs as they have power limitations. Furthermore, the target frequency bands as discussed in \revision{Section} \ref{sec:ntn-feat} for NTN are limited. To support a large number of satellite UEs, this spectrum resource appears to be scarce in NTN systems. So efficient resource (spectrum and power) allocation strategies are needed for integrating NTN into terrestrial networks. 

\subsubsection{Spectrum Sharing}

As discussed in \revision{Section} \ref{sec:ntn-feat}, the S-Band and Ka-Band are the target bands for NTN. On top of this limited spectrum allocation, we have interference from terrestrial users in these bands. In S-Band, we already have existing terrestrial communication from 4G LTE devices. With the advent of mm-wave technology, terrestrial communication is also using the Ka-band in 5G. So the satellite users will suffer from co-channel interference with the terrestrial users in both bands. To avoid this interference, we have to come up with efficient spectrum-sharing techniques to put the interference below a certain threshold ensuring proper decoding of the received signals. 

\subsubsection{Effect on Network Procedures}

Timing advances ensure synchronous uplink transmissions for all UEs. The UEs can be located at different distances from the gNB, so there is a differential propagation delay between different UEs. If the uplink reception is not synchronized, the gNB needs to make sure the allocation of resource blocks to a specific UE does not include the resource blocks already in use by other UEs, which is inefficient in terms of resource allocation. Due to the long propagation delay, the TA is much larger than the transmission time slots in NTN compared to NR. Also due to the mobility of LEO satellites, the delay is time-varying and TA needs dynamic updates for proper uplink alignments. The other processes affected by the long propagation delay are Random Access, Hybrid Automatic Repeat Requests (HARQ) procedures, etc \cite{3gpp1}. These procedures need to be modified properly to compensate for the long propagation delay. 

\subsubsection{Network Aspects}

On top of all these challenges, integration into existing terrestrial networks comes with several open research issues to be addressed. Computational offloading, which involves transferring the computational burden to satellite networks for supporting devices with low computing power, particularly in IoT applications, gets complicated due to extended propagation delay and highly mobile NGEO satellites. Network routing has been studied for a long time, and network slicing has been discussed since the implementation of 5G. However, with the emergence of NTN, the integration of terrestrial networks calls for research in this area with new effective strategies. The ever-changing network topology of mobile NTN platforms makes it challenging to solve these problems in a complex environment. 

\vspace{3mm}
\noindent\textit{Key Takeaways: 
We note that satellite-based NTNs can be extremely useful to provide ubiquitous connectivity, service continuity, and extreme reliability for diverse future 6G applications. 
Nevertheless, the extreme nature of the satellite networks, e.g., long distance between transceivers, high mobility for NGEO satellites, spectrum sharing with existing services, and high propagation loss, etc. impose a highly challenging environment to address for the research community.   
These challenges also open a new door for AI applications to move toward the future 6G revolution. 
In the following section, we discuss how AI can be incorporated so that we can address the issues for potential TNTN integration for future 6G networks.
}

\section{AI and Its Relevance to NTN Challenges}

\label{sec:ai}

AI refers to the simulation of human intelligence processes (e.g. visual reception, speech recognition, computer vision, etc.) by machines, especially computer systems. This human-level cognitive ability is achieved through either some predefined algorithms or learning from data-based approaches \cite{ai}. Many practical systems are very diverse and complex. The rule-based approaches are not very feasible for these systems because of an enormous number of scenario possibilities. As a result, the learning-based approaches show a lot more promise compared to predefined approaches in these types of real systems. As our focus for this paper is mostly on NTN which has an extremely complex and time-variant topology, we focus on learning-based approaches when we consider AI. In this section, we give an overview of these approaches to get an intuition of how these approaches can be useful in solving NTN issues discussed in the next section. 

\subsection{Machine Learning (ML)}

\label{sec:ai-ml}

Machine Learning (ML) is a special subset of AI approaches where machines learn algorithms to perform a task by generalizing from past experiences or historical data without being explicitly programmed for it \cite{ai-sam}. 
The performance of human intelligence processes can be improved with each iteration evolving through new feature extractions in ML approaches. 
Generally speaking, each ML approach has three distinctive features, namely task, performance measure, and experience \cite{ai-mit}. 
A machine is first assigned to learn to perform a specific task. 
It starts with a model with an initial set of random parameters. 
Then at every iteration, the model is recalculated based on some performance measures, essentially representing the learning process. 
Thus utilizing experiences, it can learn how to perform the task properly which is the main goal of ML approaches.



A generic ML model works in three phases utilizing various components \cite{ai-ml} - Pre-Training Phase, Training Phase, and Testing Phase. We discuss the fundamental components of these three phases as shown in Figure \ref{fig:gen-ml} below:

\begin{figure}[ht]
  \centering
  \includegraphics[width=0.5\textwidth]{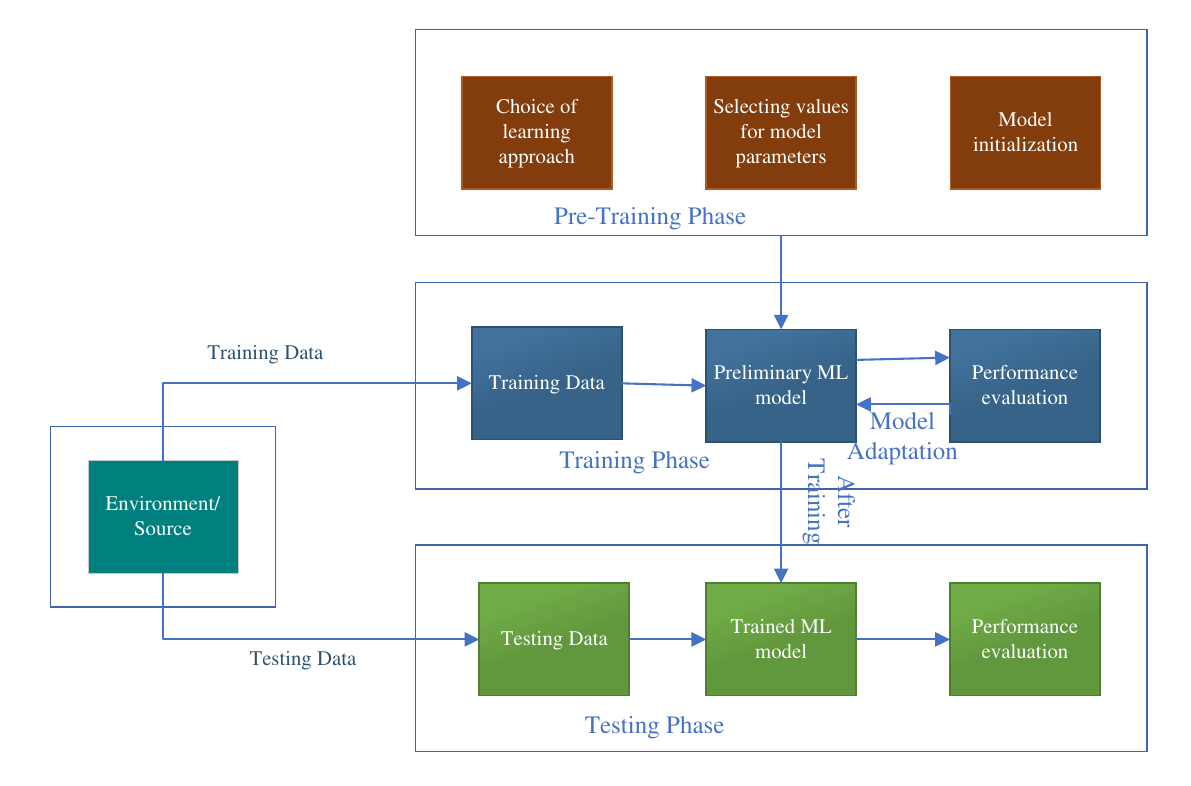}
  \caption{Generic ML model.}
  \label{fig:gen-ml}
\end{figure}

\subsubsection{Pre-Training Phase}

The Pre-Training Phase includes the choice of learning approach along with the necessary model initialization.   
The selection and design of the learning approach greatly depend on the nature of the problem for the learning systems. We show in the next few subsections different learning strategies for different problems. Each ML model generally requires some initial set of parameters and initialization that need to be carefully tuned to achieve expected performances. 

\subsubsection{Training Phase}

After setting up the preliminary model with initialization, the most important phase -- training begins. The training data is provided as input to the initial model. Typically the raw data collected for a specific problem may not be properly structured to be used for the model. Moreover, these data may contain redundant and unnecessary information which is not beneficial for learning the model. Consequently, data needs to be preprocessed in a suitable manner to have good performance. The features also need to be chosen in such a way that they can capture the correlation for empowering the learning process. The output of the model is fetched for performance evaluation. Based on the feedback from the evaluators, the model is adapted to improve its performance. This whole learning process is known as `training'. 

\begin{figure}[ht]
  \centering
  \includegraphics[width=0.5\textwidth]{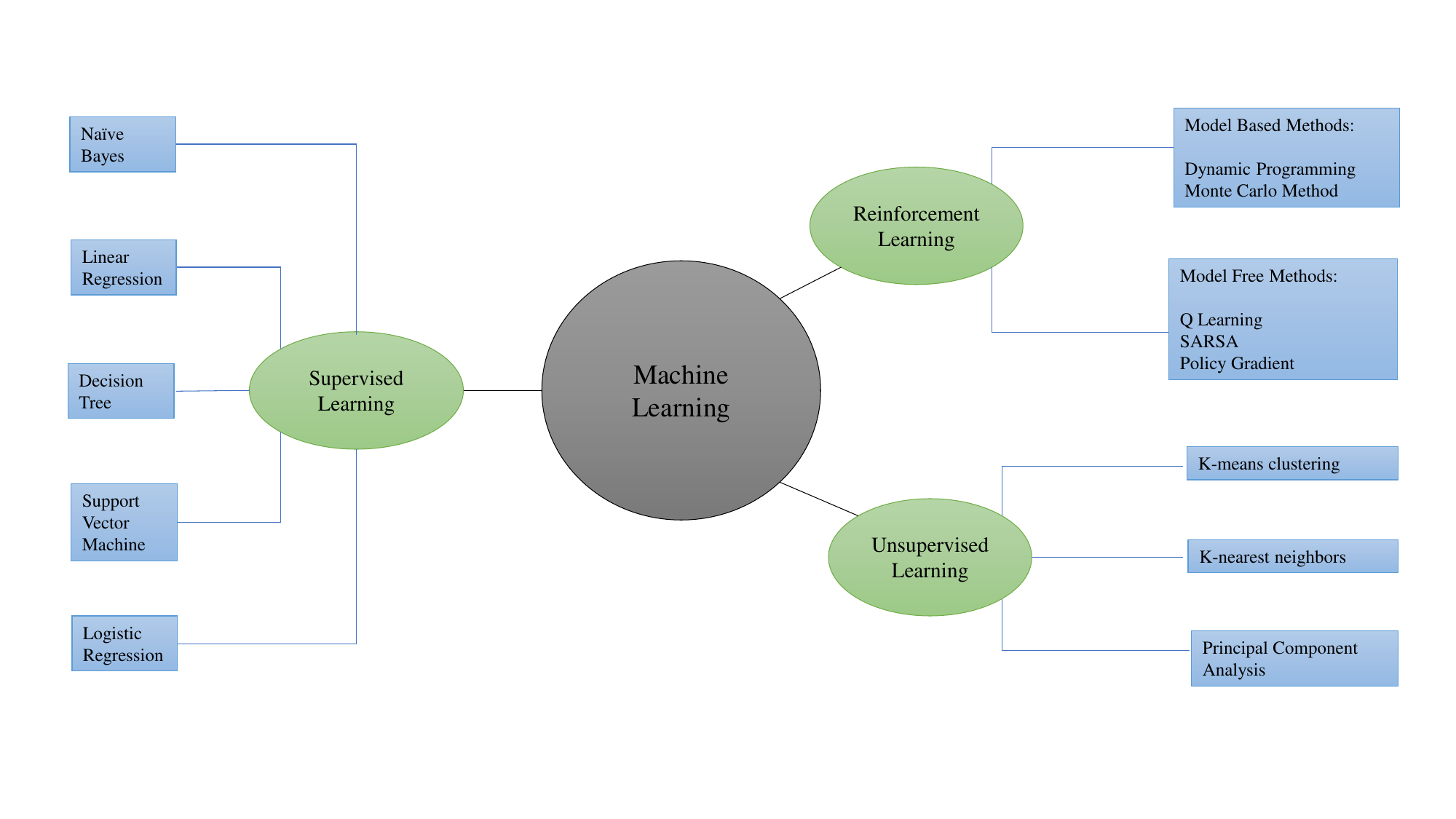}
  \caption{Taxonomy of ML approaches.}
  \label{fig:ml-types}
\end{figure}

\subsubsection{Testing Phase}

After the training, we have a trained ML model based on our provided data. This model can be used to later evaluate in the real environment. Similarly, as training data, testing data can be generated and preprocessed for evaluation. The performance evaluator provides the accuracy of the model using the testing data as inputs. This whole process is known as `testing'.
In the case of offline learning, the testing phase starts once the training is done. 
On the other hand, in the case of online training, the testing is generally executed in a parallel manner with training. 

\revision{
\subsection{Offline vs Online Learning}

\label{sec:ai-online}

Depending on the training approach, learning can be either offline or online. 
In the case of offline training, training data is generated in the pre-training phase all at once and can be used to train the model.
In this case, training continues until some predefined number of iterations or some constraints are met.
In the case of online training, training data is generated in an incremental manner instead of being generated all at once.
So the difference between the training and testing phases is blurred as discussed in Section \ref{sec:ai-ml}.
This specifically suits the fast-changing environment like wireless networks and provides benefits in terms of scalability, adaptability, and real-time learning. 
}

\subsection{Deep Learning (DL)}

\label{sec:ai-dl}

In complex real-world problems, feature extraction can turn out to be extremely challenging using generic ML models. There may be hundreds of parameters that need to be learned and the outputs may not be linearly correlated to the inputs. So general ML models may not provide satisfactory performance in learning these problems. To facilitate mapping outputs to inputs, Neural Networks (NNs) \cite{ai-nn} are widely used in ML frameworks. With the availability of a large amount of data, NNs have emerged as a key technology to be used in ML in the recent past. The learning process can be largely benefited from the introduction of NNs to deal with complicated large-scale problems. This learning process involving NNs to estimate the models is known as Deep Learning (DL) \cite{ai-dl} which is a special important subset of ML. 

NNs are inspired by the biological neural networks in the brain, more specifically the nervous system. To mimic the operation of the brain, the NNs are composed of multiple layers where each layer consists of multiple neurons followed by an activation function. Generally, the neurons in one layer are connected to the neurons in the adjacent layers. The connecting edges have weights that represent the relationship between the neurons. Each layer output can be viewed as some intermediate decisions which eventually result in the final output values. The weights are generally trained through a number of iterations using backpropagation algorithms \cite{ai-bp}. Generally, the cost function associated with the model to calculate the difference between the predicted and actual outputs is not very simple, so we use different numerical methods like Gradient Descent \cite{boydconvex}, Stochastic Gradient Descent \cite{sgd}, Mini-Batch Stochastic Gradient Descent \cite{mbsgd}, Newton's method \cite{newton} etc. and so on to estimate the gradients of the cost function with respect to corresponding weights. At each iteration, the weights are updated by an amount based on these calculated gradients and a predefined learning rate. As we move towards the gradient descent direction, it helps us to reduce the cost at every iteration. In this manner, we can map the inputs to outputs through NNs.

\subsection{Major Learning Paradigms}

\label{sec:ai-ml-types}

Depending on how an algorithm is being trained and on the basis of the availability of the output for training, learning approaches can be classified mainly into three categories: Supervised Learning (SL), Unsupervised Learning (UL), and Reinforcement Learning (RL). A short overview of different types of learning approaches is shown in Figure \ref{fig:ml-types}. These approaches are discussed below:

\subsubsection{Supervised Learning (SL)}

\label{sec:ai-sl}

In an SL model, a training dataset containing a set of features as inputs and corresponding current outputs is provided to the model. 
The model with an initial set of parameters is trained through a number of iterations for mapping inputs to outputs. 
As the output label is clearly defined, the model can improve its performance by comparing its predicted outputs with the actual outputs \cite{ai-sl}. 
\revision{
SL problems can be broadly classified into two categories depending on the type of output labels: regression and classification problems. 
Regression \cite{reg} is a statistical method that investigates the relationship between a dependent (target) variable to one or more independent (given) variables. 
In this method, the functional mapping between inputs and outputs is estimated by minimizing the error between the predicted and actual outputs. 
Here the output label can be continuous.
In classification, the output labels correspond to distinct classes arising in computer vision, image classification, etc. 
Generally, classification problems are solved by using probabilistic classifiers to map output classes from inputs.
To train complex SL problems, NNs are used to learn complicated functional mapping between inputs and outputs.
We discuss the major ML and DL approaches in the context of SL problems below:
}

\textit{ML Approaches:}
There are a number of SL algorithms to train the model. Linear regression \cite{linreg} focuses on regression problems, whereas logistic regression \cite{logreg} focuses on classification problems. Decision tree is used in classification problems by forming a tree-like structure to learn the best split at every node level based on a statistical measure like information gain \cite{dec-tree}. The classification starts at the root node and traverses down along the branches based on intermediate decisions till the leaf nodes which represent the final classification decisions. Naive Bayes Model \cite{naive} is a form of a simple probabilistic classifier that uses the Bayesian Theorem to decide the classes under the strong assumption of feature independence. It is very useful, especially in high-dimensional classification problems. Support Vector Machine (SVM) \cite{svm} is another important type of classifier that decides the splitting hyperplane between different classes by maximizing the distances between the nearest data point (in both classes) and the hyperplane. 

\begin{figure*}
\begin{minipage}[t]{0.45\textwidth}
  \includegraphics[width=\linewidth]{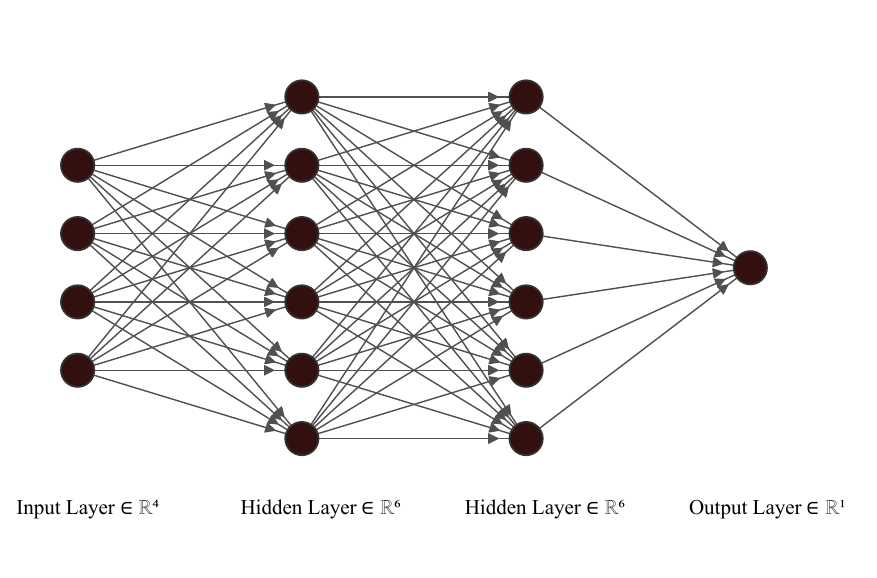}
  \caption{Fully Connected Neural Network (FCNN).}
  \label{fig:fcnn}
\end{minipage}%
\hfill
\begin{minipage}[t]{0.45\textwidth}
  \includegraphics[width=\linewidth]{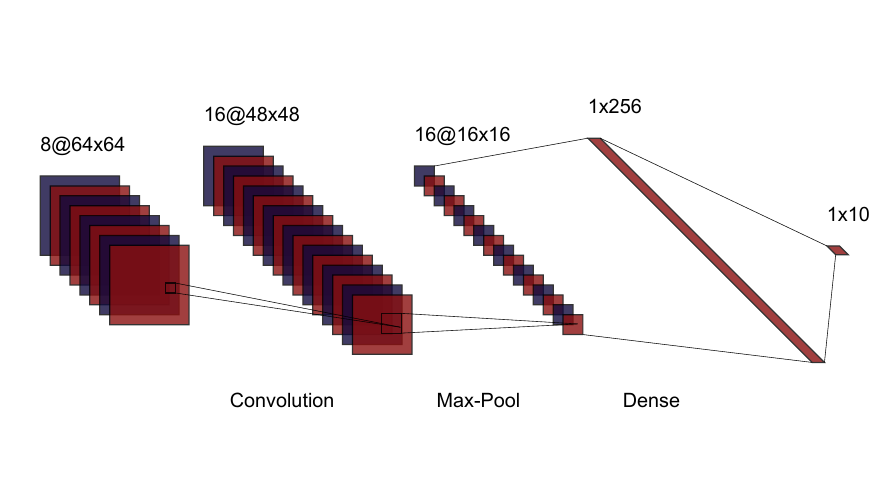}
  \caption{Convolutional Neural Network (CNN).}
  \label{fig:cnn}
\end{minipage}%
\end{figure*}

\textit{DL Approaches:}
Different DL approaches are also proposed in the literature to tackle complicated SL problems effectively. Perceptron \cite{perceptron} is one of the first NN architectures that have been proposed. It is a single-layer NN that can do binary classification like logistic regression. The main difference is to introduction of a simple activation function (step function) as a first step to more complex and advanced architectures. The simplest multi-layer NN architecture is the Fully Connected Neural Networks (FCNN) (Figure \ref{fig:fcnn}). This is also known as Multi-Layer Perceptron (MLP). It has multiple hidden layers between the input and output layers without any back loops. As the name suggests, all the neurons between two adjacent layers are connected to each other. Extreme Learning Machine (ELM) \cite{elm} is a very special type of NNs where the neurons are randomly connected and the training is done one-shot using least square fits. Another different type of NN is the Deep Residual Network (DRN) \cite{drn} with extra connections passing input from one layer to a later layer as well as the next layer. There are also Probabilistic Neural Networks (PNN) \cite{pnn} which can recognize the underlying pattern and generate the probability distribution function for different classes. 

Convolutional Neural Networks (CNN) \cite{cnn} is an important type of NN that can take multidimensional inputs like images and classify them with great accuracy by discovering spatial features (Figure \ref{fig:cnn}). The CNNs are composed of convolutional layers and subsequent pooling layers. The convolutional layers divide the whole input into smaller blocks and scan through them to learn the different features. The idea is to exploit the high correlation among neighboring cells with reduced complexity. A pooling layer is used to simplify this extraction process by getting rid of redundant features. Often CNN is accompanied by an FCNN to take care of nonlinearity and generate final classification results. A counterpart of CNN is the Deconvolutional Network (DN) \cite{dn} which takes the classes as inputs and generates CNN input features by comparing them with actual CNN inputs. 

\begin{figure*}
\begin{minipage}[t]{0.45\textwidth}
  \includegraphics[width=\linewidth]{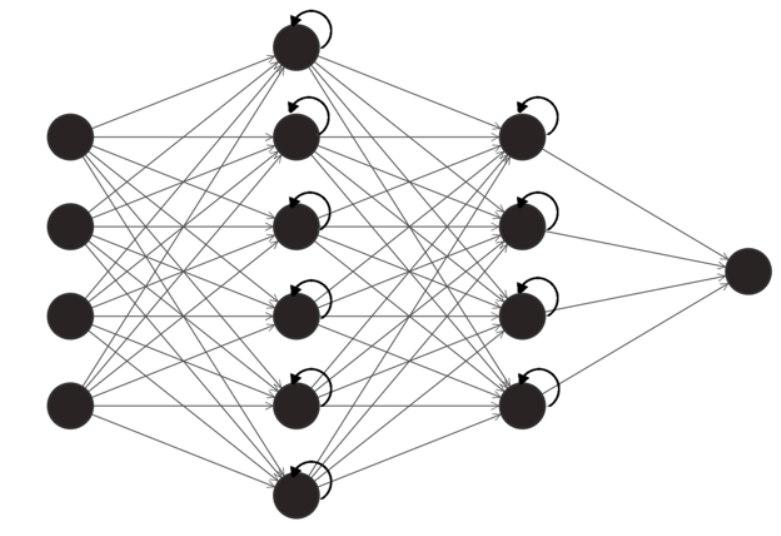}
  \caption{Recurrent Neural Network (RNN).}
  \label{fig:rnn}
\end{minipage}%
\hfill 
\begin{minipage}[t]{0.45\textwidth}
  \includegraphics[width=\linewidth]{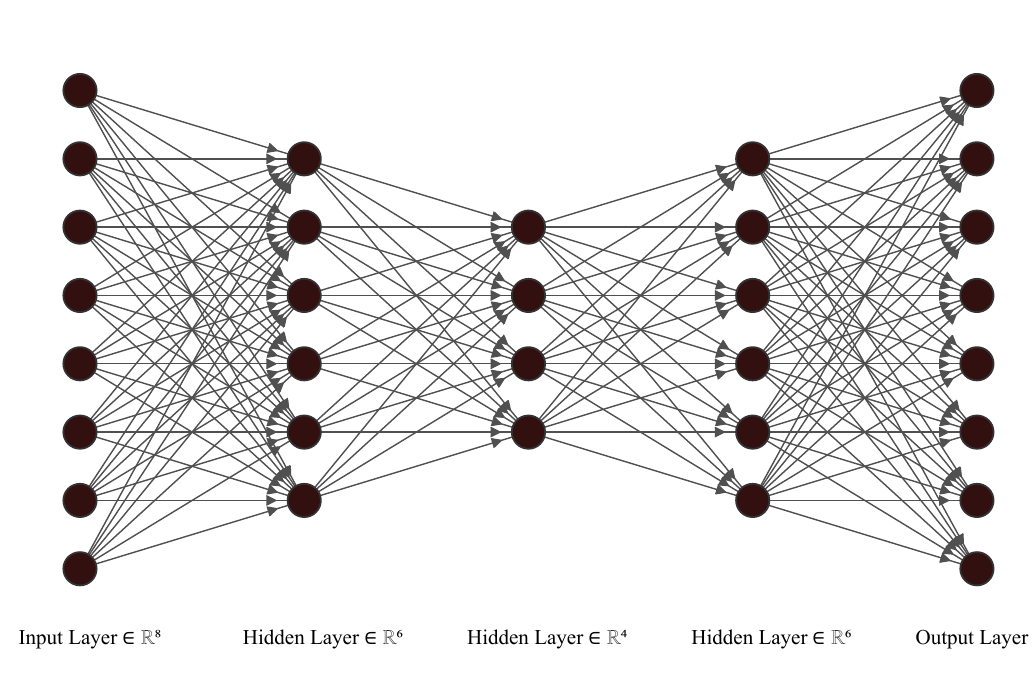}
  \caption{Autoencoder.}
  \label{fig:ae}
\end{minipage}%
\end{figure*}

Another important type of NN is Recurrent Neural Networks (RNN) \cite{rnn} with back loops. So the neurons in a layer are not only connected to previous layer neurons but also can be connected to the neurons from the subsequent layers. (Figure \ref{fig:rnn}) This allows it to capture temporal correlation among different layers and can be useful where decisions from past iterations or samples can influence current ones. However, they suffer from vanishing gradient issues due to long-term temporal dependencies \cite{vg}. To tackle this issue more sophisticated architectures like Gated recurrent units (GRU) \cite{gru} and Long-Term Short Memory (LSTM) \cite{lstm} with special memory cells and gates are introduced. Reservoir computing (RC) \cite{rc} is a low training complexity RNN framework for computation where the inputs are fed into a fixed and non-linear system, known as a reservoir, and then mapped into outputs from the reservoir neurons. Liquid State Machines (LSMs) are examples of RCs where the neurons are randomly connected receiving time-varying inputs. Echo-state networks (ESNs) are also a type of RC that uses a sparsely connected hidden layer (reservoir) with typically 1\% connectivity. The connectivity and weights of hidden neurons are fixed and randomly assigned.  

\revision{Another significant advancement in deep learning architecture, known as transformers, holds immense promise in the development of intelligent systems, particularly in communication environments. 
The transformer is a sequence-to-sequence neural network model comprising both an encoder and a decoder module, each with an identical architecture \cite{trans}. 
To streamline the input and output sequences, embedding and positional encoding layers are employed. 
Both the encoder and decoder primarily consist of a self-attention sub-layer and a position-wise sub-layer, with an additional masked attention sub-layer in the decoder. 
Each sub-layer is complemented by a residual connection and normalization module, facilitating the capture of long-range dependencies within the input data through self-attention.
}


\subsubsection{Unsupervised Learning (UL)}

\label{sec:ai-ul}

In UL, a raw unlabeled dataset is provided to discover existing patterns and features \cite{ai-ul} using some statistical learning approach. This is very useful when the data is not labeled. The algorithms find the underlying structure of the data and predict the outputs by adapting the model. Here the classes are not explicitly stated, so the classes need to be generated based on the distribution of input features in multi-dimensional spaces. It can be even used for generating labeled data to transform the original problem into an SL problem, which is usually easier to solve. 
\revision{Furthermore, clustering is another important UL problem where the model outputs different clusters based on the inherent pattern of data distribution. 
Dimensionality reduction can be also classified as a UL problem as it reduces the state space of the feature vectors in a general ML setup. }

\textit{ML Approaches:}
There are a number of unsupervised learning algorithms in the literature. 
\revision{Principal Component Analysis (PCA) \cite{pca} is primarily used for dimensionality reduction of a high dimensional dataset. 
It reduces the number of correlated features converting them into a set of uncorrelated features, which are also termed principal components, using orthogonal transformation of basis vectors. 
Reducing the dimensions of inputs also reduces the number of features to be learned, which later can be leveraged in SL techniques.
It is sometimes not considered an UL technique, but rather a preprocessing technique for data analysis with reduced dimensions.
In Probabilistic Graph Models (PGMs), the probabilistic relationship between random variables is modeled through a graph \cite{pgm}.}

K-means Clustering \cite{kmean} divides all the data points into K clusters in which each data point belongs to the cluster having the nearest mean. 
The mean of the data points in a particular cluster defines the center of the cluster. 
\revision{Another variant of K-means Clustering is called K-medoids Clustering where the centralmost data point of a cluster is defined as the center of the cluster \cite{kmed}.
Various mixture models, such as the finite mixture model, Gaussian Mixture Model (GMM) \cite{gmm}, etc. are also used for clustering. 
Hierarchical clustering can cluster data into a hierarchy of groups without predefining the number of clusters. 
It also comes with increasing computational costs compared to other clustering approaches.}
\revision{k-Nearest Neighbours (KNN)} \cite{knn} algorithm determines the k-nearest neighbors for all the data points of an unknown feature vector whose class is to be identified.

\textit{DL Approaches:}
Generally speaking, autoencoders \cite{ae} are used to help reduce the noise in data. 
In an autoencoder, first, we encode a high dimensional input, then decode it to reconstruct the input at the output again (Figure \ref{fig:ae}). 
The intermediate hidden layer neurons represent a compressed representation of the inputs getting rid of irrelevant and noisy components. 
Some other variations of this architecture are variational \cite{ae1}, noisy \cite{ae2}, and sparse \cite{ae3} autoencoders. 
\revision{In variational autoencoders, the compact representation of data is used to generate new data points sampling from the latent space.
In sparse autoencoders, the loss function is also expanded by adding another term called sparsity penalty regularization term for encouraging sparsity in the learned representations.
In denoising autoencoders, robust representations are learned from the noisy input data.

Deep Belief Networks (DBNs) \cite{dbn1} is a probabilistic generative graph model composed of hierarchical layers representing feature vectors. 
Here the top layers create undirected symmetric connections among them forming an associative memory. 
Greedy layer-wise training can be used for DBNs \cite{dbn2}.
Two symmetric DBNs can be extended to the structure of deep autoencoders for efficiently decoding the feature vectors \cite{ae-dbn}.
To use the feature extraction capability CNNs for UL, a combination of CNN and DBN is used in \cite{cnn-dbn}.
Hopfield NN \cite{hnn} is a cyclic recurrent NN architecture where all the nodes are connected to each other.
This provides an abstraction of circular shift register memory to form a global energy function and finding clusters without a supervisor. 
The Boltzmann Machine is another type of recurrent NN that has a stochastic symmetric recurrent architecture \cite{bm}.
As the convergence rate is generally slow for these NNs, a variant of this, Restricted Boltzmann Machine (RBM) is designed to learn the probability distribution over input data but in a layered manner \cite{rbm}. 

In UL, competitive learning approaches, such as Self-Organizing Maps (SOMs) \cite{som}, each neuron competes to represent an input subset. 
Here a single neuron from a group of output neurons is activated while the other neurons adjust their individual values in regard to input data distribution.}
Generative Adversarial Networks (GAN) \cite{gan} consist of any two networks, with one generating data (generative network) and the other judging the generated data (discriminating network). 
The prediction accuracy of the discriminating network is then used to evaluate the error for the generating network. 
This creates a form of competition between the discriminator and the generator to get better in their corresponding tasks. 
We can also use ensemble learning methods \cite{ai-ens} comprising multiple learning methods for better performances. 

\revision{Generative Diffusion Models (GDMs), as introduced in \cite{diff}, represents a recent breakthrough in UL leveraging DL techniques, drawing inspiration from the principles of thermodynamic diffusion. 
GDMs have gained widespread recognition due to their remarkable ability to generate high-quality data and simple implementation procedures. 
In contrast to GANs, GDMs employ a denoising network that iteratively converges to an estimate of the real sample. 
This model works in two distinct phases: the forward and reverse diffusion processes \cite{diff2}. 
In the forward diffusion phase, Gaussian noise is gradually introduced through a series of steps to create the target input for the denoising network. 
Subsequently, the denoising network is trained to reverse the noise effect for generating the original content.}

\subsubsection{Reinforcement Learning (RL)}

\label{sec:ai-rl}

\begin{figure}[ht]
  \centering
  \includegraphics[width=0.4\textwidth]{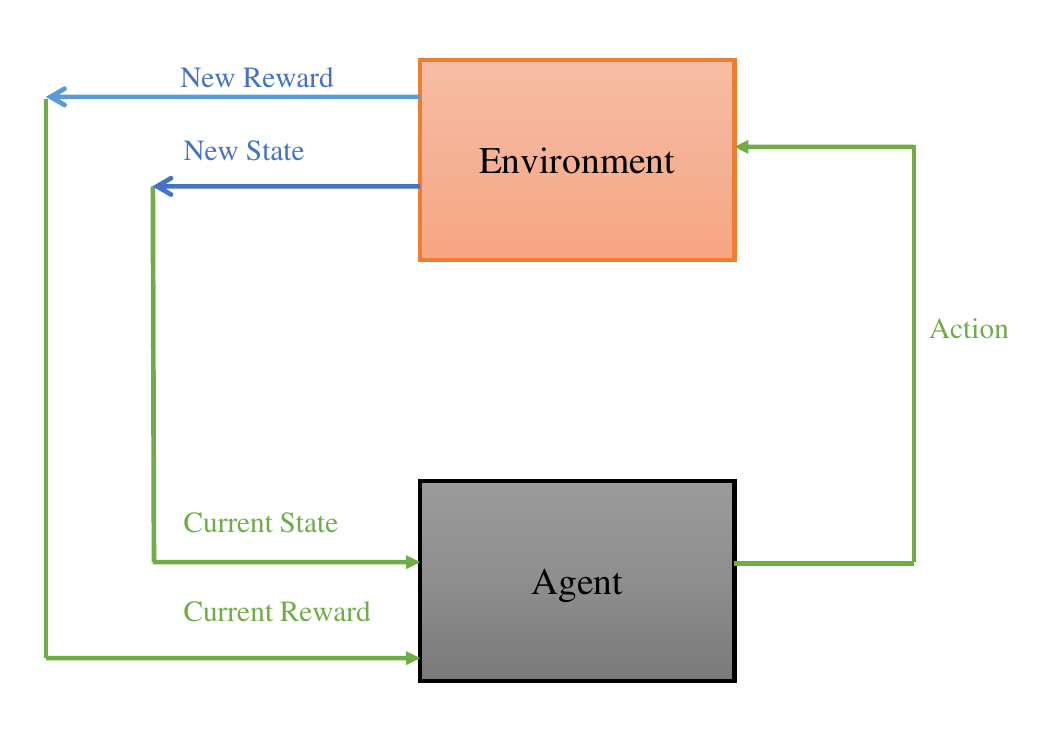}
  \caption{MDP model.}
  \label{fig:gen-rl}
\end{figure}

In RL, an agent learns to behave in a particular environment by performing thousands of actions and getting rewards or penalties based on those actions \cite{ai-rl}. This behavior (formally known as policy) is defined by the set of actions the agent learns from its experiences. The environment is defined by some mathematical models, the most common one is the Markov Decision Process (MDP) \cite{markov}. Here the feedback is neither provided using explicit labels like in SL nor the model is learned like in UL, but the behavior of an agent is learned through the rewards or penalties based on the set of actions taken by going from one state to another with a transition probability. The goal is to find out the optimum policy so that the total reward can be maximized (or the total penalty can be minimized) over a horizon of future time intervals given the current state of the agent. In Figure \ref{fig:gen-rl}, we show a generic structure of an RL framework as an MDP model.

\textit{ML Approaches:}
Depending on whether an RL model is explicitly created or not, RL can be fundamentally divided into two major categories: model-based and model-free RL methods \cite{rl-intro}. In the model-based methods, the transition probabilities between different states are assumed to be known, whereas, in model-free methods, these probabilities are learned through iterations. Dynamic Programming (DP) \cite{dp} is the most popular model-based method used in practice. However, model-free methods like the Monte Carlo (MC) method \cite{mc} are most commonly used due to their flexibility and practicality in real systems. Q-learning \cite{q} is one of the most popular model-free methods where "Q" refers to the expected rewards for an action taken in a given state over the time horizon, known as the value function. Another important counterpart of Q-learning is the State-Action-Reward-State-Action (SARSA) learning method where the agent learns the optimal policy in an online fashion \cite{sarsa}. This Q-learning is extended to the context of stochastic games \cite{sg} involving multiple agents in \cite{marl}, which is also known as the Multi-Agent Reinforcement Learning (MARL) method. Another approach for model-free RL is to learn the policy directly instead of learning the value functions, which is known as the Policy Gradient (PG) method \cite{pg}. If the policy gradient is estimated in a deterministic fashion, it is called the Deterministic Policy Gradient (DPG) method \cite{dpg}. To have the benefits of both approaches, the Actor-Critic (AC) method is proposed in \cite{ac} where the critic estimates the value function and the actor updates the policy gradient in the direction suggested by the critic. 

\textit{DL Approaches:}
General RL approaches without NNs can work well for small-scale experiments. However, when the action or state space is really large, the computation complexity exponentially increases. This phenomenon is quite common in practical systems like communication networks. To estimate the value functions or policies with RL frameworks with large state or action spaces, DL approaches can be very useful. This learning approach is also known as Deep Reinforcement Learning (DRL). The most popular and simplest 
DRL approach is the Deep Q-Network (DQN) \cite{dqn} method. In this approach, instead of an iterative approach for updating Q values in the Q-table, an NN is used to estimate the Q function value approximately. To prevent a large overestimation of action values, another DL framework is introduced on top of DQN for a fair evaluation of policies in Double Deep-Q-Network (DDQN) \cite{ddqn,ddqn2}. In Dueling Deep-Q-Network \cite{dueldqn} method, both state and action values are separately estimated. As the expected value function may be overestimated as the expected value does not capture the complete probability distribution of random variables, Distributional Deep-Q-Network \cite{disdqn} is considered to update the Q function value based on its distribution. In the case of continuous action spaces, DL aided DPG method, known as Deep Deterministic Policy Gradient (DDPG) Q-Learning \cite{ddpg} provides better results. To deal with partial observable environments, Deep Recurrent Q-Networks (DRQN) \cite{drqn} by introducing an LSTM layer in the FCNN architecture of DQN. 
Similarly, as for RL, MARL approaches can be efficiently solved using DRL architecture for each agent, namely known as a Multi-Agent Deep Reinforcement Learning (MADRL) framework. 

\revision{
\subsection{Distributed Learning Paradigms}

\label{sec:ai-dl-types}

The learning paradigms discussed in Section \ref{sec:ai-ml-types}, can be executed in various distributed approaches which will be discussed in this section. 
Unlike other fields, future 6G communications systems including satellite-based NTNs may need to incorporate a huge amount of network data from different network operators which ask for distributed approaches more.
However, there are some inherent challenges in terms of privacy and efficiency with distributed approaches as the network needs to deal with gathering data from different parties.
Three major distributed learning paradigms: Federated Learning (FL), Decentralized Learning (DcL), and Split Learning (SpL) are discussed below:

\subsubsection{Federated Learning (FL)}

FL is a distributed learning technique in which multiple network data owners collaboratively build and train a global DL model, all while ensuring data isolation and privacy, as outlined in \cite{fl1}. 
In the FL paradigm, individual data owners are initially provided with a base model by a centralized server, which they then train using their own respective data. 
Subsequently, these locally trained models are shared back with the central server, allowing it to update and maintain a global model. 
This iterative process continues until the global model reaches convergence. 
Consequently, FL enables the development of a globally trained model through distributed efforts, all the while safeguarding the privacy of the data owners.

\subsubsection{Decentralized Learning (DcL)}

DcL involves computing nodes conducting local training on their individual DL/ML models and then sharing these models with neighboring computing nodes at each iteration. 
Global convergence is achieved when all the local models have converged. 
Notably, this approach ensures that no actual data is exchanged between the computing nodes, but the local models are shared among neighboring nodes. 
One notable advantage of DcL is the absence of the need for a centralized server, which is a requirement in FL. 
An illustrative example of DcL can be found in the context of MARL, where agents collaboratively train their models in a distributed manner, as elaborated in \cite{distl}.

\subsubsection{Split Learning (SpL)}

In SpL, instead of sharing model parameters, training occurs across various computing nodes, as described in \cite{split1}. 
Each computing node has the responsibility for training multiple layers of NNs within a DL model. 
Gradients for backpropagation are exchanged among these nodes to enhance training efficiency. 
Consequently, it can yield superior privacy performance when compared to FL, as indicated in \cite{split-fl}. 
Recent endeavors have been made to combine these two approaches, aiming to harness the advantages of both methodologies, as explored in \cite{split-fl-comb}.

}

\subsection{Synergy between AI and NTN}

\label{sec:mot}

Like many other fields, NTN is expected to be a major advancement in the realm of AI applications \cite{ai-satellite}. More precise and pragmatic analytical models with reduced overhead consumption, and efficient algorithms with a lower computational complexity are the primary catalysts for the deployment of AI-enabled NTN in next-generation wireless networks. In the preceding sections, we give a concise overview of NTN and AI to introduce these two crucial aspects of this article. Now, we motivate our readers by outlining the primary motivating forces behind combining AI and NTN for future wireless networks. 

\subsubsection{Complex Task Automation}

In NTNs, the complexity of tasks and procedures involved in communication networks is significantly heightened. These tasks encompass a wide range of operations, including resource allocation, channel estimation, modulation, coding, and the intricacies of satellite management control. Attempting to perform these tasks manually is not only challenging but often unfeasible. The complexity involved in optimizing network performance for satellites, in particular, renders manual operations insufficient. Moreover, these tasks require meticulous precision to ensure uninterrupted service and mitigate potential hazards. However, the advent of ML and DL approaches has 
By harnessing ML and DL, not only can accurate actions be executed, but complex chains of procedures can also be automated seamlessly without the need for human intervention \revision{following the general ML framework as discussed in Section \ref{sec:ai-ml}.}

\subsubsection{Tractable Solutions}

The deployment of next-generation NTNs is more complex than any other previous-generation cellular network due to its multifaceted architecture. For instance, the integration of satellite networks introduces a significant number of additional parameters to consider for optimum network performance \cite{IN_AI_NTN_1}. However, this can result in computationally intractable solutions for practical networks, even if the solutions are computationally tractable, they may be very inefficient. Resource management in TNTN networks is a prime example of this, as resource optimization in TNTN networks often turns into non-convex optimization problems, where only suboptimal or heuristic solutions can be obtained using numerical techniques \cite{bh-it1,bh-gop}. \revision{Fortunately, DL techniques can approximate complicated functions involving a large number of input variables with the help of NNs, as discussed in \revision{Section} \ref{sec:ai-dl}.} As a result, complicated network functionalities can be characterized with NNs and resource management issues can be solved in a tractable manner \cite{bh-dl, bh-dl2}.

\subsubsection{Data-Driven Decision Making}

Although probabilistic and deterministic models can be used to model NTN functionalities, these models are often derived using very strong assumptions to get the general closed-form expressions, resulting in significant deviations in performances in simulations compared to real networks. In contrast, ML models are obtained based on \revision{real} data, which means different scenarios are taken into account during training, without the need for making any assumptions. 
\revision{For instance, resource scheduling for users or network slices in a cellular network is typically decided based on the channel condition of the corresponding users or user groups. 
However, the channel is highly time-variant, so the decision feedback needs to be in real-time to incorporate optimal scheduling decisions for all the users in the network.
For NTNs, the scenario is worse due to the extremely time-variant nature due to the high mobility of NGSO satellites and dynamic propagation environments. 
Various AI models, on the other hand, have shown great promise in dealing with this kind of challenging problem due to their potential to capture real scenarios with more precision than theoretical models with a reasonable amount of computation complexity.}


\subsubsection{Adaptability and Learning} 

AI algorithms can adapt to changing network conditions and learn from experience. Through ML techniques, AI can continually improve its performance, optimize network operations, and adapt to evolving user demands. By leveraging AI techniques such as RL and predictive modeling, NTNs can adaptively allocate resources, optimize network parameters, and proactively detect and mitigate faults \revision{through online learning as discussed in Section \ref{sec:ai-online}}. AI enables NTNs to dynamically respond to changing network conditions, enhance operational efficiency, and ensure uninterrupted service delivery. The ability to learn from data and make intelligent decisions without human intervention empowers NTNs to continually improve their performance, optimize resource utilization, and deliver reliable connectivity in complex and evolving environments.

\subsubsection{Reduced Computation Complexity}

Obtaining optimal algorithms for various challenges in NTNs can be a daunting task. Even if such algorithms are derived for complex systems, their computational complexity often renders them impractical for real-world implementation. This complexity arises from the vast number of variables that govern different network procedures in NTNs. However, data-driven AI techniques offer a promising solution by reducing the dimensions of high-dimensional data through feature learning. Particularly, DL approaches have demonstrated remarkable effectiveness in extracting implicit features from complex systems. As a result, these techniques prove highly valuable in addressing the diverse challenges encountered in NTN environments.

\subsubsection{Reduced Transmission Overhead}

In some cases, traditional methods heavily rely on the exchange of information between various network participants, such as satellites and users. This might lead to a large overhead in communication channels, resulting in a decrease in the overall throughput of the network. AI can be used to reduce the control overhead of NTNs significantly. For example, to calculate the Doppler shift, the UEs must be provided with the latest ephemeris information of the satellites \cite{dop-acc}. However, this would cause an immense overhead and a decline in the achievable data rate for the UEs. Alternative DL techniques can be employed to estimate the Doppler shift without requiring any ephemeris information from the satellites \cite{dop-ml-rsrp}. This leads to a significant decrease in transmission overhead over communication channels, resulting in superior network throughput. 

\subsubsection{Real-time Implementation}

Network optimization and management decisions in NTNs usually require real-time implementation, usually in the order of milliseconds to tens of milliseconds. Consequently, complex algorithms cannot be used to obtain these real-time decisions. In most cases, the algorithms become either heuristic or offline. To have an online adaptable approach, AI techniques can be considered as a suitable option. For example, an online DRL-based approach \revision{as discussed in Section \ref{sec:ai-online}} can be used to obtain resource management decisions in real-time and ensure proper utilization of available resources in NTNs \cite{ntn-ra-leo-rl1}. This is particularly valuable in latency-sensitive decision-making, such as scheduling, handover decisions, etc.

\subsubsection{Leveraging CSI}

In communication networks, CSI is fed back to the BS from the UE to assist in selecting different schemes - such as modulation, channel coding, etc. - for improved network performance. Leveraging this data, which contains the general state of the channel, different ML approaches can be benefited. For example, RL approaches can use this data to train models. This implies that we do not need to modify the information segments sent from the UE to the BS for deploying these RL schemes, but rather can rely on feedback already existing in the communication networks. This again illustrates the capability of AI to integrate into traditional communication networks without any additional overhead costs.
\revision{For NTNs, this is more important as the spectrum is more scarce and expensive; utilizing traditional CSI feedback for learning becomes another motivating factor for AI approaches to NTNs.}

\vspace{3mm}
\noindent\textit{Key Takeaways: The data-driven ML and DL approaches are the major AI technologies for empowering satellite-based NTNs for the next-generation 6G networks. 
Due to their inherent capability of capturing practical scenarios with real-time tractable solutions, different learning paradigms, such as SL, UL, and RL can be extremely beneficial in addressing various challenges associated with future NTN-empowered 6G networks.
Consequently, there have been a lot of research activities to deal with these challenges in the literature.
In the following section, we explore various current research thrusts for incorporating AI into NTN in greater detail to get insight into potential research scopes.}

\revision{\section{Related Works}

\label{sec:related}

\begin{table*}[ht]
\centering
\caption{Related papers on AI approaches for Satellite-based NTNs in 6G.}
\label{tab:ntn-related-works}
\begin{tabular}{|p{.02\linewidth}|p{.03\linewidth}|p{.1\linewidth}|p{.1\linewidth}|p{.15\linewidth}|
p{.1\linewidth}|p{.1\linewidth}|p{.15\linewidth}|}
    \hline
    \textbf{Ref.}           & \textbf{Pub.} 
                            & \multicolumn{2}{c|}{\textbf{Background}} 
                            & \multicolumn{4}{c|}{\textbf{Discussion on AI-Enabled NTN in 6G}} \\ \cline{3-8}

                            & \textbf{year}
                            & NTN challenges
                            & AI relevance
                            & Research thrusts
                            & 6G perspective
                            & Current efforts
                            & Practical challenges \\ \hline






    \cite{COMST_AI_NTN_2}   &  2021 
                            & \checkmark 
                            &   
                            &  
                            & \checkmark
                            &  
                            & \\  \hline 

    \cite{5G-space}         &  2021 
                            & \checkmark 
                            &   
                            &  
                            & \checkmark
                            &  
                            & \\  \hline

    \cite{NTN_8}            &  2021 
                            & \checkmark 
                            &   
                            &  
                            & \checkmark
                            &  
                            & \\  \hline    

    \cite{NTN_7}            &  2022 
                            & \checkmark 
                            &   
                            &  
                            & \checkmark
                            &  
                            & \\  \hline  

    \cite{NTN_4}            &  2022 
                            & \checkmark 
                            &   
                            &  
                            & \checkmark
                            &  
                            & \\  \hline     

    \cite{COMST_AI_NTN_1}   &  2022 
                            & \checkmark 
                            &   
                            &  
                            & \checkmark
                            &  
                            & \\  \hline                               

    \cite{COMST_NTN_3}      &  2023 
                            & \checkmark 
                            &   
                            &  
                            & \checkmark
                            &  
                            & \\  \hline


    \cite{ai-6g1}           &  2020 
                            & 
                            & \checkmark   
                            &  
                            & \checkmark
                            &  
                            & \\  \hline  

    \cite{ai-6g2}           &  2020 
                            & 
                            & \checkmark   
                            &  
                            & \checkmark
                            &  
                            & \\  \hline          

    \cite{ai-6g3}           &  2020 
                            & 
                            & \checkmark   
                            &  
                            & \checkmark
                            &  
                            & \\  \hline  

    \cite{ai-6g4}           &  2021 
                            & 
                            & \checkmark   
                            &  
                            & \checkmark
                            &  
                            & \\  \hline 

    \cite{ai-6g6}           &  2022 
                            & 
                            & \checkmark   
                            &  
                            & \checkmark
                            &  
                            & \\  \hline                             

    \cite{ai-6g5}           &  2023 
                            & 
                            & \checkmark   
                            &  
                            & \checkmark
                            &  
                            & \\  \hline


    \cite{IWC_AI_NTN_2}     &  2019 
                            & \checkmark 
                            & \checkmark   
                            & Short  
                            & \checkmark
                            &  
                            & \\  \hline
                            
    \cite{ai-satellite}     &  2019 
                            & \checkmark 
                            & \checkmark   
                            & Short  
                            & \checkmark
                            &  
                            & Short \\  \hline
                            
    \cite{ai-iiot-ntn}      &  2020 
                            & \checkmark 
                            & \checkmark   
                            & Comprehensive, but only covers IoT applications  
                            & \checkmark
                            & 
                            & Short and only covers IoT applications\\  \hline
                            
    \cite{COMST_AI_NTN_1}   & 2021 
                            & \checkmark 
                            & \checkmark   
                            & Short   
                            & \checkmark 
                            & Complete 
                            & \\  \hline
                          
    \cite{COMST_AI_NTN_2}   &  2021 
                            & \checkmark 
                            & \checkmark   
                            & Short uncategorized   
                            & 
                            & Does not cover AI for NTN 
                            & \\  \hline
                            
    \cite{AI_NTN_MAIN_1}    &  2021 
                            & \checkmark 
                            & \checkmark   
                            & Comprehensive, but does not cover the 6G perspective  
                            & 
                            &  
                            & \\  \hline

    \cite{IN_AI_NTN_1}      & 2022 
                            & \checkmark 
                            & \checkmark   
                            & Short  
                            & \checkmark
                            &  
                            & Short \\  \hline

    \cite{ml-ntn}           &  2023 
                            & \checkmark 
                            & \checkmark   
                            & Short  
                            & 
                            & 
                            & \\  \hline

    \cite{rl-ntn}           &  2023 
                            & \checkmark 
                            & \checkmark   
                            & Comprehensive, but only covers RL  
                            & \checkmark
                            & Only covers RL
                            & \\  \hline
                            
    \cite{ml-dl-sat}        &  2023 
                            & \checkmark 
                            & \checkmark   
                            & Comprehensive, but does not cover the 6G perspective  
                            & 
                            & 
                            & \\ \hline
\end{tabular}
\end{table*}

The possibility of potential integration of NTNs into 5G-Advanced \cite{NTN_2} and future 6G networks to support various future high-demanding use cases has attracted significant attention from the research community in recent times. 
This emerging area of research has spurred numerous investigations to address the unique challenges and opportunities posed by NTN integration. 
\cite{COMST_NTN_3} discusses the potential integration aspects for satellites, which is an integral part of NTNs, into future communication networks. 
\cite{5G-space} presents a summary of 3GPP efforts towards supporting NTNs in the 5G-Advanced networks. 
\cite{COMST_AI_NTN_1} presents the real system prototypes along with the general overview discussion on NTNs.
\cite{COMST_AI_NTN_2} presents the challenges from the aspects of different communication layers to provide better insights for addressing these issues.
In \cite{NTN_1, NTN_2}, a concise discussion on various NTN components, use cases, technological enablers, and challenges for realizing NTN in 6G is presented. 
In \cite{NTN_3}, a detailed survey on the evolution of satellite networks towards the convergence with terrestrial networks from 3G to 6G along with the proposed architectures, use cases, and challenges is presented. 
In \cite{NTN_4}, future architectural options, use cases along the challenges associated with NTN-integrated 6G networks are explored. 
In \cite{NTN_5}, the necessary architectural evolution for integrating NTNs into 6G networks along with the challenges is discussed. 
\cite{NTN_6} specifically focuses on the integrated Space-Air-Ground Integrated Network (SAGIN) in 6G while discussing the above topics in the context of NTNs.
Another short magazine paper, \cite{NTN_7} on NTN architectures, motivational use cases in 6G, necessary 5G NR modifications and future research directions is also in the literature.  
In \cite{NTN_8}, a detailed discussion on architectural options for integrating NTN into future 6G and the challenges associated with it is presented. 

Likewise, AI has been acting as a driving force for various applications in wireless environments, especially in the last couple of decades; many surveys have been published on these topics recently \cite{ai-w2,ai-w3,ai-w4,ai-w5}.
To facilitate the potential of AI in the 5G-Advanced and 6G environments several research articles and surveys are in the literature \cite{ai-w1, ai-w6, ai-wireless, fed-wireless}. 
In \cite{ai-6g1,ai-6g3,ai-6g5}, some short surveys on the role of AI enabling 6G networks focusing on the vision, research opportunities, and challenges are presented. 
\cite{ai-6g2} discusses the explainability of AI to address various 6G challenges. 
In \cite{ai-6g4,ai-6g6}, comprehensive surveys on vision, enabling technologies, and applications for AI on 6G are presented. 
Some relevant surveys are also published focusing on different aspects of AI-enabled 6G like pervasive network intelligence \cite{ni-6g}, green communications \cite{gc-6g}, privacy \cite{pr-ml-6g, pr-ml-ed-6g}, network access and routing \cite{nr-ml-6g}. 
As NTNs are expected to be integrated into the existing terrestrial environment for the development of 6G networks, it is clear that AI is expected to play a crucial role in this process. 
To unleash the full potential of AI to enable NTN in 6G, we need to have a clear understanding of the potential issues of NTN, we can gain insight into what AI tools can be useful down the road to resolve those issues. 

There have been a few research articles capturing the key aspects of AI as an enabling technology for NTN in 6G in the recent past. 
In \cite{COMST_AI_NTN_1, COMST_AI_NTN_2}, a short discussion on important applications of AI/ML in satellite-based NTN communication for 6G is provided along with the general discussion on NTN. 
In \cite{IN_AI_NTN_1}, several potential AI approaches for sustainable integrated Terrestrial and Non-Terrestrial Networks (TNTN) with a focus on maritime networking are discussed in a concise manner. 
In \cite{IWC_AI_NTN_2}, a brief discussion of ML approaches to tackle different potential problems associated with integrated TNTNs is presented. 
In \cite{ai-satellite}, it provides a short discussion on ML approaches for a limited number of issues related to next-generation mega-satellite networks.
In \cite{ml-ntn}, a compact discussion on different ML and DL techniques at various layers of the Open Systems Interconnection (OSI) model for NTN integration into existing 5G infrastructures is presented. 
Even though the above-mentioned works attempt to capture the role of AI in future 6G networks for enabling integrated TNTN environments, they are generally brief and do not provide a comprehensive overview of works in this particular domain.
In \cite{ai-iiot-ntn}, the potential role of AI techniques in the provision of NTN-based Intelligent Internet of Things (IoT) services is discussed; they do not focus on cellular environments for future integrated TNTN 6G networks.
In \cite{AI_NTN_MAIN_1}, reviews of potential AI approaches for both broadcasting and communication satellites are provided. However, they do not focus on the issues related to NTN-integrated 6G networks, rather only focus on general satellite communication. 
In \cite{rl-ntn}, a comprehensive review of the control approaches like coverage, spectrum, interference, and mobility management required by NTN platforms that are solved using RL formulations is presented, but they do not focus on other AI approaches related to prediction and estimation.
A very recent comprehensive survey paper on ML and DL applications on satellite communications is published \cite{ml-dl-sat}. However, they do not discuss the current research efforts from the integrated 6G perspective and the potential challenges of applying ML and DL techniques in this domain. 

Most existing articles either concentrate on analyzing the architecture and challenges within Non-Terrestrial Networks (NTNs) or take a broader perspective on AI applications in wireless communications. While a few research articles touch upon potential research directions for AI-driven NTNs, these discussions are often not exhaustive or do not fully grasp the role of AI in 6G networks integrated with NTNs. Additionally, the current state of research and the practical complexities tied to AI-empowered NTN-integrated 6G networks remain largely unexplored. This survey article attempts to offer a comprehensive overview of various AI techniques employed to address the distinct challenges encountered in NTN technology.
The list of related articles along with the key features is provided in Table \ref{tab:ntn-related-works}.

}

\section{AI-NTN: Current Research Thrusts}

\label{sec:ai-ntn}

AI is considered to be one of the major driving forces for empowering next-generation NTNs. To unleash the great potential of AI in this field, exploring potential research thrusts of AI-NTN integration is extremely important. The scarce network resources, high mobility, and complex and time-varying hierarchical network topology give rise to different unique challenges in realizing NTNs for future wireless networks. Conventional optimization and estimation approaches are not always feasible for practical deployment in real networks. Various data-driven AI techniques are being explored by researchers due to their inherent capability of learning the surrounding environment and providing superior performances in practical scenarios. In this section, we discuss the current research thrusts for AI applications into NTNs.

\begin{figure}[ht]
  \centering
  \includegraphics[width=0.5\textwidth]{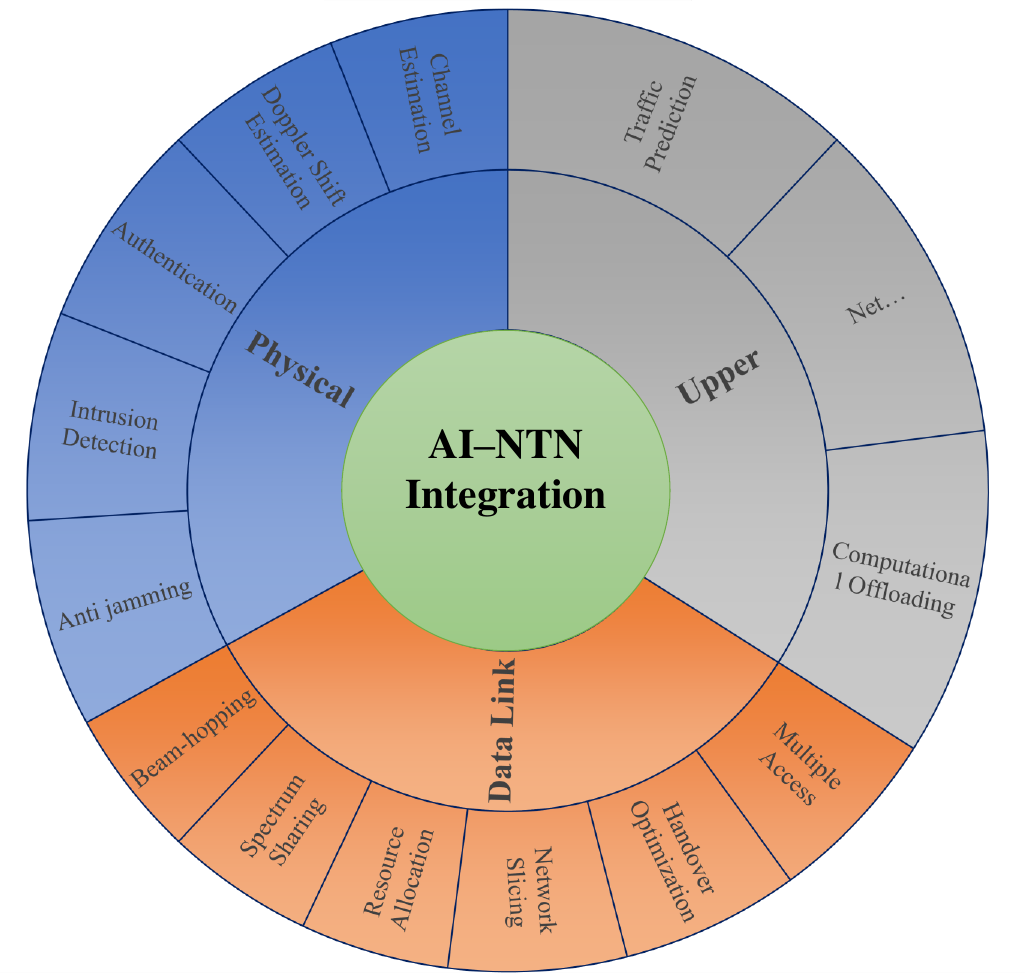}
  \caption{Taxonomy of research thrusts.}
  \label{fig:taxonomy}
\end{figure}

\newrevision{
\subsection{Taxonomy of Research Thrusts}
We categorize current research areas according to the distinct challenges encountered across various communication layers, facilitating a clearer understanding of the current AI-NTN research landscape. 
NTN, owing to its dynamic propagation environment and the high mobility of NGSO satellites, presents inherent challenges that span all the layers of communication systems. 
As the lower layers, namely, the physical and data link layers are highly affected by the new impairments, we discuss various challenges associated with these layers in the next two separate subsections. 
Following this, we group traditional network and higher-layer challenges in a subsequent subsection.
Within each subsection, we provide insights into the problem description, existing conventional methods, and the application of AI-based approaches to tackle these issues. 
While discussing AI methods, we cover SL, UL, and RL approaches, encompassing perspectives from both ML and DL for each research focus within their respective subsections.
For a visual representation of this classification scheme, please refer to Figure \ref{fig:taxonomy}.
}

\subsection{Physical Layer Aspects}

\subsubsection{Channel Estimation}

\revision{Channel estimation is an important aspect of NTNs, serving a dual role in encompassing comprehensive network planning and managing interference, similar to other wireless networks. 
This entails the technique of estimating the impacts of the channel through which a transmitted signal traverses in a wireless environment. 
Conventionally, the channel effect is encapsulated in an information block termed CSI in modern communication systems. 
While conventional methods like MMSE or Least Squares are employed for CSI estimation, they often entail high computational costs and may not always align with the demands of real networks. 
Furthermore, obtaining timely CSI information gets more challenging due to extended propagation delays and fast-changing propagation environments in NTN conditions.}

Therefore, ML-based methods are increasingly being adopted by the research community and vendors, as a promising alternative for channel prediction. This channel estimation can potentially be turned into an SL problem by considering channel features such as distance, time delay, received power, azimuth Angle of Arrival (AoA) and Departure (AoD), elevation angle, Root Mean Square (RMS) Delay Spread, and frequency as inputs and CSI as output labels. In \cite{ntn-pl2}, the reciprocity property of the downlink and uplink channels in Time Division Duplexing (TDD) systems is considered. So the downlink channel is estimated from uplink CSI using an LSTM-based DL model. In \cite{ntn-pl1}, CSI is estimated from historical CSI data using a CNN-LSTM model. However, as channel estimation is a near-real-time process, low-complexity NNs such as ESNs need to be explored for realistic implementations. 
\revision{In \cite{csi-dncnn}, a denoising CNN is used to reduce the LS channel estimation error. 
In \cite{csi-gru}, a CSI prediction scheme is presented without utilizing any ephemeris information, rather only using past CSI feedback information leveraging GRUs with low prediction error. 
In \cite{csi-arima-lstm}, an auto-regressive integrated moving average of past CSIs is utilized to predict future CSIs where the order of the past ones is determined by an LSTM network. 
In \cite{csi-graph}, graph attention networks are used for cascaded channel estimation for Reflective Intelligent Surface (RIS) assisted satellite networks in IoT communications. 
In \cite{csi-ml}, future CSI information is predicted using k-Nearest Neighbour and MLP-based algorithms from past CSI and some correlated network metrics such as latency, terminal velocity, weather, and environment state, etc. which is later used to adapt the modulation and coding scheme for next timestamp. 
In \cite{channel-rnn}, an RNN-based CSI compression technique is presented especially focusing on future SAGIN networks. 
An ANN is trained to estimate the fading at 40 GHz band exploiting the knowledge of its previous channel states in \cite{csi-ann}.
In \cite{csi-lstm}, an LSTM-based CSI prediction framework is discussed to provide in future NTN-integrated 6G networks.} 
 
\subsubsection{Doppler Shift Estimation}

\begin{table*}[htb]
\centering
\caption{Summary of AI approaches for Doppler shift estimation in satellite-based NTN.}
\label{tab:dop}
\begin{tabular}{|p{.06\linewidth}|p{.26\linewidth}|p{.05\linewidth}|p{.05\linewidth}|p{.05\linewidth}|p{.05\linewidth}|p{.1\linewidth}|p{.1\linewidth}|}
    \hline
    {\textbf{Reference}} &\textbf{Input to the ML model} &  \multicolumn{3}{c|}{\textbf{Learning Approach}} &\textbf{DL Tool} & \multicolumn{2}{c|}{\textbf{Comments on Models}} \\ \cline {3-5} \cline {7-8}
    &  &\textbf{SL} & \textbf{UL} & \textbf{RL} && RL Model & DL Model \\ \hline
    
    \cite{dop-ml1-ch} &  Averaged PSD & \checkmark &  &  & \checkmark & & FCNN \\ \hline
    \cite{dop-ml2-ch} &  Modulation scheme, delay profiles, SNR & \checkmark &  &  & \checkmark & & CNN-LSTM \\ \hline
    \cite{dop-ml-rsrp} & RSRP & \checkmark &  &  & \checkmark & & FCNN \\ \hline
    \cite{dop-ml-sat} & CSI & \checkmark &  &  & \checkmark & & CNN \\ \hline

\end{tabular}
\end{table*}

As the LEO satellites move around the Earth typically at a very high speed, both the satellite and ground user transceivers experience a large Doppler effect due to their relative velocity. If the transmitter moves towards (or away from) the receiver, the emitted signal from the transmitter may take less (or more) time to reach the receiver depending on the direction of the movement, hence the frequency of the signal increases (or decreases). This shift in signal frequency due to the motion of the transmitter, the receiver, or both refers to the Doppler shift. If the original frequency is $f_0$, the Doppler shift due to the motion of transceivers towards some specific direction with some specific relative velocity can be given by:

\begin{equation*}
    \delta f = f_0 \times \frac{v}{c} \times cos(\theta)
\end{equation*}
\noindent Here \\
$v$ = The relative velocity of the transceiver\\
$\theta$ = The angle between the direction of the transceiver and the direction of the propagating signal 

For LEO satellites, due to high mobility, This frequency offset is pretty significant (48 kHz with a center frequency of 2GHz \cite{3gpp1}). Due to these frequency offsets, UEs tune to some different carrier frequencies from their originally assigned carrier frequencies. This may lead to ICI between multiple UEs as discussed in \revision{Section} \ref{sec:ntn-chal}. 

There have been significant efforts to characterize the Doppler effect for LEO satellites since the launching of communication satellites. In \cite{dop-sim}, an equation for Doppler shift is derived for the simple case of LEO satellites with circular orbits in the equatorial plane and ground observing points on the equator. In \cite{dop-char}, the Doppler shift is analytically derived assuming the trajectory of the satellite with respect to the earth by a great circle arc and the speed of the satellite as constant. In \cite{dop-acc}, the Doppler shift is characterized by considering a new orbit generator using different orbital parameters through a rigorous analysis. UEs with Global Navigation Satellite System (GNSS) can get the global positioning of satellites and estimate the amount of Doppler shift needed to be addressed for the next transmission slot \cite{dop-gnss}. However, this increases the cost and complexity which may not be feasible for ground UEs \cite{3gpp2}. Additionally, The GNSS signals are weak, not ubiquitous, and susceptible to interference and spoofing. Recently, there have been also efforts to estimate the Doppler shift in LEO satellite systems using various other approaches, such as stochastic geometry \cite{dop-stoc}, Maximum A Posteriori (MAP) \cite{dop-map}, algebraic solutions \cite{dop-alg}, two-stage estimators consisting of time-varying Burg spectral analyzer and alpha-beta filter \cite{dop-blind} etc. In \cite{dop-lin}, the Doppler shift is estimated using reference signals in more than one frequency position in Orthogonal Frequency Division Multiplexing (OFDM) carrier in a 5G integrated NTN system. 

These different theoretical approaches can estimate the Doppler shift with a certain accuracy in different scenarios. However, the methods are generally very cumbersome due to the complexity associated with the orbital mechanics of the satellites. Most of these methods come with simplifying assumptions to keep the approach feasible for practical systems, thereby affecting accuracy. Moreover, due to the constant high-speed movement of LEO satellites, the wireless environment associated with it becomes time-variant. The computation complexity increases more to model these temporal variations using traditional estimation approaches. Additionally, the UEs may need the ephemeris information of the satellites to compute the Doppler shift associated with its motion, which creates large additional overheads in the communication channels. To characterize this Doppler effect, ML-based algorithms seem to appear as potential practical alternatives to the research community.

In wireless communication systems, due to the mobility of the transceivers, the channel between the transceivers changes significantly resulting in received signal power variation and Doppler shift. So, intuitively, the CSI of this channel should contain information about the Doppler shift. This idea has been already explored in terrestrial networks to generate a model using ML\cite{dop-ml1-ch, dop-ml-rsrp, dop-ml2-ch}. The ground truth values or the labels are usually generated using the ephemeris information. Different channel characteristic variables like Rician K factor, azimuth AoA width, mean azimuth AoA and channel estimation errors are generated randomly, and averaged Power Spectral Density (PSD) is used as inputs with some preprocessing to a multi-layered FCNN to estimate the Doppler shift in \cite{dop-ml1-ch}. In \cite{dop-ml-rsrp}, RSRP values mapped from an ambiguity reducer are used to generate the weights for an MLP.  In \cite{dop-ml2-ch}, different time and frequency domain signals with various modulation schemes, delay profiles, and Signal to Noise Ratio (SNR) have been used as inputs to a hybrid CNN-LSTM model to estimate the Doppler shifts. In NTN, the research in this domain is still at the early stage The estimated CSI is used as input to a CNN model to estimate the Doppler shift in \cite{dop-ml-sat}. In the future, other potentially efficient SL models can be also explored to generate the real-time accurate Doppler shift in an online manner. In table \ref{tab:dop}, we summarize the AI approaches for Doppler shift estimation in NTN. Even though the DL techniques are found to be useful in estimating Doppler shift using channel parameters, Doppler shift can be also estimated by analyzing the predictable trajectory of the satellites. Complexity analysis is required to justify the  
applicability of these DL architectures replacing the state of art methods in real systems. 
\revision{
\subsubsection{Security - Physical Layer Authentication}
Due to the new interfaces introduced by satellite-integrated terrestrial architectures, various spoofing and replay attacks can be launched using these interfaces.
Spoofing attacks involve an attacker satellite impersonating a legitimate one, while replay attacks involve the retransmission of previously intercepted messages to deceive users.
Generally, in terrestrial networks, these kinds of attacks are detected and mitigated by using standard cryptographic techniques, a concept also investigated in satellite communications \cite{ntn-pla-gen1,ntn-pla-gen2}.
However, when it comes to NTN-integrated future 6G networks, these conventional cryptographic methods face several challenges. 
Firstly, these techniques are computationally intensive and, thereby challenging to implement in satellites due to their limited onboarding capabilities. 
Secondly, the highly dynamic and massive scale network topology of NTNs, particularly for enabling IoT devices, necessitates significant modifications in network protocol design and introduces overheads that may not be practical to manage with existing architectures
Also, these cryptographic techniques often assume that attackers lack the computational resources to break the encryption. 
However, with ongoing advances in quantum computing research, these assumptions may no longer hold in the future, presenting yet another challenge that needs to be addressed.

Physical layer authentication offers a promising alternative to these conventional techniques.
In \cite{pla-basic}, Wyner introduced the concept of physical layer authentication where the message was encoded in such a way that the mutual information between the legitimate channel and wiretap channel is maximized. 
This encoding generally captures the unique characteristics of the channel between the user and the legitimate transmitter, serving as a means to verify the transmitter's identity.
This technique has been already explored in terrestrial networks using the CSI information as radio signatures for the transmitter devices \cite{pla-ter1,pla-ter2,pla-ter3}.
However, the prevalence of Line of Sight (LoS) paths in satellite networks makes radio fingerprinting using channel fading information from CSI impractical.
Furthermore, due to the high mobility of NGSO satellites, as discussed in Section \ref{sec:ntn-chal}, the high Doppler Shift is introduced in received signals, which can be used to verify the identity of the legitimate satellites.
In \cite{ntn-pla-nml1}, a maximum likelihood estimation and uniform quantizer are used to obtain the secret key bits from the Doppler frequency shifts, which is used in the authentication of legitimate satellites \cite{ntn-pla-nml2}
In \cite{ntn-pla-orbit}, an orbital information -- time difference of arrival-based authentication mechanism is introduced providing low false authentication rates.

In the recent past, various DL techniques have shown lofty promises in the field of extracting features from noisy data, which is also leveraged in this field using different ML models.
In \cite{ntn-pla2}, CNNs and Autoencoders are used to extract the necessary channel features for physical layer authentication of legitimate satellites. 
In \cite{ntn-pla1,ntn-pla3,ntn-pla4}, both the received signal power and Doppler shift are used for radio fingerprinting using SVMs providing improved authentication rates.

\subsubsection{Security - Intrusion Detection}

In modern satellite-terrestrial integrated networks,  the majority of satellite communication systems rely on elementary security threat detection mechanisms.
Typically, these mechanisms operate by flagging an anomaly if the received signal frequency deviates from the baseline spectrum by a predetermined threshold. 
However, this simplistic approach frequently leads to a significant number of false positives.
On top of that many anomalies represent unusual behavioral patterns, exhibiting temporal correlations that escape detection by these simple detectors. 
Consequently, these conventional methods often struggle to effectively identify and respond to sophisticated security threats.

To address these challenges, DL techniques are explored to efficiently detect security threats using various innovative approaches.
In the study presented in \cite{ntn-id-rfmlp}, an ensemble model combining Random Forest (RF) and MLP is developed to improve the performance of security threat detection across diverse datasets for satellite communications.
\cite{ntn-id-rfdl} leverages critical feature selection driven by RF to streamline complexity and enhance the relevance of features before the detection phase.
These features are then forwarded to different NN architectures, including LSTM, GRU, RF, and ANN enabling robust security threat detection.
These models are tested on different datasets where GRU-empowered threat detection models exhibit superior performances by capturing temporal behavioral patterns.
In \cite{ntn-id-lstm-ul}, a UL approach using LSTM networks is explored, which can not only detect unforeseen security threats but also does not need any labeled data.
In another study as shown in \cite{ntn-id-lstm-rnn}, two SL and five UL approaches are considered for threat detection to show the effectiveness of ML techniques. 
In \cite{ntn-id-drl}, a DDPG-based DRL framework is considered where the agents decide whether the aerial platform is malicious or not (actions) based on their behavior (states) and the system condition (rewards) for threat detection.
Recognizing the computational constraints of satellites and Internet of Things (IoT) devices, federated learning approaches are also investigated as detailed in \cite{ntn-id-fed1,ntn-id-fed2,ntn-id-fed3} for threat detection.

\subsubsection{Security - Anti-jamming}

Satellites are vulnerable to jamming threats due to their predictable and periodic visibility in NTNs, so anti-jamming approaches are important to tackle this challenge.
Conventional spread spectrum techniques are used in anti-jamming for satellite networks.
However, they are not very useful in dealing with new smart jamming attacks which can adjust their actions based on the network feedback.
Various RL techniques are adopted to tackle these problems in an efficient manner.
In \cite{ntn-aj1,ntn-aj3}, a hierarchical anti-jamming Stackelberg game is introduced for routing anti-jamming problems which is later solved by providing fast anti-jamming decisions using a DRL-based routing algorithm for satellites. 
in \cite{ntn-aj4}, a DL-based jamming detection algorithm is proposed for satellite navigation systems. 
In \cite{ntn-aj2}, an anti-jamming coalition game is formed to decrease energy consumption, and suboptimal jamming policies are obtained by RL approaches.
In \cite{ntn-aj5}, ML-aided cognitive anti-jamming communication is designed, developed, and tested on real satellite-ground links. 
}

\subsection{Data Link Layer Aspects}
\subsubsection{Beam Hopping}

\begin{figure}[ht]
  \includegraphics[width=0.5\textwidth]{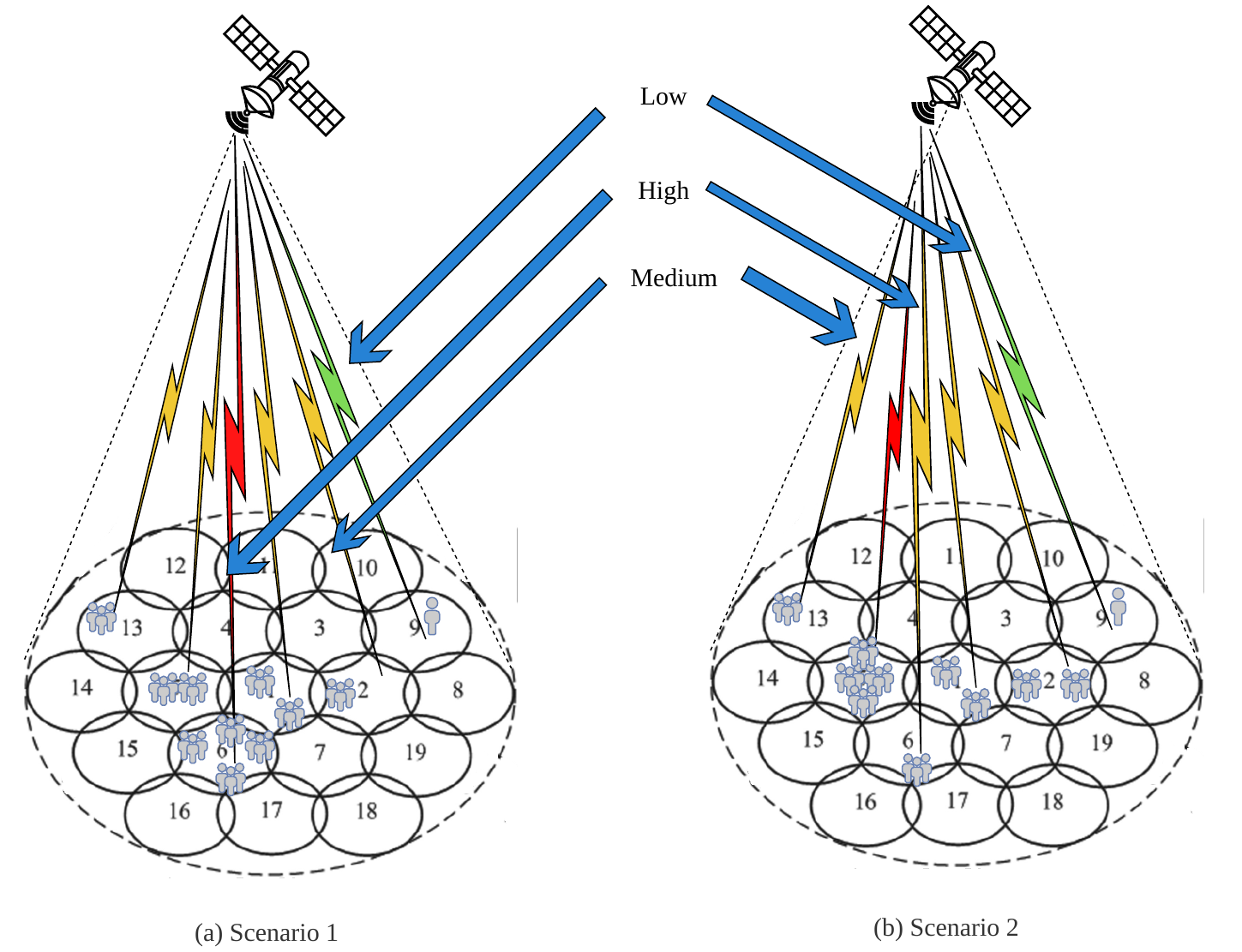}
  \caption{A simple beam hopping example in satellite-based NTN.}
  \label{fig:beam-hop}
\end{figure}

Modern communication satellites form multiple beams to support a large number of users over a large area through spatial multiplexing into different NTN cells. Each satellite can effectively reuse the allocated spectrum with very low co-channel interference as well as provide strong signals at the ground user terminals with relatively low transmission power using beamforming techniques. However, due to the high cost and low availability of onboard processing computing resources in satellite systems, mostly simple fixed beam allocation policies are used in traditional satellite communication. These strategies lack the flexibility to adapt to the temporal and spatial variation of traffic demands in real satellite networks. Beam hopping is a technique for allocating beams in a flexible manner so that these changes can be addressed efficiently. It refers to a procedure for activating different beams according to the current demands of an NTN cell covered by those beams, so effectively hopping the set of active beams from one combination to another \cite{bh-ov}. In Figure \ref{fig:beam-hop}, a simple beam-hopping scenario is depicted, where we have different NTN cells with varying demands. We classify the cells into three different categories, e.g., high, medium, and low, based on their traffic demands. In the first scenario,  the low-demand NTN cells, e.g. cell $9$, have less number of active beams than high-demand NTN cells, e.g. cell $6$, even lesser than moderate-demand NTN cells, e,g, cell $1,2,5$ or cell $13$. However, due to mobility or change in traffic patterns, the traffic demand in cell $13$ reduces and in cell $5$ increases. As a result, we can see the intensity of the beams also changes accordingly in these two cells at a later time, and a new beam-hopping pattern emerges. 

\begin{table*}[htb]
\centering
\caption{Summary of AI approaches for beam-hopping in satellite-based NTN.}
\label{tab:bh}
\begin{tabular}
{|p{.06\linewidth}|p{.06\linewidth}|p{.06\linewidth}|p{.06\linewidth}|p{.06\linewidth}|p{.05\linewidth}|p{.05\linewidth}|p{.05\linewidth}|p{.05\linewidth}|p{.1\linewidth}|p{.1\linewidth}|}
    \hline
    {\textbf{Reference}} &\multicolumn{4}{c|}{\textbf{Target Optimization Objectives}} &  \multicolumn{3}{c|}{\textbf{Learning Approach}} &\textbf{DL Tool} & \multicolumn{2}{c|}{\textbf{Comments on Models}} \\ \cline {2-8} \cline {10-11}
    & {Throughput}  & {Tx Delay} & {Delay Fairness} & {Capacity-Demand Ratio} &\textbf{SL} & \textbf{UL} & \textbf{RL} && RL Model & DL Model \\ \hline
    
    \cite{bh-dl} & &&&\checkmark & \checkmark & & & \checkmark& & FCNN \\ \hline
    \cite{bh-dl2} & &&&\checkmark & \checkmark & & & \checkmark& & FCNN \\ \hline
    \cite{bh-drl-d1} & &\checkmark&&  &  & & \checkmark & \checkmark& DQN & FCNN \\ \hline
    \cite{bh-drl-d2} & &\checkmark&&  &  & & \checkmark & \checkmark& DQN & CNN \\ \hline
    \cite{bh-drl-ga} & \checkmark&\checkmark&\checkmark& &  & & \checkmark& \checkmark &  Proximal Policy Optimization & FCNN  \\ \hline
    \cite{bh-drl-mo1} & \checkmark&\checkmark&\checkmark& &  & & \checkmark & \checkmark& DQN & CNN \\ \hline
    \cite{bh-drl-mo2}& \checkmark&\checkmark&\checkmark& &  & & \checkmark & \checkmark& DQN & CNN \\ \hline
    \cite{bh-marl}& \checkmark&\checkmark&& &  & & \checkmark & \checkmark& DDQN & FCNN \\ \hline
\end{tabular}
\end{table*}

The key question of beam hopping is to find out which beams need to be activated when and for how long while maximizing the network performances given the capacity constraints \cite{bh-ch}. This can be effectively formulated as an optimization problem considering different network performance metrics such as system throughput, delay, fairness, etc. as the objective(s) along with power and spectrum constraints. In \cite{bh-op}, a convex optimization framework with an objective to match the system capacity to traffic demand along with power allocation constraints is considered. This yields a close-form solution giving insights into resource allocation policies for maximizing network performances from different perspectives. However, from the perspective of real networks, the convex objective function is not very realistic, so the results are not applicable to real networks in a straightforward manner. Assuming the non-convexity of the problem, obtaining a globally optimal solution with efficient algorithms gets difficult. In \cite{bh-gop}, the steepest gradient descent algorithm is chosen to get the sub-optimal solution using the optimal set of precoding vectors. Some heuristic iterative approaches are also proposed in \cite{bh-it1}, \cite{bh-it2}, \cite{bh-it3} to tackle these non-convex problems in a practical and feasible manner. Different meta-heuristic approaches like Genetic Algorithm (GA) \cite{bh-ga}, Simulated Annealing (SA) Algorithm \cite{bh-sa}, Particle Swarm Optimization (PSO) \cite{bh-pso}, and combined metaheuristic approaches like GA-SA \cite{bh-gasa} have been considered to generate suboptimal solutions with a reduced amount of computational complexity. 

The main challenge in designing a beam-hopping pattern in an optimization framework lies in the large search space associated with an optimal solution. The size of the search space for finding out an optimal beam hopping pattern scales exponentially with the number of beams in the satellite networks. Modern satellites can have hundreds to thousands of beams depending on their coverage area, so the computational complexity becomes pretty high, and the computation time becomes pretty large to find out the exact solutions. The low-complexity suboptimal solutions using iterative and metaheuristic approaches achieving satisfactory performances in real networks are not very abundant. In this context, the DL approaches turn out to be a suitable alternative for this problem. 

In \cite{bh-dl}, \cite{bh-dl2}, an SL approach is considered by forming labeled datasets with beam hopping patterns as outputs and channel matrix, transmission power, and traffic demand as inputs. First, a mixed integer linear problem formulation for matching the offered capacity to traffic demands is reduced to a simple linear programming problem. A training dataset is generated using conventional optimization algorithms and a DL model is trained on this dataset by considering beam hopping patterns as labels. Furthermore, the optimization framework can be potentially transformed into an RL problem to capture the optimal beam-hopping pattern in a learning environment. In \cite{bh-drl-d1},\cite{bh-drl-d2} the transmission delay is minimized considering the power and beam allocation constraints using a DRL approach. The state space consists of the average transmission delay and the buffer length with beam hopping pattern as actions and the negative Hadamard product of the current states, the negative of total queuing delay as the reward function. In \cite{bh-drl-ga}, a combined DRL-metaheuristic approach is considered to optimize both the throughput and delay fairness while at the same time designing different reward functions for the two cases. In \cite{bh-drl-mo1}, \cite{bh-drl-mo2}, a network consisting of real-time and non-real-time traffic is considered. A multi-objective problem minimizing the transmission delay for real-time traffics, maximizing the throughput for non-real-time traffics as well as overall delay fairness is considered. Individual reward functions are designed to capture each of the goals. In \cite{bh-marl}, a cooperative multi-agent framework is considered to dynamically allocate the power and bandwidth to illuminating beams optimizing throughput and delay fairness using a DDQN. In table \ref{tab:bh}, we summarize the AI approaches for beam-hopping in NTN. As traffic demand changes with time, recursive architectures such as RNN, ESN, etc. should be also explored to design the NN for DL architectures used to address beam-hopping issues. Also, distributed learning architectures can be useful to design efficient beam-hopping schemes. 
\subsubsection{Spectrum Sharing}

\begin{table*}[htb]
\centering
\caption{Summary of AI approaches for spectrum sharing in TNTN.}
\label{tab:ss}
\begin{tabular}{|p{.06\linewidth}|p{.08\linewidth}|p{.08\linewidth}|p{.08\linewidth}|p{.05\linewidth}|p{.05\linewidth}|p{.05\linewidth}|p{.05\linewidth}|p{.1\linewidth}|p{.1\linewidth}|}
    \hline
    {\textbf{Reference}} &\multicolumn{3}{c|}{\textbf{Problem Insight}} &  \multicolumn{3}{c|}{\textbf{Learning Approach}} &\textbf{DL Tool} & \multicolumn{2}{c|}{\textbf{Comments on Models}} \\ \cline {2-7} \cline {9-10}
    & {Spectrum Sensing}  & {Spectrum Occupancy Prediction} & {Spectrum Access}  &\textbf{SL} & \textbf{UL} & \textbf{RL} && RL Model & DL Model \\ \hline
    
    \cite{ntn-sp-dl} &  &\checkmark& & \checkmark &  &  & \checkmark & & CNN-BiLSTM \\ \hline
    \cite{ntn-sr-dcnn} &  &\checkmark&  & \checkmark &  &  & \checkmark & & CNN \\ \hline
    \cite{ntn-ss-ccnls} &  &\checkmark&  & \checkmark &  &  & \checkmark & & CNN-LSTM \\ \hline
    \cite{ntn-spa-cnls} & \checkmark &\checkmark&  & \checkmark &  &  & \checkmark & & CNN-LSTM \\ \hline
    \cite{ntn-ss-drl1} & &&\checkmark &  &  & \checkmark  &  & Modified Q-Learning & \\ \hline
    \cite{ntn-ss3} & \checkmark &\checkmark&  & \checkmark &  &   & \checkmark &  & SVM-CNN\\ \hline
        \cite{ntn-ra-madrl2} & &   & \checkmark & &  & \checkmark & \checkmark & MADDPG & FCNN \\ \hline

\end{tabular}
\end{table*}

In traditional communication systems, satellite, and terrestrial cellular networks generally occupy different frequency bands, so they do not interfere with each other. However, the satellites in the new integrated TNTN environment for 6G are expected to use the same S and Ka-Band as discussed in \revision{Section} \ref{sec:ntn-feat}. This improves the overall spectral efficiency of the integrated networks as well as provides a better QoE for the users. However, as both the satellite and the ground network use the same frequency band, the signals 
\begin{figure}[ht]
  \centering
  \includegraphics[width=0.45\textwidth]{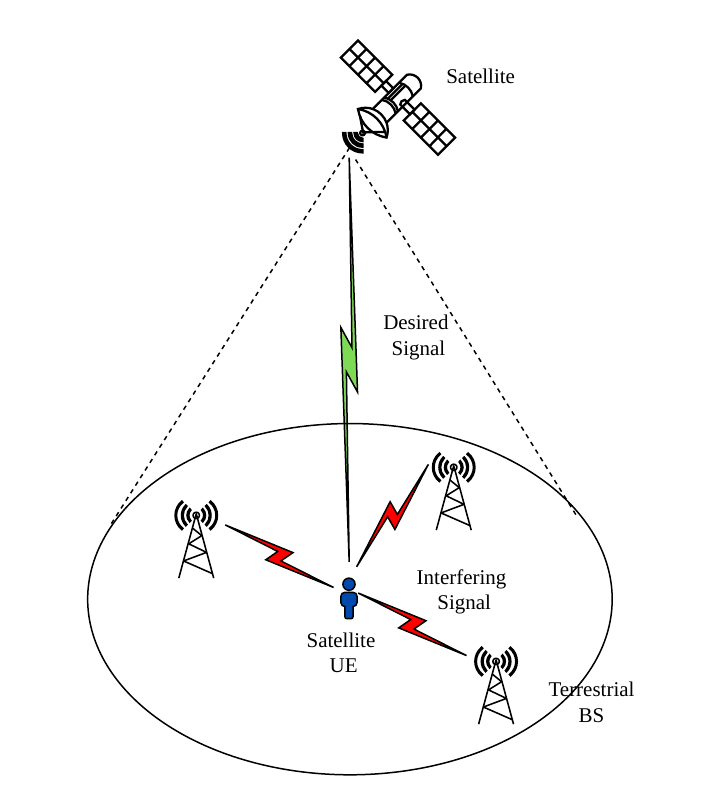}
  \caption{General spectrum sharing scenario in TNTN.}
  \label{fig:NTN-spec_shar}
\end{figure}
produced by them interfere with each other, i.e. cause Co-Channel Interference (CCI) to each other.  In Figure \ref{fig:NTN-spec_shar}, a simple spectrum-sharing scenario in the downlink channel in an integrated TNTN network is shown. The satellite user is connected to a satellite and the downlink channel is indicated using the green link. There are three more terrestrial BSs using the same channel as the satellite provide CCI to the satellite user (indicated by red links).

In TNTN, the spectrum-sharing phenomenon needs more attention because we have a hierarchical network scenario consisting of non-terrestrial and terrestrial BSs as shown in Figure \ref{fig:NTN-arch}. To support this complex topology in a single framework, we need to come up with efficient spectrum-sharing strategies causing low interference to the users \cite{ntn-ss-gen1}. In conventional spectrum sharing methods, we use efficient frequency reuse, leveraging directional antennas, adaptive power control, etc. methods to mitigate the effect of CCI. However, traditional four-color frequency reuse can effectively reduce the level of interference at the expense of more spectrum. beamforming approach can reduce the interference greatly, but that too comes at the cost of increasing complexity. 

To tackle this situation, a process called spectrum sensing is introduced in cognitive radio networks, where the unlicensed users can sense the occupancy status of the target band using some radio sensing method \cite{cr-mit}. The popular spectrum sensing methods are Energy Detection (ED)\cite{cr-ed}, Cyclo-Stationary Detection (CSD) \cite{cr-cyc}, Eigen Value-based Detection (EVD) \cite{cr-evd} etc. However, these methods either are simple with poor performance in low SNR scenarios (ED) or provide good performance but with more computational complexity (CSD and EVD). For these reasons, ML has been adopted for spectrum sharing to capture the correlation with a reduced computational complexity which can be extended to integrated satellite-terrestrial network scenarios \cite{ntn-ss-gen1}.

Different intelligent learning approaches are adopted to tackle the spectrum-sharing problem for next-generation TNTN networks \cite{ntn-ss1}. In \cite{ntn-sp-dl}, a spectrum-sharing strategy is developed for LEO satellites from the GEO satellite spectrum historical occupancy data using a CNN-BiLSTM model. Here the LEO satellite users are considered as unlicensed secondary users and the GEO satellite users are considered as the licensed primary users. In \cite{ntn-ss-ccnls}, a CNN-LSTM-based spectrum sensing method is introduced for satellites to capture the spatial and temporal correlation effectively for spectrum occupancy of satellite systems. In \cite{ntn-spa-cnls}, a CNN-LSTM model is introduced to predict the frequency assignment for satellites based on historical data. In \cite{ntn-ss-drl1}, a modified Q-Learning algorithm is used in an RL setup for the adaptive selection of access and modulation schemes for NGSO satellites in an NGSO-GEO system. In \cite{ntn-ss3}, an SVM model is first used for low complexity spectrum sensing, then a CNN-based spectrum prediction model based on historical data is developed. In \cite{ntn-ra-madrl2}, a cooperative MADRL framework is considered for bandwidth management in a game-theoretic model minimizing inter-beam interference. In \cite{ntn-sr-dcnn}, a CNN-based spectrum reconstruction method from incomplete data is discussed for satellite networks.  In table \ref{tab:ss}, we summarize the AI approaches for spectrum sharing in TNTN. However, these spectrum sensing decisions need to be in real-time to increase the overall throughput of the secondary users; this means the conventional LSTM architectures need to be replaced by efficient low-complexity ESN architectures to tackle this in an online manner. Furthermore, spatial spectrum sharing scenarios need to be also considered along with the state of art temporal spectrum sharing scenarios leveraging the benefits of future 3D SAGIN networks. 
\subsubsection{Resource Allocation}

\begin{table*}[htb]
\centering
\caption{Summary of AI approaches for resource allocation in satellite-based NTN.}
\label{tab:ra}
\begin{tabular}{|p{.06\linewidth}|p{.12\linewidth}|p{.12\linewidth}|p{.05\linewidth}|p{.05\linewidth}|p{.05\linewidth}|p{.05\linewidth}|p{.1\linewidth}|p{.1\linewidth}|}
    \hline
    {\textbf{Reference}} &\multicolumn{2}{c|}{\textbf{Target Objective}} &  \multicolumn{3}{c|}{\textbf{Learning Approach}} &\textbf{DL Tool} & \multicolumn{2}{c|}{\textbf{Comments on Models}} \\ \cline {2-6} \cline {8-9}
    & {Spectral Efficiency }  & {Energy Efficiency} &\textbf{SL} & \textbf{UL} & \textbf{RL} &&  RL Model & DL Model \\ \hline
    
    \cite{ntn-ra-leo-rl1} & \checkmark & &  &  & \checkmark &  & AC &  \\ \hline
    \cite{ntn-ra-mul-rl2} & \checkmark & &  &  & \checkmark &  & Q-learning &  \\ \hline
    \cite{ntn-ra-drl1} & \checkmark & \checkmark &  &  & \checkmark & \checkmark & DQN & FCNN \\ \hline
    \cite{ntn-ra-drl3} & \checkmark &  &  &  & \checkmark & \checkmark & DQN & CNN \\ \hline
    \cite{ntn-ra-drl4} & \checkmark &  &  &  & \checkmark & \checkmark & DQN & FCNN \\ \hline
    \cite{ntn-ra-madrl1} & \checkmark &  &  &  & \checkmark &  & MARL & \\ \hline
    \cite{ntn-ra-modrl-ens} & \checkmark & \checkmark  &  &  & \checkmark & \checkmark & DQN & Ensembles of FCNN \\ \hline
    \cite{ntn-ra-dl1} & \checkmark & \checkmark  & \checkmark &  &  & \checkmark & & FCNN \\ \hline
    \cite{ntn-ra-dl2} & \checkmark & \checkmark  & \checkmark &  &  & \checkmark & & FCNN \\ \hline
\end{tabular}
\end{table*}

Power and spectrum are the two fundamental resources for any type of wireless network, and NTN is also not an exception. The spectrum allocation is typically performed by the assignment of carriers with equal width from the allocated spectrum for that service. Hence, the number of assigned carriers and their positions are optimized to achieve good signal quality with the minimum resources. Often, the carrier assignment is achieved by the orthogonal splitting of the spectrum resources, which is also known as frequency reuse. However, the strict orthogonality of the frequency bands cannot be always achieved to achieve better spectral efficiency. In case of lack of orthogonality of spectrum resources used by different transceivers can also introduce CCI. The interfering signal can be effectively suppressed by increasing transmission power for the original signals. However, as power is also a scarce resource, we cannot increase the transmission power indefinitely and increasing transmission power will result in a decrease in energy efficiency. For better resource utilization, a more robust radio resource management needs to be designed by controlling both power and spectrum resources \cite{ntn-ra-gen2}. 

Generally, an optimization framework can be considered to optimize the system performance with bandwidth and power constraints. In most cases, such optimization problems are non-linear and non-convex due to objective function nonlinearity and complex constraints involving Signal to Interference and Noise Ratio (SINR) \cite{ntn-ra-gen1}. Furthermore, the carrier assignment indicator variables result in a mixed-integer programming problem \cite{ntn-ra-gen1}. Hence, no optimal solution can be determined using the known methods of convex optimization with low computation complexity. Instead, suboptimal and metaheuristic approaches are proposed, which tackle parts of the problem separately and then iteratively tune the parameters \cite{ntn-ra-gen1}. Different suboptimal approaches are adopted to optimize resource allocations \cite{ntn-ra-opt1,ntn-ra-opt2,ntn-ra-opt3,ntn-ra-opt4,ntn-ra-opt5} for satellite systems. However to reduce computation complexity several heuristic \cite{ntn-ra-heu1} and metaheuristic approaches like GA \cite{ntn-ra-ga}, PSO \cite{ntn-ra-pso} are explored to reach the suboptimal solutions within a shorter computation time.

To tackle this resource allocation issue in real satellite networks in a practical manner, ML approaches are being started to be adopted by the research community. A DL framework is combined with conventional optimization algorithms to overcome the computation complexity issue of the conventional approach in \cite{ntn-ra-dl2,ntn-ra-dl1} by reducing the feature space. A model-free DRL framework is adopted for power allocation of high throughput satellites in \cite{ntn-ra-drl4}. A Q-learning-based long-term capacity allocation algorithm in an RL framework is introduced for a heterogeneous satellite network in \cite{ntn-ra-mul-rl2}. In \cite{ntn-ra-leo-rl1}, an Actor-Critic and Critic Only based RL framework is considered for optimal resource allocation for LEO satellite networks. Different advanced RL frameworks like DRL \cite{ntn-ra-drl1,ntn-ra-drl2, ntn-ra-drl3}, Multi-objective DRL \cite{ntn-ra-modrl-ens} and MADRL \cite{ntn-ra-madrl1} are also proposed to solve the resource allocation issue for satellites.  In table \ref{tab:ra}, we summarize the AI approaches for resource allocation in TNTN. As both power and spectrum are equally important and scarce resources for NTNs, new DL architectures need to be explored to jointly allocate these resources in an efficient manner for NTNs.

\subsubsection{Network Slicing}

\begin{figure}[ht]
  \centering
  \includegraphics[width=0.45\textwidth]{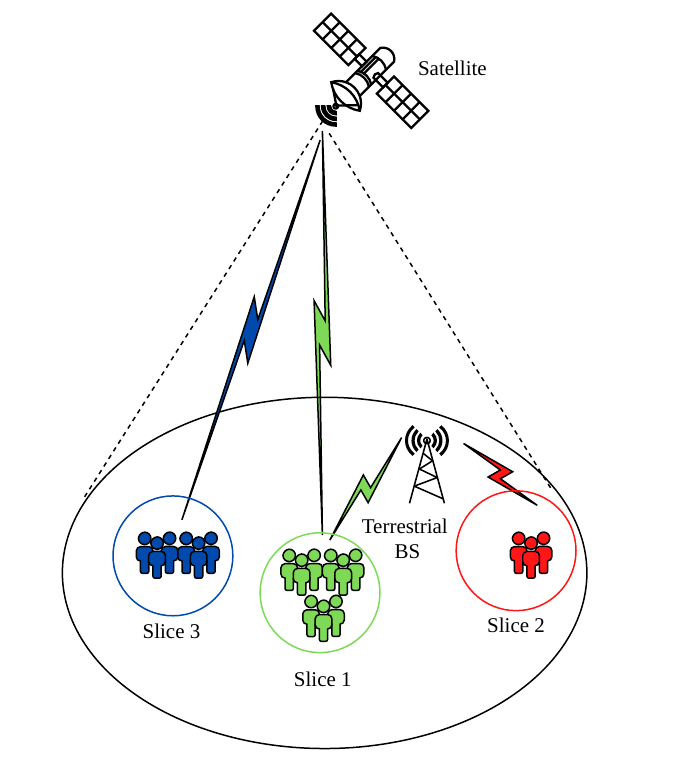}
  \caption{Network slicing in satellite-terrestrial integrated networks.}
  \label{fig:NTN-slicing}
\end{figure}

Network slicing refers to the process of virtually partitioning the physical network into different \textit{network slices} corresponding to different service requirements. The slices are allocated with radio resources as per the demand of the users belonging to the slices. Slicing is useful for wireless networks as each slice can share the same physical network infrastructures while receiving necessary radio resources for guaranteeing a minimum level of service to the users. Also, network slicing provides the flexibility to switch users between slices with different amounts of allocated resources responding to changes in traffic conditions. Integrated TNTN networks appear to be an excellent candidate for applying the concept of network slicing due to their diversified traffic patterns. In fact, different use cases like mMTC and eMBB applications can be extensively benefited through the network slicing in these networks. In Figure \ref{fig:NTN-slicing}, a simple network slicing scenario is shown. Here the network consists of a satellite and a terrestrial BS which form 3 slices in a combined manner. Slice 1 is for high-priority users, they share network resources from the satellite and the terrestrial BS (depicted by green links). Slice 2 is for users with low latency requirements, the terrestrial BS provides resources to the users (shown by red links). Slice 3 is for the remote users who can only be served by the satellite (shown by blue links). 

In a general network slicing framework, a composite utility function consisting of different network performance characteristics like average throughput and other costs like slice reconfiguration cost, resource reservation cost, etc. is formulated as an objective function that needs to be minimized. The constraints are generally the minimum service level 
To ensure real-time implementation, simple heuristic approaches are tested on real platforms \cite{ns-gen1,ns-gen2,ns-gen3}. 
requirements depending on the type of services for particular slices.
\revision{In \cite{ns-gen1}, an extensible 5G network slicing framework in conjunction with satellite networks is discussed to facilitate the integration of satellite services into 5G. In \cite{ns-gen4}, a multi-objective optimization problem comprising latency, computational, and power requirements in an edge-computing scenario is formulated to find suitable slice scheduling strategies based on numerical methods.
However, these approaches do not guarantee optimal performance guarantee. To tackle this issue, different AI-based approaches are explored as it is done in the case of traditional 5G terrestrial networks. 

In \cite{ns-ai-tn}, RL-based network slicing frameworks for satellite-integrated future 6G networks are discussed along with experimental results for simple networks.
In \cite{ns-ai-vntn}, AI-based network slicing for space-air-ground integrated vehicular networks is discussed from the perspective of slice creation, user association, and resource scheduling.
In \cite{ns-vnf, ns-vnf2}, satellite-terrestrial network slice resource allocation frameworks utilizing network function virtualization are presented which can be leveraged for applying advanced AI-based methods. 
In \cite{ns-dl}, FCNNs are used to train a suitable set of network parameters that can produce latency similar to a non-linear optimizer-based network slicer.} In \cite{ns-ai}, a general Radio Access Network (RAN) slicing problem is considered where the objective function is a weighted function of bandwidth and spectrum consumption satisfying QoS and inter-slice isolation constraints. In a simple 2-slice satellite-terrestrial integrated network, different DL architectures are tested. In \cite{ns-ai2}, an ML approach similar to the meta-heuristic ACO approach is considered to realize network slicing in a TNTN environment. An air-ground integrated network is considered in a DRL framework in \cite{ns-ddrl}, later solved by the DDPG algorithm. Here both the actor and critic networks are FCNNs consisting of four layers. Distributed learning architectures can be potentially explored in future works for real network implementations.
\subsubsection{Handover Optimization}




\begin{figure*}
  \centering
  \includegraphics[width=\linewidth]{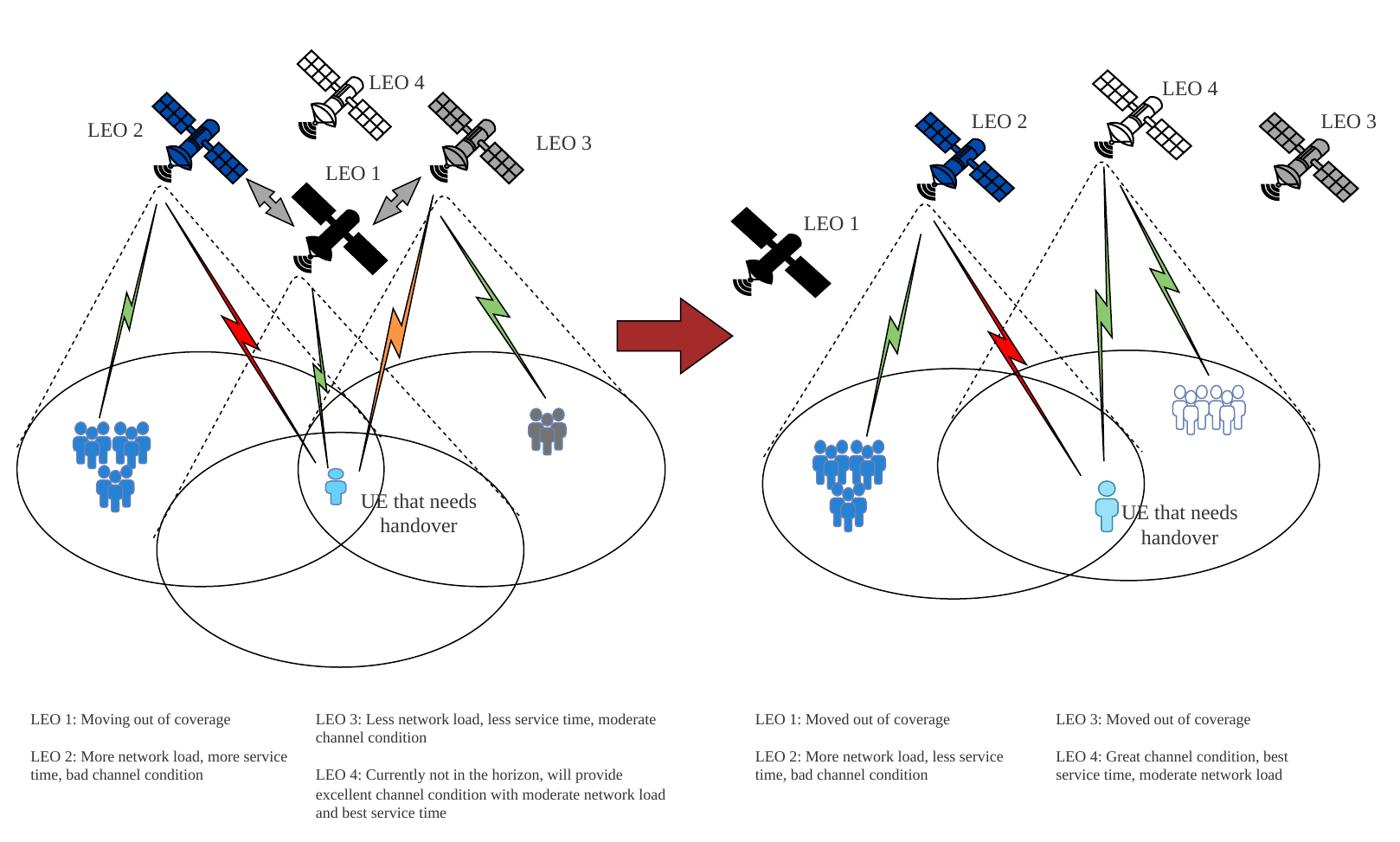}
  \caption{A typical handover scenario in LEO satellite-based NTN.}
  \label{fig:ho}
\end{figure*}

\begin{table*}[htb]
\centering
\caption{Summary of AI approaches for handover optimization in satellite-based NTN.}
\label{tab:ho}
\begin{tabular}{|p{.06\linewidth}|p{.06\linewidth}|p{.06\linewidth}|p{.06\linewidth}|p{.06\linewidth}|p{.05\linewidth}|p{.05\linewidth}|p{.05\linewidth}|p{.05\linewidth}|p{.1\linewidth}|p{.1\linewidth}|}
    \hline
    {\textbf{Reference}} &\multicolumn{4}{c|}{\textbf{Target Optimization Objectives}} &  \multicolumn{3}{c|}{\textbf{Learning Approach}} &\textbf{DL Tool} & \multicolumn{2}{c|}{\textbf{Comments on Models}} \\ \cline {2-8} \cline {10-11}
    & {Signal Quality}  & {Network Load} & {Service Time} & {Tx Delay} &\textbf{SL} & \textbf{UL} & \textbf{RL} && RL Model & DL Model \\ \hline
    
    \cite{ho-sq} & \checkmark & &\checkmark &   &  & & \checkmark & &  Q-Learning & \\ \hline
    \cite{ho-qoe} & \checkmark &\checkmark &\checkmark &\checkmark   &  & & \checkmark & & Online Q-Learning & \\ \hline
     \cite{ho-marl} & \checkmark &\checkmark &\checkmark &  &  & & \checkmark & & Multi-Agent Q-Learning & \\ \hline
     \cite{ho-drl} & \checkmark &\checkmark &\checkmark &  &  & & \checkmark & \checkmark & DQN & CNN \\ \hline
     \cite{ho-ai} & \checkmark &&&   & \checkmark & &  & \checkmark &  & CNN \\ \hline
     \cite{ho-auc-dl} & \checkmark & &\checkmark &   &  & &  & \checkmark &  Auction based approach (Game theory) & FCNN \\ \hline
     \cite{ho-ddqn} & \checkmark &\checkmark & &  &  & & \checkmark & \checkmark & DDQN & FCNN \\ \hline
     \cite{ho-sdqn} & \checkmark &\checkmark & &  &  & & \checkmark & \checkmark & Successive DQN & FCNN \\ \hline
\end{tabular}
\end{table*}

In order to maintain an orbital path around the Earth at a lower altitude, an LEO satellite needs to move at a much higher velocity (around $7.8$ km/s) compared to a GEO satellite. 
So these satellites orbit around the Earth typically within $2$ hours \cite{3gpp1}. 
Due to the smaller orbital period, any LEO satellite remains visible to a ground UE for only several minutes which poses a great challenge for integrating these satellites into the traditional terrestrial networks. 
The UE needs to undergo multiple handovers within a short span of time interval regardless of its mobility status for seamless continuation of the data sessions \cite{NTN_2}. 
This frequent handover phenomenon in LEO satellite networks creates a lot of overhead in communication channels and results in overall degradation in network performances. 
Moreover, due to lower altitudes, the coverage area of an LEO satellite is much smaller compared to a GEO satellite. 
Typically a large number of LEO satellites are needed to maintain global coverage across the Earth with complex constellations. 
In the case of an ultra-dense constellation of LEO satellites (like Starlink), each UE is generally covered by multiple satellites, so the UE can choose the best one from the list of suitable candidate LEO satellites.  
This problem can be potentially solved in an optimization framework jointly considering different handover decision criteria.

In traditional terrestrial communication networks, a UE chooses to attach to a BS based on periodic signal power and quality measurements, such as Reference Signal Received Power (RSRP), and Reference Signal Received Quality (RSRQ) for the link between the BS and the UE. 
Moreover, load balancing is also important to ensure no BS gets overloaded or underloaded as a result of initial attachment or handover procedures.
However, for LEO satellite networks, choosing a satellite BS merely based on signal measurements and network load information is not enough due to the limited visibility time of these satellites. 
So the UE also needs to take the potential service time into account before attaching to a satellite. 
In Figure \ref{fig:ho}, a simple general handover scenario involving multiple LEO satellites and a single UE is shown. 
Here initially, the UE is connected to an LEO satellite, indicated as LEO 2, and it needs an immediate handover to some other neighboring satellite covering the UE, either to LEO 1 or LEO 3 as it will soon lose the coverage of LEO 2. 
As shown in Figure  \ref{fig:ho}, LEO 2 has more network load and bad channel condition, but offers more service time; LEO 3 has less network load, moderate channel condition, but offers less service time. 
Furthermore, a new satellite, LEO 4 becomes available for providing coverage to the UE with excellent channel conditions, great service time with moderate network load. 
So even for a simple case involving 4 LEO satellites, the handover decision is not straightforward for a single UE. 
So finding a suitable handover strategy for a UE jointly considering all handover criteria becomes a complicated problem to be solved.

Different simple greedy strategies, like Maximum Service Time (MST), Maximum Signal Quality (MSQ), or Minimum Network Load (MNL) \cite{ho-gen} are adopted to solve the problem in a simple heuristic manner but none of these approaches provide the optimal solution. 
The satellite handover scenario can be also modeled as a directed graph between different satellites for a single user where the weights can be set by different handover criteria like Quality of Service (QoS), service time, etc. \cite{ho-graph, ho-graph3}. 
A bipartite graph matching problem between the satellites and the users \cite{ho-graph2} is also considered in the literature to provide the optimal handover decision for satellites. In addition, a network flow-based cost minimization approach is considered in \cite{ho-nf} by weighting each edge as the QoS perceived by the user. A handover strategy based on a potential game in a bipartite graph is considered in \cite{ho-game}. Different heuristic algorithms are also proposed to solve the problem \cite{ho-t1, ho-t2, ho-t3}. A dynamic optimization problem is considered to be solved based on forecasting in \cite{ho-for}. Channel reservation is also associated to design an efficient handover algorithm while balancing the load for satellites in \cite{ho-chlo}.

An RL framework can be naturally adopted for solving this problem considering the handover criteria as states and UEs as agents who act by selecting a suitable LEO satellite and collecting a reward based on the network performances. 
In \cite{ho-sq}, only the overall signal quality of the network is maximized using the RL approach without considering any other criteria. 
In \cite{ho-drl, ho-qoe}, a multi-objective optimization problem considering satellite load and signal quality constraints is solved using the DRL approach. 
In real networks, we have a large number of UEs; the handover decision for one UE can affect another UE, so the handover problem needs to be solved in a cooperative manner. 
In \cite{ho-marl}, a MARL framework is considered where multiple UEs cooperatively optimize the number of handovers in the whole network considering different handover criteria. 
In \cite{ho-ai}, using graph matching, a database of optimum handover decisions in satellite networks is produced and later it is used to predict handover decisions using a CNN model. 
Advanced DL architectures like Auction based DL \cite{ho-auc-dl}, DDQN \cite{ho-ddqn}, Successive DQN \cite{ho-sdqn}, etc. are also considered to provide optimal handover decisions.  In table \ref{tab:ho}, we summarize the AI approaches for handover optimization in NTN involving LEO satellites. 
However, as all the system models consider the agents located at the UE side, it does not comply with the current standards where the handover decision is generally controlled by the BSs (satellites in this case). 
Furthermore, the distributed multi-agent learning architectures give rise to stability issues in real implementations.
The handover criteria also need to be carefully investigated to provide the agents with the necessary information to learn the mobility behavior of the environment.
These issues need to be resolved in an efficient manner for future research work in this domain.
\revision{
\subsubsection{Multiple Access}

Multiple access is a vital technique that enables multiple users to efficiently share network resources like spectrum and time. 
In traditional satellite networks, orthogonal multiple access schemes like Time Division Multiple Access (TDMA), Frequency Division Multiple Access (FDMA), Code Division Multiple Access (CDMA), and Space Division Multiple Access (SDMA) are employed. 
These schemes allocate distinct time slots, spectra, codes, or spatial divisions to users, ensuring their orthogonality in resource utilization. 
However, the performance of these conventional methods is constrained by the inherent limitations of these network resources. 
To meet the extremely high data rate and low latency demands of future 6G networks, innovative and efficient multiple access techniques, such as Non-Orthogonal Multiple Access (NOMA) and Rate Splitting Multiple Access (RSMA), have emerged in satellite network research. 
These advanced techniques deliver superior performance, achieving higher spectral efficiency and reduced latency, thereby paving the way for the evolution of future 6G networks.

In contrast to other conventional approaches, NOMA allows multiple users to share the same time-frequency resource block by allocating different power levels to users based on their respective channel conditions. 
The users are assigned the transmit power levels in an inverse manner with respect to their channel conditions, i.e. users with poorer channel conditions are allotted more transmit power whereas users with better channel conditions are allotted lower transmit power.
These signals are subsequently encoded, combined using superposition coding, and transmitted to the receiver.
Then Successive Interference Cancellation (SIC) technique is employed to decode signals for different users.
Starting with the user granted the highest transmission power, SIC successively extracts signals while treating others as interference. 
This process continues down to the user with the lowest transmission power, efficiently enabling multiple users to share the resource block and enhancing spectral efficiency. 
However, it's important to note that SIC's computational complexity increases with this approach.

The optimum power allocation problem in NOMA for NTNs can be formulated as a non-convex problem which is often difficult to solve using conventional approaches. 
In\cite{ntn-noma1}, a long-term power allocation scheme for NOMA in satellite-IoT networks is solved by deriving the optimal decoding order leveraging DL techniques.
A DQN-based DRL approach is investigated in \cite{ntn-noma2} for optimum power allocation in satellite-IoT networks under different channel conditions and delay-QoS requirements of NOMA users.
In another study as discussed in \cite{ntn-noma3}, the non-convex problem involving integer variables is later reformulated as a mixed-integer convex problem which was later solved by two DL techniques instead of conventional iterative solutions.
Some studies also focus on the non-convex user selection problem for given power allocations based on the CSI feedback.
Such a study, \cite{ntn-noma4} used DQN to find out the suitable user pairing for delay-limited NOMA-based satellite networks considering the channel conditions and delay constraints as states. 
k-means UL approach is also considered to find out the pair of terrestrial users to be simultaneously served by space and aerial BSs adopting NOMA in \cite{ntn-noma5}.
Q-learning is adopted in \cite{ntn-noma6} to allocate the time slots and communications channels for IoT-satellite terrestrial relay networks.

Another significant multiple access method, RSMA, is also explored in the context of satellite networks to enhance spectral efficiency. 
In RSMA, user messages are partitioned into two segments: common and private. 
The common signals are collectively encoded and merged into a unified data stream intended for all users, while each user linearly precodes their private messages. 
On the receiver side, the common component is extracted while treating the private signals as noise, employing the SIC technique. 
Subsequently, each user extracts their respective private signals.
This provides the users with another way to share the same resource blocks with an increase in spectral efficiency. 
Generally maximizing the sum rate for both parts is a complicated non-convex problem and can be solved by Weighted MMSE (WMMSE) problem which is difficult to implement in practical hardware. 
A successive convex approximation as well as KarushKuhnTucker (KKT) conditions are used to calculate the transmit power in RSMA power for different beams in satellite networks in \cite{ntn-rsma4}.
However, DL techniques can be extremely useful in modeling the solution framework with low complexity as shown in \cite{ntn-rsma3,ntn-rsma5}.
Here a deep unfolding technique is used to implement the WMMSE algorithm using a deep NN and momentum-accelerated Projection Gradient Descent algorithm.
A DRL framework using Proximal Policy Optimization is used to maximize the sum rate in \cite{ntn-rsma2}.
Here each BS works as an agent, the channel state information i.e. SINR of the private and common messages is used as the states, the action is to find the suitable power allocation whereas the reward is the achieved sum rate. 
}

\subsection{Upper Layer Aspects}
\subsubsection{Computation Offloading}

\begin{table*}[htb]
\centering
\caption{Summary of AI approaches for task offloading in TNTN.}
\label{tab:to}
\begin{tabular}{|p{.06\linewidth}|p{.08\linewidth}|p{.08\linewidth}|p{.08\linewidth}|p{.05\linewidth}|p{.05\linewidth}|p{.05\linewidth}|p{.05\linewidth}|p{.1\linewidth}|p{.1\linewidth}|}
    \hline
    {\textbf{Reference}} &\multicolumn{3}{c|}{\textbf{Problem Insight}} &  \multicolumn{3}{c|}{\textbf{Learning Approach}} &\textbf{DL Tool} & \multicolumn{2}{c|}{\textbf{Comments on Models}} \\ \cline {2-7} \cline {9-10}
    & {Energy Consumption}  & {Tx Delay} & {Computational Resources}  &\textbf{SL} & \textbf{UL} & \textbf{RL} && RL Model & DL Model \\ \hline

    \cite{to-rl} & \checkmark & \checkmark & & &  & \checkmark & \checkmark & DDPG &  FCNN \\ \hline
    \cite{to-drl1} & \checkmark & \checkmark &  & &  & \checkmark & \checkmark & Model-Free &  FCNN \\ \hline
    \cite{to-drl2} & \checkmark & \checkmark &  \checkmark & &  & \checkmark & \checkmark & Value Iteration, DQN, DDQN &  FCNN \\ \hline
    \cite{to-drl2} & \checkmark & \checkmark &  \checkmark  & &  & \checkmark & \checkmark & Dueling DDQN &  FCNN \\ \hline
    \cite{to-marl} & \checkmark & \checkmark & \checkmark & &  & \checkmark &  & MARL &   \\ \hline
    \cite{to-madrl} & \checkmark & \checkmark &  & &  & \checkmark & \checkmark & DQN based MARL & FCNN  \\ \hline
    \cite{to-dl1} & \checkmark & \checkmark &  \checkmark  & \checkmark &  &  & \checkmark &  &  LSTM \\ \hline
    \cite{to-dl2} & \checkmark & \checkmark &  \checkmark  &  &  & \checkmark & \checkmark & DQN &  FCNN \\ \hline
    \cite{to-ddl1} & \checkmark & \checkmark &   &  &  & \checkmark  & \checkmark &  &  Distributed FCNN (For solving Optimization) \\ \hline
    \cite{to-ddl2} & \checkmark & \checkmark &  \checkmark  & \checkmark &  &  & \checkmark &  &  Distributed FCNN (For solving Optimization) \\ \hline

\end{tabular}
\end{table*}

\begin{figure}[ht]
  \centering
  \includegraphics[width=0.5\textwidth]{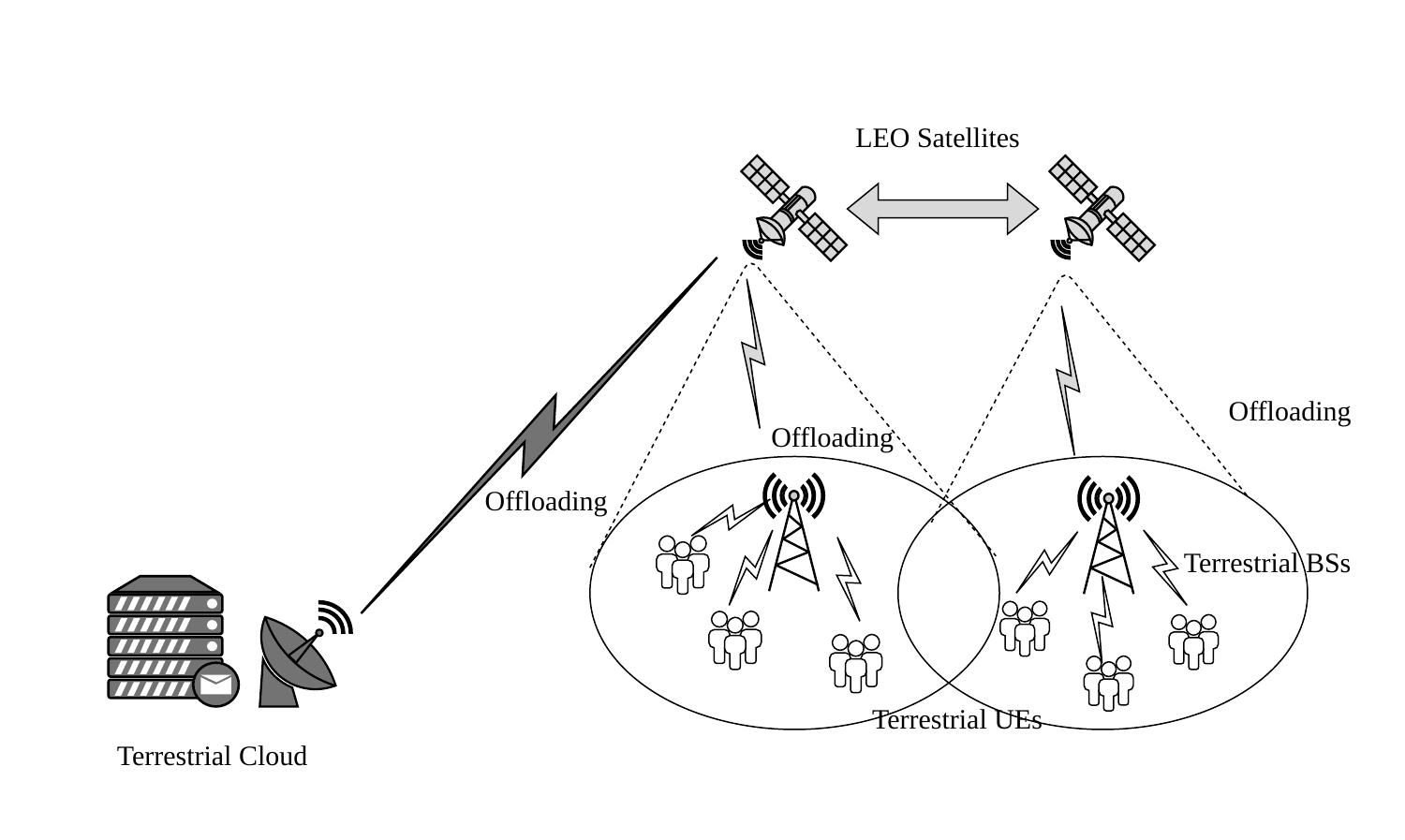}
  \caption{Task offloading in satellite-terrestrial integrated networks.}
  \label{fig:NTN-task-off}
\end{figure}

One of the most important applications of satellite-terrestrial integrated networks is enhancing the computation capabilities of existing terrestrial network architectures leveraging satellites. With traditional terrestrial networks, supporting a diverse set of new applications like AR, VR, etc. with high data processing  and extremely low latency requirement can get very challenging. Generally, terrestrial BSs are deployed sparsely due to high infrastructure and maintenance costs. Due to resource constraints, in case of the high demand for data processing for these types of applications, the BSs need to offload the computation tasks to the terrestrial cloud via the satellites \cite{to-ov1}. However, due to longer propagation delay, the latency requirements set by the applications are difficult to be met \cite{to-ov2}. Nevertheless, due to the emergence of LEO satellites with comparatively low propagation delay, the overall delay is considerably reduced. Also instead of acting as relays, the satellite can now also do the processing works acting as edge-servers. So we can consider a three-level hierarchical architecture comprising of ground UEs connected to terrestrial BSs, LEO satellites, and terrestrial cloud as shown in Figure \ref{fig:NTN-task-off} where the terrestrial BSs can offload the computational tasks to LEO satellites and to terrestrial clouds via the LEO satellites. 

The main challenges in task offloading problems lie in meeting the delay constraints for low-latency applications while minimizing the energy consumption for the satellites. So this can be formulated as an optimization problem to come up with an efficient offloading approach for integrated TNTN architecture. Such an optimization problem is solved using different conventional approaches like 3D hypergraph matching  \cite{to-gen1}, game theory \cite{to-game}, stochastic approach \cite{to-stoch}, efficient algorithms \cite{to-gen2} in the existing literature. In \cite{to-joint}, a joint optimization framework comprising task offloading and resource allocation is also considered in an integrated satellite-terrestrial environment. Although these algorithms work well in theory for particular scenarios, in real networks, the feasibility of these algorithms is compromised due to different issues. Some of these works do not consider the cooperation among terrestrial cloud and LEO satellite servers which result in sub-optimal approaches \cite{to-gen1,to-game}. Also, these approaches are some predefined models highly dependent on different network states causing a large overhead in networks. Moreover, they usually converge to the solutions after a large number of iterations causing high computational complexity. 

To tackle these issues, different ML approaches are proposed in the literature to solve task offloading problems. In \cite{to-drl1}, a DRL-based task offloading framework dependent on channel state information is proposed. A similar DRL-based framework is also considered in \cite{to-drl3} with additional consideration of the dynamic queue condition in satellites. In \cite{to-drl2}, both DQN and DDQN are explored to solve the task offloading problem in a decentralized manner. DDPG algorithm is considered to solve the optimization problem in a DQN framework in \cite{to-rl} while taking the potential security issues into account. An LSTM model is used to solve the task offloading problem while considering channel conditions and energy dynamics in \cite{to-dl1}. A DL-based caching strategy is considered in satellite edge networks in \cite{to-dl2}. As we have multiple satellites in the real networks, to improve the overall system performances, different multi-agent architectures are considered both in a distributed \cite{to-ddl1} and cooperative environment \cite{to-marl,to-madrl}. Distributed architectures for generating discrete offloading decisions in a supervised manner are also considered in \cite{to-ddl2}.  In table \ref{tab:to}, we summarize the AI approaches for task offloading in TNTN. As the delay constraints vary with network traffic types, the offloading decisions need to be derived taking network traffic types into account. Potential research works can show how computational offloading can be done for various network traffics and show superior network performances.

\subsubsection{Network Routing}

\begin{table*}[htb]
\centering
\caption{Summary of AI approaches for network routing in TNTN.}
\label{tab:nr}
\begin{tabular}{|p{.06\linewidth}|p{.08\linewidth}|p{.08\linewidth}|p{.08\linewidth}|p{.05\linewidth}|p{.05\linewidth}|p{.05\linewidth}|p{.05\linewidth}|p{.1\linewidth}|p{.1\linewidth}|}
    \hline
    {\textbf{Reference}} &\multicolumn{3}{c|}{\textbf{Target Optimization Objective}} &  \multicolumn{3}{c|}{\textbf{Learning Approach}} &\textbf{DL Tool} & \multicolumn{2}{c|}{\textbf{Comments on Models}} \\ \cline {2-7} \cline {9-10}
    & {Throughput}  & {Tx Delay/jitter} & {Error Rate}  &\textbf{SL} & \textbf{UL} & \textbf{RL} && RL Model & DL Model \\ \hline

    \cite{ntn-rt-rl1} & \checkmark & \checkmark & \checkmark & &  & \checkmark &  & Q-Learning &  \\ \hline
    \cite{leo-rt-rl1} & \checkmark &  & \checkmark & &  & \checkmark &  & Q-Learning &  \\ \hline
    \cite{leo-distributed-rt-rl2} & \checkmark & \checkmark &  & &  & \checkmark &  & Distributed Q-Learning &  \\ \hline
    \cite{ntn-rt-fuzzy-cnn} & \checkmark & \checkmark &  & &  & &  \checkmark &  & Fuzzy-CNN \\ \hline
    \cite{IWC_AI_NTN_2} & \checkmark &  & \checkmark &  \checkmark &  & & \checkmark  &  & CNN \\ \hline
    \cite{ntn-rt-dl2} & \checkmark & \checkmark &  &  \checkmark &  & & \checkmark  &  & FCNN \\ \hline
    \cite{ntn-rt-dist-ml1} & \checkmark & \checkmark & \checkmark &  \checkmark &  & & \checkmark  &  & ELM \\ \hline
    \cite{leo-rt-gnn} &  & \checkmark & \checkmark & \checkmark   &  & & \checkmark  &  & Graph Neural Network (GNN) + FCNN \\ \hline
    \cite{ntn-rt-drl1} &  & \checkmark & \checkmark &   &  & \checkmark & \checkmark  & DDPG & LSTM \\ \hline

\end{tabular}
\end{table*}

\begin{figure}[ht]
  \centering
  \includegraphics[width=0.5\textwidth]{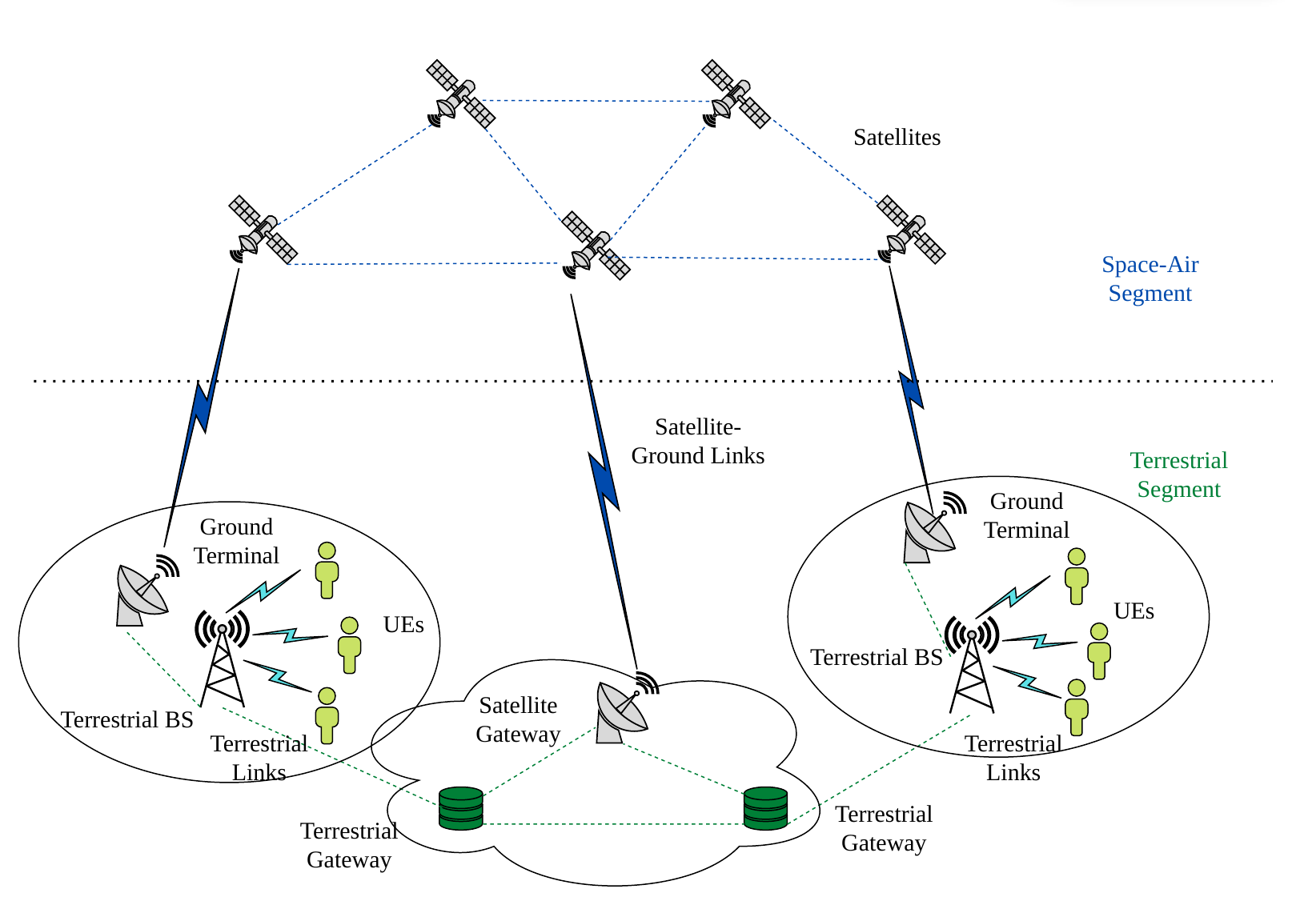}
  \caption{Network routing in satellite-terrestrial integrated networks.}
  \label{fig:NTN-routing}
\end{figure}

In wireless networks, depending on the traffic and channel conditions, the network traffics are routed to different paths among different network nodes so that the overall network performance can be improved. In any network with static channel and traffic conditions, this routing problem can be transformed into the well-known shortest path problem and solved by Dijkstra's algorithm \cite{rt-dj}. Here the network nodes can be considered as nodes in the graph and the edges can represent the links between different nodes. The weights of the edges can be defined based on the target network performance metrics like delay, jitter, throughput, packet loss, etc. However, the topology of the real satellite-terrestrial integrated networks (shown in Figure \ref{fig:NTN-routing}) are very complex and dynamic due to hierarchical network architecture and uncertain channel and traffic conditions, respectively. So simple Djikstra's algorithm cannot be directly applied to meet the performance requirements in these networks.

In real satellite networks, Asynchronous Transfer Mode (ATM) routing is introduced in \cite{ntn-rt-atm}. The well-known Open Shortest Path First (OSPF) \cite{ospf} protocol-based Internet Protocol (IP) routing is also adapted to the dynamic satellite environment in \cite{ntn-rt-ip}. However, the dynamics of satellite-terrestrial networks are very different from traditional terrestrial networks due to their highly dynamic network topology, link status, and traffic conditions. Depending on the instantaneous network topology, a static-dynamic combined routing scheme is considered in \cite{ntn-rt-adapt}. An ant-colony-based optimization (ACO) framework is considered in \cite{ntn-rt-aco}. In \cite{ntn-rt-kfwca}, a Kalman filter-based Wolf Colony Optimization algorithm is used to solve the local optimal solution issue in \cite{ntn-rt-aco}. An improved ACO framework is considered to find out the optimal set of links with multiple network constraints in \cite{ntn-rt-opt}. A Coordinate Graph (CG) model-based network routing approach for three-dimensional TNTN  is considered in \cite{ntn-rt-tvgraph}. In \cite{ntn-rt-set}, minimum flow maximum residual path-based network flow algorithm is used to find out the optimal network routing path for satellites. In \cite{ntn-rt-hygeo}, a 3-dimensional network mapping using hyperbolic geometry is considered for integrated satellite-terrestrial networks. 

To cope with the dynamic environment in integrated satellite-terrestrial networks, different DL architectures are proposed in the literature. In \cite{ntn-rt-fuzzy-cnn}, fuzzy logic is used to evaluate task requirements to improve the CNN output for optimal path allocation. The network routing optimization problem can be put into an RL framework. In \cite{ntn-rt-rl1}, a speed-up Q-learning algorithm is used to find out the optimal routing strategy for TNTN. A similar Q-learning-based RL framework is also considered to solve the routing problem for LEO satellites in \cite{leo-rt-rl1}. To tackle the complexity issue, a DRL framework is used to generate optimal routing strategies in TNTN \cite{ntn-rt-drl1} and LEO satellite networks \cite{leo-rt-rl1}. FCNN \cite{ntn-rt-dl2}, CNN \cite{IWC_AI_NTN_2}, etc. architectures are used to solve the routing problem in a supervised manner. Other ML frameworks like GNN \cite{leo-rt-gnn} and ELM \cite{ntn-rt-dist-ml1} are also considered to solve the routing problem in NTNs. In table \ref{tab:nr}, we summarize the AI approaches for network routing in TNTN. Recursive NN architectures need to be also explored for capturing the temporal behavior in network routing decisions. Furthermore, the channels are extremely dynamic and time-varying in the case of TNTNs; the channel conditions can be also considered in the learning criteria of RL frameworks. 

\revision{
\subsubsection{Traffic Prediction}
Traffic prediction is very critical in modern communication systems to ensure high-speed low latency communications.
Particularly, in NTNs, accurate traffic prediction is extremely crucial due to the highly dynamic network topology as well as diverse user requirements. 
At its core, traffic prediction involves forecasting future network traffic based on past usage patterns.
Conventional approaches such as Auto Regressive Moving Average (ARMA), Auto Regressive Integrated Moving Average (ARIMA) \cite{tf-arima, tf-arima-2}, etc. are typically used for these predictions.
However, the DL approaches have emerged as more effective alternatives by providing improved performances in recent times due to their inherent capability to capture spatial and temporal correlations. 

In \cite{ntn-tf1}, Radial Basis Functions (RBF) neural network-based short-term traffic flow forecasting is proposed. 
In \cite{ntn-tp-lstm}, an LSTM-based architecture is utilized for traffic prediction due to its temporal characteristics handling capability where the attention mechanism is used to balance the effect of inputs on outputs properly.
The RNN architectures suffer from gradient explosion issues.
To overcome this issue, GRU architectures are explored for traffic prediction in \cite{ntn-tp-gru, ntn-tp-gcn_gru, ntn-tp-mod-gru}.
In \cite{ntn-tp-gru}, the transfer learning approach and particle filter online training algorithm are combined to address the lack of online training data and reduce the training time complexity. 
In \cite{ntn-tp-gcn_gru}, GNNs are used to extract the spatial features of the satellite network traffic from the input network topology, which is later used as an input to a GRU network for traffic prediction.
In \cite{ntn-tp-mod-gru}, on top of the attention mechanism and GRU models, PSO is used to obtain the best set of hyperparameters for the network. 
}

\begin{figure}[ht]
  \centering
  \includegraphics[width=0.5\textwidth]{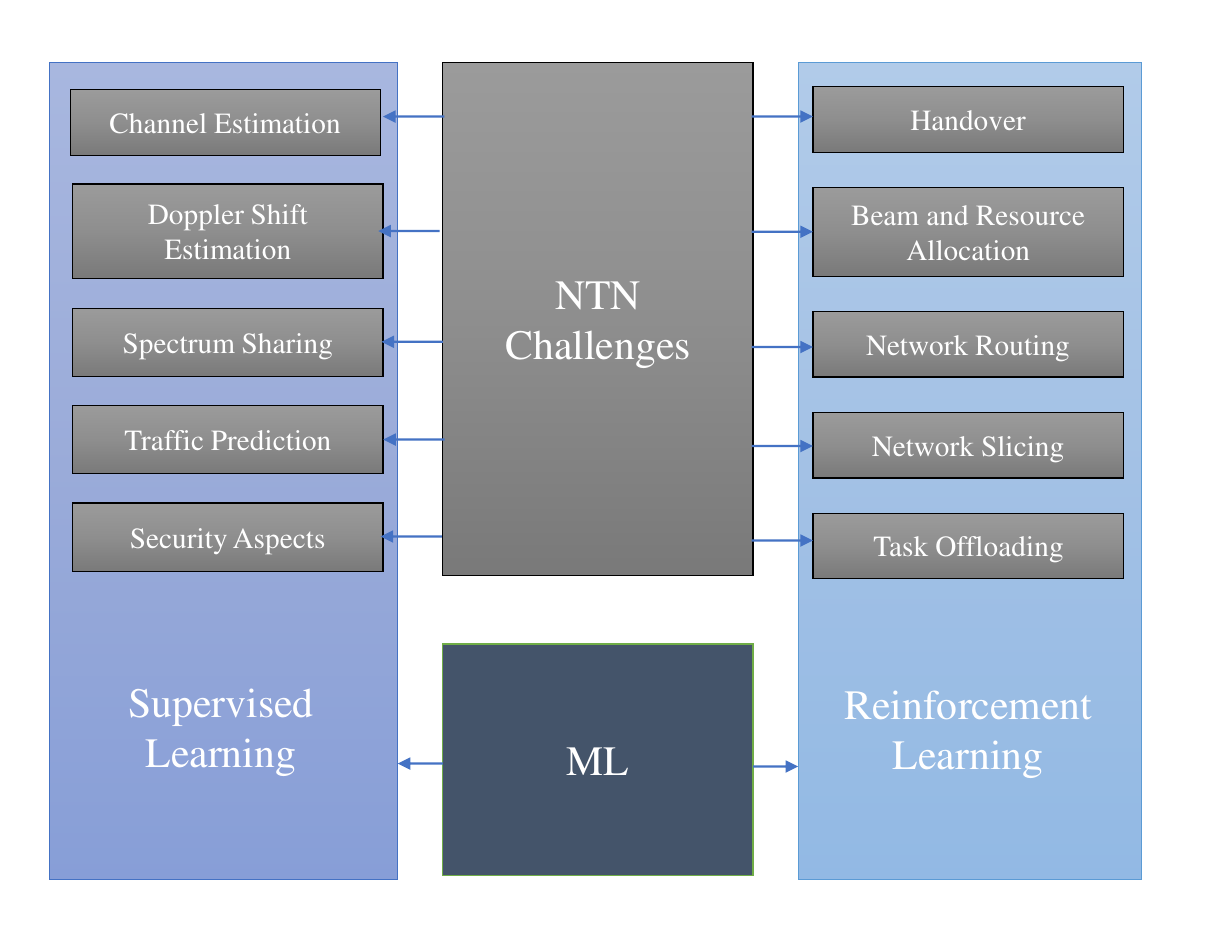}
  \caption{Relationship diagram different AI techniques and NTN challenges.}
  \label{fig:AI-NTN-relation}
\end{figure}
\vspace{3mm}
\noindent\textit{Key Takeaways:
As evidenced by the above discussion, various RL techniques are used to examine network optimization problems such as handover, beam, and resource allocation, task offloading, network routing, and network slicing, while SL techniques are employed to tackle estimation problems, such as Doppler shift, channel state, and spectrum sensing. UL techniques have not been extensively covered in the literature due to the ambiguity and difficulty of applying them in real networks. To further illustrate the interrelations between different NTN challenges and AI techniques, we present Figure \ref{fig:AI-NTN-relation}.}

\section{AI-NTN Integration: Current Status}

\label{sec:progress}

In the previous section, we discuss how AI can be beneficial for us in resolving potential NTN issues for the next-generation 6G networks. 
In this section, how AI can be applied to real systems to resolve the challenges associated with NTNs.
We begin our discussion by discussing the current ML testbeds for satellites. 
Then we discuss how AI can be potentially applied to future 6G networks by utilizing the RAN Intelligent Controller (RIC) embedded in the Open Radio Access Network (O-RAN) framework \cite{oran} to overcome the inflexibility of the monolithic cellular networks. 
Finally, we provide a discussion on current research efforts toward realizing the Software Defined Radio (SDR) based 5G-NTN platform development in the O-RAN framework. 

\subsection{ML Testbeds for Satellite Networks}

\label{sec:ml-satellite-testbed}

MultI-layer awaRe SDN-based testbed for SAtellite-Terrestrial networks (MIRSAT) testbed \cite{mirsat} provides a Software Defined Network (SDN) based experimentation platform for testing network slicing algorithms on NGSO constellations. The European Space Agency (ESA) has numerous completed projects focusing on the applicability of AI techniques in satellite networks such as MLSAT \cite{mlsat} and SATAI \cite{satai}. There are also several other ongoing projects of ESA focusing on AI-satellite issues like AI integrated 5G-Satellite testbed \cite{anchor}, AI-based interference detection \cite{skymon}, AI-based signal processing \cite{spaice}, etc. All these testbeds show promises for AI to be an integral part of future satellite-terrestrial integrated networks.

\begin{figure*}[ht]
  \centering
  \includegraphics[width=0.8\textwidth]{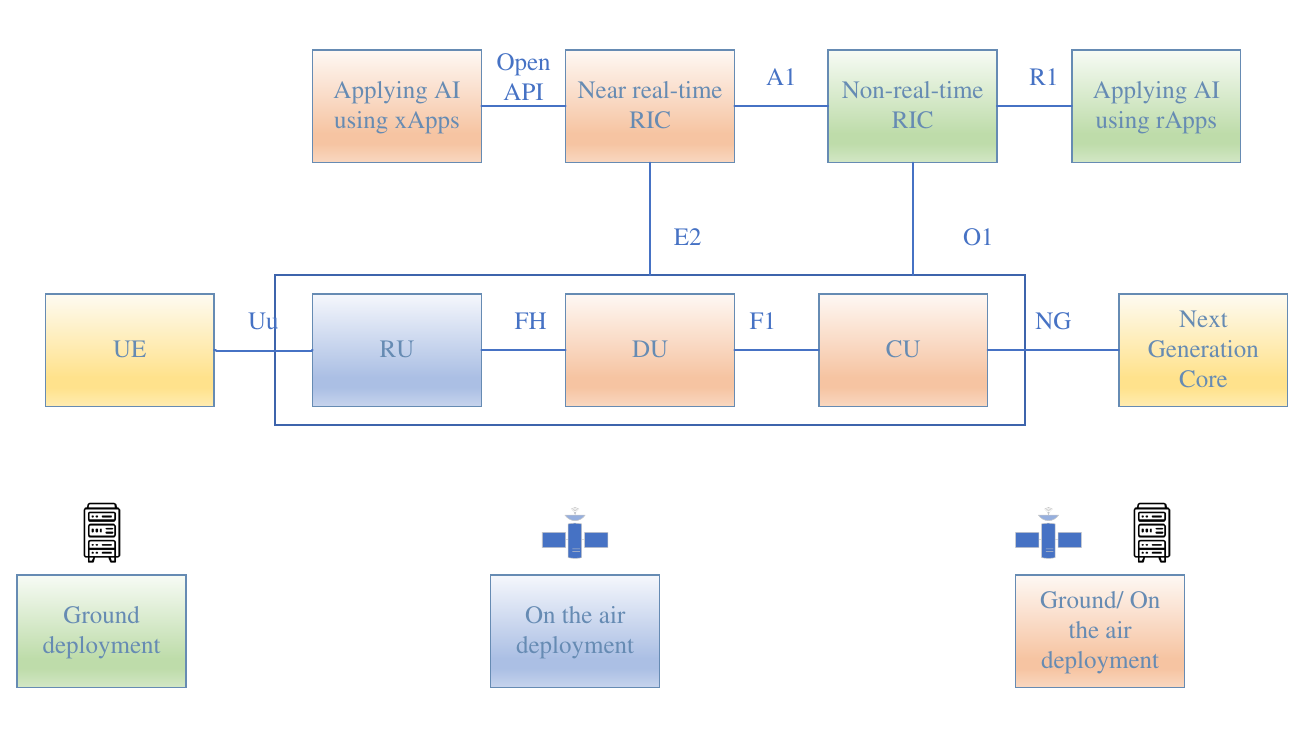}
  \caption{Various architectural deployments for NTN in the O-RAN framework for regenerative payload.}
  \label{fig:ntn-oran}
\end{figure*}

\subsection{AI-NTN Integration through O-RAN-based RIC}

\label{sec:oran}

Traditional 5G networks with little or no reconfiguration capabilities suffer from a wide variety of challenges to satisfy the heterogeneity and variability of the networks and meet the strict application requirements \cite{oran-2}.
Even though there has been a significant amount of research on addressing different issues in 5G cellular networks, an open interface for the deployment of AI algorithms is required. 
O-RAN offers a general framework for the deployment of AI algorithms in 5G-Advanced networks \cite{oran-3}. 
It achieves this by facilitating an open interface that enables the exchange of network KPIs and control information between the RAN and the AI controller. 
This integration allows for the implementation of a closed-loop control framework for the RAN using different AI approaches \cite{oai-ml-1, oai-ml-2}. As a potential integral part of future 6G networks, NTNs are expected to be deployed in the O-RAN framework to leverage AI capabilities effectively. 

In 5G, a base station, namely gNB has multiple functional splits, namely:
\begin{enumerate}
    \item Central Unit (CU): responsible for higher layers such as non-real-time link and network layer functionalities.
    \item Distributed Unit (DU): responsible for lower layer such as near real-time link and upper PHY layer functionalities.
    \item Radio Unit (RU): responsible for low PHY layer functionalities.
\end{enumerate}
A new central controller entity called RIC is introduced in the O-RAN architecture which can provide network monitoring and control functionalities in near real-time and non-real-time through external and internal applications, called xApps and rApps respectively, for the purpose of network optimization.
Evidently, these xApps and rApps can provide us with an effective way of deploying AI algorithms extracting network KPIs, and sending control commands for optimizing network performance.

To realize the NTN architecture in the O-RAN framework, the satellites can be used for either transparent or regenerative payloads as discussed in \revision{Section} \ref{sec:ntn-arch}. 
In the case of regenerative payloads, where the NTN platforms work as BSs, there are multiple options for potential O-RAN-based NTN deployment.
There can be three different architectural deployments for NTN gNBs in the regenerative architecture:
\begin{enumerate}
    \item RU in the space/air, CU and DU on the ground,
    \item Both RU and DU in the space/air, CU on the ground,
    \item CU, DU, and RU in the space/air.
\end{enumerate}

The non-real-time which does not need to consider latency requirements, is expected to be deployed on the ground considering power, onboard capability, and mobility constraints.
However, the near real-time RIC needs to be close to DUs to provide near real-time control functionalities which provide two different options for its deployments with corresponding pros and cons.
The near real-time RIC should be on the ground when only RU is in the air, whereas it should be also in the air in the other two cases. 
There is a clear trade-off between the latency and power, mobility, and onboard capability constraints. If the near-real-time RIC is in the air, the latency for control commands will be low, whereas the cost will be high for hosting it in the air.
In Figure \ref{fig:ntn-oran}, the potential framework for AI-Enabled NTN deployment in O-RAN framework as specified in \cite{oran} is illustrated.
Depending on the deployment scenarios of the near-real-time RIC, the xApps can be also deployed also in the air or on the ground, and so do the AI control algorithms.

\subsection{Current Research Efforts for AI-NTN Integration through O-RAN-based RIC}

\label{sec:ntn-oai}

To facilitate the integration of NTN into 6G networks, there have been already some technical advancements and experimental research works towards developing real prototypes for testing and evaluation of proof-of-concept methods.
OpenAirInterface (OAI) is an open-source 3GPP compliant SDR-based protocol stacks that are widely used across the research community for experimentation with 5G networks \cite{oai-5g}.
As specified in the O-RAN framework, OAI protocol stacks adopt the notion of RIC by enabling service-oriented controllers using an efficient Software Development Kit, called Flexible RIC (FlexRIC) \cite{flexric}. 
This RIC provides an interface for applying AI algorithms in order to optimize the network performances through xApps as discussed before. 
This enables us to perform experiments for testing diverse AI approaches for optimizing the performances of real 5G networks. 

OAI has been adopted for developing experimental prototypes with 5G-NTN adaptations due to its efficient and flexible design and structure \cite{oai-5g-ntn}. 
Currently, there are several research projects on 5G-NTN that are leveraging OAI protocol stacks to perform experiments with NTN adaptations for both in-lab validation and over-the-satellite testing. 
5G AgiLe and flexible integration of SaTellite And cellulaR (5G-ALLSTAR) \cite{5g-allstar} and 5G New Radio EMUlation over SATellite (5G-EmuSat) \cite{5g-emusat} project developed a 5G-NTN platform with necessary PHY and MAC layer 5G-NR adaptations on top of OAI 5G protocol stacks and a satellite-channel emulator for in-lab validation. 
5G-EmuSat even has also demonstrated its over-the-satellite capability by having direct access to a UE using a satellite channel. 
5G Space Communications lab also has performed in-lab validation experiments extending OAI-4G protocol stacks for NTN along with ISL implementation using SDR \cite{5g-space-com}. 
Two current ongoing projects focusing on GEO and LEO satellites, named 5G-GOA \cite{5g-goa} and 5G-LEO \cite{5g-leo} respectively, are currently working on implementing necessary 3GPP NTN adaptations extending from 5G protocol stacks of OAI. Even though current implementations are mostly for demonstration purposes, integration of NTN into OAI 5G protocol stacks paves the way for deploying AI algorithms through xApps in the future.

\vspace{3mm}
\noindent\textit{Key Takeaways: Currently, there are some deployed ML testbeds specifically designed for satellite networks. Moreover, O-RAN is envisioned to unleash the great potential of AI in enabling the future 6G networks via satellite-based NTNs by addressing various challenges associated with it. 
Nevertheless, both O-RAN and NTN standardization aspects are still in the development process, and different SDR-based 5G protocol stacks, such as OAI, are being incorporated with NTN adaptations.}

\section{AI-NTN Integration: Challenges}

\label{sec:chal}

NTNs come with an intrinsic set of challenges when it comes to deploying AI models. 
Even though there is a significant decrease in the launching and maintenance cost of various NTN platforms, especially satellites, cost optimization is still one of the major limiting factors of realizing NTNs for 6G communication on a large scale. 
With that being the case, these platforms have limited power, spectrum, and computational resources which limits the performance of the AI models. 
The unavoidable long propagation delay along with the complex and time-varying nature of the NTN environment introduces additional challenges for AI models to be trained and deployed in real-time. 
In this section, we discuss these open research issues to get an insight into designing an efficient AI-based non-terrestrial system with robust and superior performance.

\subsection{Limited Onboard Capability}

\revision{
Advanced AI applications necessitate specialized AI-capable embedded chipsets developed by leading technology companies, including NVIDIA, AMD, Intel, and Qualcomm \cite{ai-chip}. 
Typically, these chipsets feature on-chip accelerators like CUDA and Tensor, enabling highly efficient parallel processing of intricate data tasks, especially those involving extensive matrix operations as utilized in DL methodologies.
}
The performance of these AI algorithms highly depends on the availability of the computational resources and data processing capabilities of the AI hardware blocks. 
More computational capability typically means more power consumption as well as more physical space which increases the overall maintenance cost for the non-terrestrial platforms. 
Most non-terrestrial platforms such as satellites have a very limited amount of computational resources due to cost optimization. 
\revision{
The efficiency of computing devices, particularly those used for onboard purposes, is often quantified by evaluating the ratio of computational power to the product of power consumption, total mass, and associated cost. 
This metric, highlighted in \cite{ai-chal}, necessitates a notably high value for onboard computing devices in satellites.
All these AI-capable chipsets must undergo meticulous design and rigorous testing procedures to ensure they satisfy the minimum computing efficiency standards mandated for space platforms.
}
A lot more advancements in miniaturization and power efficiency are necessary to ensure the adaption of AI models and algorithms into non-terrestrial platforms with adequate onboard capabilities \revision{as well}.  

\subsection{Aging of Information}

The long propagation delay is a great challenge in the way of AI-based NTN deployment. 
Online RL frameworks are very promising for solving different NTN challenges due to their inherent capability of adapting to fast time-varying environments as we show in the previous sections. 
However, the performance of these algorithms is highly dependent on the feedback received from the environment. 
As the network changes very rapidly, the feedback has to be real-time or near real-time to ensure the integrity of the information embedded in the feedback. 
\revision{For example, the resource allocation for different network slices and users needs to be near-real-time (in the order of 1-10 ms) and real-time (less than 1 ms), respectively.}
For NTNs, as we know from \revision{Section} \ref{sec:ntn-feat}, the propagation delay is extremely high due to the long distance between the transmitter and the receiver. 
Thus the feedback exchange time intervals are quite high compared to terrestrial environments which hampers the online training approach greatly. 
Furthermore, to adapt to the highly time-varying environment, AI models usually need to send an appropriate chain of control commands to the network components. 
\revision{Due to the rapid channel variations, the channel coherence time is significantly reduced, leading to potential issues where both received information and transmitted control commands may become outdated, resulting in reduced effectiveness for resource allocation decisions from AI algorithms.}

\subsection{Additional Communication Overheads}

On top of long propagation delay, non-terrestrial platforms also have limited bandwidth due to the scarcity of spectrum resources and ensure no additional interference to the licensed services.
As the generic RL frameworks depend on the feedback received from the environment, the additional overhead introduced by the network parameters results in undesirable network resource consumption. 
Even though CSI feedback in 5G networks contains a set of network parameters, this may not be enough for all different network problems.
\revision{For instance, when dealing with the handover optimization problem, many solution techniques operate under the assumption that mobility state information for satellites and users is readily available, which is not typically included in the CSI feedback. 
Consequently, this assumption introduces additional feedback overhead alongside the existing format.}
This additional communication overhead puts an additional burden on the limited spectrum of resources allocated for the non-terrestrial platforms.

\subsection{Security Aspects}

Applying AI in NTNs introduces more vulnerability to various security attacks by introducing new attack surfaces and less transparency. 
Adversarial attacks, data poisoning, and model evasion involving the manipulation of input data to AI models can cause degradation in network performance and reliability \cite{ai-chal-adv}. 
\revision{Since this input data is gathered from the NTN environment, featuring various attack interfaces for potential attackers, there is a risk that the training data used for AI models could be compromised by these attackers.}
Denial-of-Service (DoS) attacks can cause interruptions in crucial network operations by overwhelming the network with too much resource consumption \cite{ai-chal-dos}.
\revision{As an illustrative example within the context of NTNs, an attacker could conceivably gain access to one of the network slices. 
They could then exploit this access to excessively consume network resources by setting up extreme requirements, potentially leading to network congestion and subsequently causing a decline in overall network performance.}
\revision{As discussed before, the AI controller and the network environment, especially in the online setup, needs to exchange control information during training of the models.}
While carrying the control information between the AI controller and the network, an attacker can intercept and possibly modify this information, which is known as a Man-in-the-Middle (MiM) attack \cite{ai-chal-mim}.
\revision{ This can often result in a degradation in network performances due to the compromised control information.
Therefore, A security-constrained framework for deploying AI models needs to be designed carefully in such a way that they can detect and mitigate these attacks while maintaining the overall network performance.}

\subsection{Environmental Conditions}

NTN platforms, especially satellites are generally deployed in pretty hostile environments with extreme radiation, extreme temperatures, and other extreme environmental conditions. 
The computational hardware for AI models is very susceptible to radiation as they are built on customized circuitry \cite{ai-chal}. 
\revision{The satellites need to consider both single-event effects caused by ionizing particles \cite{ai-chal-ion} as well as the effects due to long-term radiation \cite{ai-chal-rad}.
These effects may result in bit flips both in registers and memories introducing errors in the control logic of their hosting hardware platforms.
}
Also, in space, these circuitry elements need to withstand extreme temperatures for a long period of time \cite{ai-chal-temp}.
To ensure the proper performance of these models, the hosting hardware components are required to be more advanced and have rigorous testing to ensure extreme environmental tolerance, which increases the cost of the satellite operation.

\subsection{Scalability Issue}

RL frameworks can be naturally applied to address a variety of NTN problems with control objectives as we discuss in the previous section. 
However, for large-scale 6G satellite-terrestrial integrated wireless networks, the complex network topology entails high dimensional state and action spaces that can lead to high computational complexity for RL models \cite{rl-comp}.
In the case of MARL frameworks, the state space grows exponentially with the increase in the number of agents \cite{marl-chal}, making this approach infeasible for large-scale real networks. 
Although DRL approaches can be helpful in reducing the state space \cite{drl}, more research is needed to effectively address this challenge in order to successfully deploy RL approaches in TNTNs.

\subsection{Lack of Convergence}

An important challenge to be addressed when applying the distributed RL framework in real networks for solving various important NTN challenges like handover optimization is to deal with its uncertainty in convergence \cite{marl}. 
In this framework, multiple agents try to optimize their goals based on the rewards received from the environment. 
In a competitive environment, when all the agents are attempting to maximize their long-term returns, they may take conflicting actions, resulting in a non-stationary environment with no convergence to an optimum state \cite{marl-chal}. 
As a result, no optimum policy can be obtained for the system as a whole. 
As highlighted in \cite{ho-marl}, this convergence issue has limited the number of UEs (agents) that can be considered in the simulation environment, thus hindering the potential of this approach.

\subsection{Scarcity of Quality Data}

All ML approaches are data-driven, so the availability of suitable training data is of paramount importance for the improved performance of these methods. 
However, in satellite-terrestrial integrated networks, the generation of quality data can be sometimes very costly and inefficient, even impossible at times due to spectrum and intermittent connectivity constraints. 
Due to this inherent data generation issue, applying different ML approaches can get extremely challenging. 
Additionally, the data distribution and characteristics in non-terrestrial environments may differ significantly from terrestrial environments, requiring careful consideration during model training and adaptation.
As a result, the training procedure can be greatly hampered resulting in performance degradation of these approaches in real networks. 

\subsection{Complicated Hyperparameter Settings}

The complexity of satellite-terrestrial networks, such as their topology and time-varying nature, can make traditional ML approaches less effective. 
As a result, DL methods have become increasingly popular due to their powerful feature extraction capabilities through NNs. 
The performance of any NN is reliant on the hyperparameter settings, such as the number of layers, activation functions, number of neurons in a layer, and learning rate. 
However, there is no way of deriving an optimal set of these parameters for any given problem to provide the best performance.
In fact, tuning these parameters to provide satisfactory performance for a particular problem is not any straightforward process, but rather dependent on empirical speculations. 
This means the training process is not a one-time event, but rather a trial-and-error process that involves multiple attempts to determine the most suitable parameters. 
Moreover, depending on the nature of the problem, it can be challenging to determine possible candidates for the parameters to begin with. 
This results in a very uncertain and time-consuming training process.
For NTNs, these issues are more severe due to their high network complexity resulting in a more complex set of hyperparameters. 

\subsection{Lack of Generalization}

As data-driven ML approaches are used to train ML algorithms, it can be difficult to generalize these algorithms to different scenarios. 
Trained models are able to capture the characteristics of the training data, but this does not always guarantee successful performance with test data due to the varying nature of NTNs. 
A model trained for a specific scenario may not be successful in another, and may not be able to adapt to different NTN scenarios. 
Even if the model has not encountered certain scenario features during training, it is desirable to have a model that is generalizable and performs well in any context. 
Developing such models is one of the biggest challenges of the NTN domain due to their high network complexity.
As there are no theoretical performance bounds for these empirical ML models, unpredictable performance drops can occur while deploying in the real environment.



\vspace{3mm}
\noindent\textit{Key Takeaways: The cost-limited on-board computation, highly dynamic environmental conditions, and long propagation delay introduce a diverse set of challenges to realize the AI-enabled NTN environment for future 6G networks. These challenges need to be addressed with efficient solutions to ensure superior network performances in real NTN deployments. 
}
\section {Insights and \revision{Potential Future Studies}}

\label{sec:insights}

In this comprehensive study, we delve into the realm of NTNs and their relationship with AI techniques, establishing a solid background for our exploration. We explore the synergy between NTNs and AI, highlighting how these two domains intersect and complement each other.
Moving forward, we shift our focus to the current research thrusts in the field, examining ongoing efforts to bring these concepts to fruition in real-world networks. While highlighting these advancements, we also address the potential challenges that must be overcome to realize the full potential of NTNs in the context of future 6G networks.
Within this section, we provide an in-depth discussion of valuable insights and \revision{potentianl future studies} for leveraging various AI techniques in the context of satellite-based NTNs. 

\subsection{\revision{Insights}}

\revision{In this section, we present a summary of the lessons learned and insights gained from our paper's discussion. These insights are intended to serve as valuable guidance and information for the implementation and integration of AI in shaping the future landscape of 6G networks. }

\subsubsection{Existing Learning Approaches}

Upon examining the \revision{contents} presented in \revision{Section} \ref{sec:ai-ntn}, it becomes evident that SL and RL approaches take center stage in addressing the diverse array of challenges faced by satellite-based NTNs in future 6G networks, primarily due to the availability of real-world data and feedback mechanisms within existing networks. 
\revision{In the context of SL, having access to well-labeled data is of utmost importance, especially in scenarios involving estimation problems like channel estimation and Doppler Shift estimation.
On the contrary, RL shines when dealing with problems lacking clear labels but featuring a notion of reward functions. 
Furthermore, RL techniques are extremely suitable and efficient for problems where supervision is lacking which is usually the case for many NTN problems such as resource allocation, beam hopping, and network routing as illustrated in Section \ref{sec:ai-ntn} of the paper. 
Meanwhile, RL techniques can be effectively employed across a wide spectrum of problems using general network feedback, such as CSI, acknowledgments, and more. Consequently, a significant portion of research efforts tends to leverage RL frameworks to address their specific challenges.

\subsubsection{Leveraging Deep Neural Networks}

The emergence of Deep NNs and their effectiveness in addressing intricate challenges in fields like computer vision and natural language processing has piqued the interest of the research community in applying these architectures to network-related issues.
Satellite-based NTNs introduce a unique set of challenges, characterized by highly dynamic network conditions and a multitude of variables influencing network performance. 
Traditional ML) approaches often fall short in comprehensively addressing these complex problems, frequently limited to small-scale issues.
As a result, DL techniques have gained significant popularity within the research community, proving to be a more adept choice for tackling the multifaceted challenges encountered in satellite-based NTNs, as elaborated in Section \ref{sec:ai-ntn}.
}

\subsubsection{Potential Learning approaches}

The nature of UL approaches presents a unique set of challenges in the context of highly dynamic and time-varying NTNs. Understanding and capturing the intrinsic behavioral patterns within such networks prove to be particularly hard. However, it is important to note that UL approaches still hold the potential to derive the distribution of crucial network parameters that may not be readily accessible in real networks. These derived parameters can play a pivotal role in addressing various NTN challenges.
\revision{The distributed learning approaches such as FL can be also beneficial for future satellite-based NTNs as the computing capabilities requirements can be reduced to a minimum enhancing practical feasibility.}

\subsubsection{Enabling O-RAN-based RIC}

Currently, there are some ongoing research efforts focused on developing SDR-based prototypes for NTNs with adaptations to OAI 4G and 5G protocol stacks, as discussed in Section \ref{sec:ntn-oai}. However, to fully unlock the potential of AI in NTN for future 6G networks, the integration of the RIC into these implementations is crucial. This integration is particularly important given that the immense benefits of AI in addressing NTN deployment challenges for future 6G networks are demonstrated in \revision{Section} \ref{sec:ai-ntn} but current 5G networks lack a dedicated interface for applying AI algorithms. By enabling the O-RAN framework with RIC, the deployment of AI algorithms in real NTN networks can be efficiently performed, paving the way for advanced capabilities and improved performance.

\revision{
\subsubsection{Practical Implications}

The cost limitations on onboard computation, the extreme environmental conditions, and the extensive propagation delays form a multifaceted array of challenges when endeavoring to bring AI-driven NTNs to fruition in anticipation of the forthcoming 6G network era, as elaborated in Section \ref{sec:chal}.
These formidable challenges necessitate the development of innovative, resourceful solutions to ensure superior network performance in practical NTN deployments.
Notably, three key factors come to the fore when considering the limitations imposed on AI capabilities for satellites and other NTN platforms: power, bandwidth, security, and physical space.
Advancements in miniaturization, secured system design, energy-efficient design principles, and the judicious utilization of available bandwidth resources serve as the driving forces enabling AI technologies within satellite-based NTNs.

\subsection{Potential Future Studies}

In the preceding sections, we have observed how a multitude of ML and DL approaches has played an important role in shaping the trajectory of future NTN-enabled 6G networks. 
Nonetheless, we have also encountered certain limitations that enforce the requirement for exploring alternative, more efficient methodologies.
Furthermore, the integration of AI into NTNs introduces a set of inherent challenges to be addressed carefully. 
In this section, our attention shifts to these prospective areas of future research, aiming to establish a resilient framework for the forthcoming era of 6G networks powered by AI techniques.

}

\subsubsection{Interrelated Issues}

 \revision{Section} \ref{sec:ai-ntn} sheds light on the interconnected nature of the various issues encountered in NTNs. It is crucial to recognize that addressing a singular problem can serve as an initial step toward resolving larger, more complex challenges inherent in TNTNs. However, when transitioning these solutions into real-world networks, it becomes imperative to acknowledge and account for the intricate interdependencies among various aspects. An illustrative example of such interrelations lies in the dynamic nature of network load status following a user's attachment to a satellite. In this scenario, integrating resource allocation strategies into the handover decisions can yield enhanced network performance. MIMO systems can be also beneficial for the single-user and multi-user cases for NTNs as in LTE-Advanced \cite{mimo1}. By considering the broader context and understanding how different aspects influence one another, we can develop more holistic and effective approaches for real-world NTN implementations.

\subsubsection{Recurrent Learning Architectures}

Presently, the majority of DL algorithms deployed to tackle the time-varying nature of NTNs in beam-hopping, resource allocation, network slicing, etc. rely on feed-forward NNs. While these architectures have proven successful in computer vision applications such as image detection and classification, they may not effectively capture the temporal behavior inherent in these NTN problems.
Unlike feed-forward networks, recurrent architectures possess the ability to capture and process temporal dependencies within the problem domain. By leveraging RNNs or other similar architectures, we can effectively model and solve the corresponding NTN challenges in a more comprehensive and accurate manner.
\revision{In particular, for dynamic spectrum access and sharing approaches low complexity NN architectures such as ESNs can be very useful for NTNs as illustrated in \cite{rc-drl-dsa, deqn-dss}.}

\subsubsection{Online Implementation}

One major limitation of the current works in the domain is the limited consideration given to online implementation and the associated computational complexity when designing algorithms for various control operations in NTNs. This oversight poses a significant hurdle to the practical application of these algorithms in real NTNs as many control decisions in NTN systems must be made in real-time, and the use of complex deep feed-forward NNs becomes impractical. To address this challenge, exploring alternative options becomes imperative. One such option involves investigating low-complexity architectures such as ESNs and ELMs or combining them with traditional feedforward-NNs. These low-complexity architectures offer a more viable solution for online implementation, enabling the deployment of DL algorithms in real NTN networks in a timely and efficient manner.

\subsubsection{Distributed Learning Models}

In the context of integrated satellite-terrestrial networks, the adoption of distributed learning models can significantly enhance scalability. These models involve distributing the training and inference processes of machine learning algorithms across multiple computing nodes, resulting in accelerated computation and improved efficiency. Various distributed approaches, such as data parallelism, model parallelism, ensemble learning, and federated learning, offer promising solutions to address the diverse challenges faced by NTNs in extended network environments \cite{dml}. By leveraging these distributed approaches, NTN systems can effectively harness the power of parallel computing and collaborative learning to overcome constraints and achieve optimal performance.

\subsubsection{Control Feedback Design}

One of the major motivating factors for implementing feedback-based learning, such as RL methods, in NTNs, is the inherent feedback system of the current cellular networks. CSI information is readily available for the BSs which can be helpful in network optimization approaches. However, with the emergence of NTNs, new challenges arise, necessitating the efficient design of feedback mechanisms to minimize the overall overhead while improving network performance. This consideration is crucial, as AI approaches for addressing various issues may require similar types of feedback. The utilization of combined feedback can prove highly beneficial in optimizing network performance and achieving efficient resource allocation, thus enhancing the overall effectiveness of AI algorithms in NTNs.

\subsubsection{Development in Miniaturization}

The limited availability of computational resources currently poses a challenge to the onboard capability of satellites, especially when deploying AI algorithms. However, the miniaturization of satellite components and equipment has emerged as a solution to this issue. By reducing the weight and size of equipment, miniaturization enables the integration of more powerful processors and larger memory devices within the limited space available on satellites. This advancement in computational resources greatly facilitates the deployment of AI algorithms, unlocking new possibilities for satellite applications. Achieving miniaturization in satellite technology requires innovations in material science, efficient Integrated Circuit (IC) design, advancements in IC fabrication technologies, System-on-Chip (SoC) integration, and Micro-Electro-Mechanical Systems (MEMS) design, among others. The development of miniaturization is particularly crucial for NTNs, as it enhances the onboard capability of satellites and enables the realization of advanced technologies and functionalities in space-based systems.

\subsubsection{Energy Efficiency}

The launching and maintenance of satellites require substantial power consumption, which imposes limitations on the onboard capability of satellites. Consequently, efficient energy system design becomes a critical criterion for NTNs. To address this, various aspects need to be considered, including lightweight component design, advanced power management techniques, efficient power conversion, optimized propulsion system design, effective energy storage systems, etc. By focusing on these factors, satellite systems can achieve higher energy efficiency, which is essential for the successful deployment of advanced AI algorithms. The performance of these algorithms relies on the availability of computational resources, making energy efficiency a crucial aspect to maximize the satellite's capabilities within the given power constraints.

\subsubsection{Secured System Design}

As highlighted in \revision{Section} \ref{sec:chal}, security concerns in NTNs can be highly significant, introducing new attack vectors and vulnerabilities. NTNs are susceptible to a range of security attacks, including adversarial attacks, data poisoning, DoS attacks, Fuzzy attacks, MiM attacks, and more. These attacks have the potential to severely impact network performance and compromise the integrity and confidentiality of data. To address these challenges, it is essential to design efficient intrusion detection and prevention systems specifically tailored for secure NTNs. By continuously monitoring a set of relevant network parameters and detecting anomalies in the network's behavioral patterns, mitigation techniques can be promptly deployed to ensure optimal network performance and safeguard against potential degradation caused by security breaches.
\section{Conclusion}

\label{sec:conc}

NTN is considered the driver of ubiquitous, reliable, and scalable 6G wireless networks. It adds new dimensions to the existing traditional terrestrial communication systems by providing connections to remote and isolated areas subject to geographical constraints and offloading the primary links during traffic peaks. However, diverse unique challenges are accompanied by the deployment of NTN in existing communication systems. The long propagation delay, high Doppler effect, spectrum sharing, complicated resource allocation, and fast and frequent handover are the major problems associated with NTN deployment. Integration to existing terrestrial networks presents a set of new problems such as task offloading, network routing, network slicing, etc. to be addressed in an efficient manner. The convergence of AI and NTN allows for the building of sustainable AI-based Non-Terrestrial Networks addressing many of these challenges. Depending on the characteristics of the problem at hand, various learning approaches can be employed. When dealing with prediction and estimation problems, SL techniques appear to be a more suitable choice. On the other hand, for tasks involving closed-loop control, RL techniques show greater promise. By tailoring the learning approach to the specific problem, we can effectively leverage the strengths of each technique and achieve optimal results.

However, the integration of AI into NTNs presents certain challenges that need to be addressed. Both the industry and research community are collaborating to ensure the successful implementation of AI-based NTNs in next-generation wireless networks. This includes the establishment of ML testbeds specifically designed for satellite networks and the adaptation of SDR-based OAI 4G/5G protocol stacks for NTN applications. In order to realize satellite-based NTNs in future 6G networks, several practical challenges must be overcome. These challenges include addressing the constraints of cost-limited onboard capabilities, managing the highly time-varying nature of satellite networks, and mitigating the effects of long propagation delays. It is important to consider these interconnected issues and develop joint solutions to enhance overall network performance. Furthermore, exploring low-complexity and distributed learning architectures that incorporate efficient control feedback mechanisms is essential for enabling real-time, online implementation. Additionally, ensuring the secure, compact, and energy-efficient design of NTN platforms is integral to the successful deployment of satellite-based NTNs in the 6G era.

\bibliography{references/IEEEabrv, references/intro, references/ntn, references/ai, references/research/beamhopping, references/research/handover, references/research/doppler, references/research/spectrum, references/research/resource, references/research/channel, references/research/security, references/research/traffic, references/research/routing, references/research/offloading, references/research/slicing, references/research/multiple_access, references/progress }
\bibliographystyle{IEEEtran}

\end{document}